\documentclass[rmp,aps,twocolumn,showpacs,nofootinbib,nobibnotes,superscriptaddress,secnumarabic]{revtex4-1}

\usepackage{amssymb}
\usepackage{graphicx}
\usepackage{epsfig}
\usepackage{amsmath}
\usepackage{amsfonts}
\usepackage{color}

\begin{document}

\title{Quantum Spin Liquid States}
\date{\today}

\author{Yi Zhou}
\affiliation{Department of Physics and Zhejiang Institute of Modern Physics,
Zhejiang University, Hangzhou, 310027, P. R. China}
\affiliation{Collaborative Innovation Center of Advanced Microstructures, Nanjing 210093,
China}

\author{Kazushi Kanoda}
\affiliation{Department of Applied Physics, University of Tokyo,
Hongo 7-3-1, Bunkyo-ku, Tokyo 113-8656, Japan}

\author{Tai-Kai Ng}
\affiliation{Department of Physics, Hong Kong University of Science and
Technology, Clear Water Bay Road, Kowloon, Hong Kong, China}

\begin{abstract}
   This article is an introductory review of the physics of quantum spin liquid (QSL) states. Quantum magnetism is a rapidly evolving field, and recent developments reveal that the ground states and low-energy physics of frustrated spin systems may develop many exotic behaviors once we leave the regime of semi-classical approaches. The purpose of this article is to introduce these developments. The article begins by explaining how semi-classical approaches fail once quantum mechanics become important and then describes the alternative approaches for addressing the problem. We discuss mainly spin $1/2$ systems, and we spend most of our time in this article on one particular set of plausible spin liquid states in which spins are represented by fermions. These states are spin-singlet states and may be viewed as an extension of Fermi liquid states to Mott insulators, and they are usually classified in the category of so-called $SU(2)$, $U(1)$ or $Z_2$ spin liquid states. We review the basic theory regarding these states and the extensions of these states to include the effect of spin-orbit coupling and to higher spin ($S>1/2$) systems. Two other important approaches with strong influences on the understanding of spin liquid states are also introduced: (i) matrix product states and projected entangled pair states and (ii) the Kitaev honeycomb model. Experimental progress concerning spin liquid states in realistic materials, including anisotropic triangular lattice systems ($\kappa$-(ET)$_{2}$Cu$_{2}$(CN)$_{3}$ and EtMe$_{3}$Sb[(Pd(dmit)$_{2}$]$_{2}$), kagome lattice systems (ZnCu$_{3}$(OH)$_{6}$Cl$_{2}$) and hyperkagome lattice systems (Na$_{4}$Ir$_{3}$O$_{8}$), is reviewed and compared against the corresponding theories.

\end{abstract}


\maketitle

\tableofcontents

\bibliographystyle{apsrmp4-1}

\section{Introduction}
 Quantum spin liquid (QSL) states in dimensions of $d>1$ have been a long-sought dream in condensed matter
 physics. The general idea is that when acting on spin systems, quantum mechanics may lead to exotic ground states and low-energy behaviors that
 cannot be captured by traditional semi-classical approaches. The difficulty in implementing this idea is that we have no natural place to start once we have left the comfort zone of semi-classical approaches, at least in dimensions larger than one. Except for a few exactly solvable models, we must rely heavily on numerical or variational approaches to ``guess" the correct ground state wavefunctions and on a
 combination of sophisticated numerical and analytical techniques to understand the corresponding low-energy excitations.

  Several excellent reviews are available on QSLs \cite{Lee08,Balents10} and frustrated magnetism \cite{BookDiep,BookLacroix}.
  This article complements those mentioned above by providing a pedagogical introduction to this subject and reviews the current status of the field. We explain, at an introductory level, why sophisticated approaches are needed to study QSL states, how these approaches are implemented in practice, and what new physics may be expected to appear. The experimental side of the story and the drawbacks or pitfalls of the theoretical approaches are also discussed. We
 concentrate mainly on spin $1/2$ systems and study in detail one particular set of plausible spin liquid states that are usually termed resonating valence bond (RVB) states. The spins are treated as fermions in these states, which may be viewed as an extension of Fermi liquid states to Mott insulators. They are usually classified in the category of $SU(2)$, $U(1)$ or $Z_2$ spin liquid states. Because of the intrinsic limitations of the fermionic RVB approach, many other approaches to spin liquid states have been developed by different authors. These approaches often lead to other exotic possibilities not covered by the simple fermionic approach. Two of these approaches are introduced in this article for completeness: (i) matrix product states and projected entangled pair states and (ii) the Kitaev honeycomb model.

  The article is organized as follows. In section II, we introduce the semi-classical approach to simple quantum antiferromagnets,
  and we explain the importance of the spin Berry phase and how one can include it in a semi-classical description to obtain the correct theory. In particular, we show how it leads to the celebrated Haldane conjecture. The existence of end excitations as a natural consequence of the
  low-energy effective theory of these systems is discussed. One-dimensional quantum spin systems are of great interest at present
  because they provide some of the simplest realizations of symmetry-protected topological (SPT) phases in strongly correlated systems.

  The limitations of the semi-classical approach when applied to systems with frustrated interactions are discussed in section III,
  where we introduce the alternative idea of constructing variational wavefunctions directly. We introduce Anderson's famous idea of the
  RVB wavefunction for spin $1/2$ systems and discuss how this idea can be implemented in practice. The difficulty of incorporating the $SU(2)$ spin algebra in the usual many-body perturbation theory is noted, and the trick of representing spins by particles (fermions or bosons) with {\em constraints} to avoid this difficulty is introduced. The non-trivial $SU(2)$ gauge structure in the fermion representation of RVB states and the resulting rich structure of the low-energy effective field theories for these spin states ($SU(2)$, $U(1)$ and $Z_2$ spin liquids) are discussed. An interesting linkage of the $U(1)$ spin liquid state to the (metallic) Fermi liquid state through a Mott metal-insulator transition is introduced.

  The difficulty of finding controllable approaches for studying spin liquid states has led to an extension of the RVB approach and a search for alternative approaches. Some of these approaches are reviewed briefly in section IV, including (i) the extension of the RVB approach to include the effect of spin-orbit coupling and to higher spin ($S>1/2$) systems, (ii) the concepts of matrix product states and projected entangled pair states, and (iii) the Kitaev honeycomb model. The main message of this section is that a larger variety of exotic spin
 states become possible when we leave the paradigm of spin $1/2$ systems with rotational symmetry. The $U(1)$ and $Z_2$ spin liquid states belong to merely a very small subset of the plausible exotic states once we leave the paradigm of semi-classical approaches.

  Section V is devoted to a survey of experimental research on spin liquid states. Special attention is paid to the $U(1)$ spin liquid state, on which most
  experimental efforts have been focused. The best studied examples are a family of organic compounds denoted by $\kappa$-(ET)$_{2}$Cu$_{2}$(CN)$_{3}$
  (ET) \cite{Kanoda03} and Pd(dmit)$_{2}$(EtMe$_{3}$Sb) (dmit salts) \cite{Itou08}. Both materials are Mott insulators
  near the metal-insulator transition and become superconducting (ET) or metallic (dmit) under modest pressure. Despite the large magnetic exchange $J\approx 250$~K observed in these systems, there is no experimental indication of long-range magnetic ordering down to a temperature of $\sim30$ mK. A linear temperature dependence of the specific heat and a Pauli-like spin susceptibility have been found in both materials at
 low temperature, suggesting that the low-energy excitations are spin-$1/2$ fermions with a Fermi surface \cite{SYamashita,Matsuda12}. This Fermi-liquid-like behavior is further supported by their Wilson ratios, which are close to one. In addition to ET and dmit salts, the kagome compound
 ZnCu$_{3}$(OH)$_{6}$Cl$_{2}$ \cite{Kagome07} and the three-dimensional hyperkagome material Na$_{4}$Ir$_{3}$O$_{8}$ \cite{Na4Ir3O8}
 are also considered to be candidates for QSLs with gapless excitations. Experimental surveys on these QSL candidate materials are presented in this article, including their thermodynamics, thermal transport and various spin spectra.
 We also briefly introduce the discoveries of a few new materials and discuss the existing discrepancies between experiments and theories.
 The paper is summarized in section VI.

 \section{From semi-classical to non-linear-$\sigma$ model
 approaches for quantum antiferromagnets}

  Here, we consider simple Heisenberg antiferromagnets on bipartite lattices (with sublattices $A$ and $B$) with the Hamiltonian
  \begin{equation}
  \label{hm1}
  H=J\sum_{\langle i,j\rangle}\mathbf{S}_i\cdot\mathbf{S}_j,
  \end{equation}
  where $J>0$ and $\langle i,j\rangle$ describes a pair of nearest neighbor sites in the bipartite lattice. In a bipartite lattice, any two nearest neighbor sites
  always belong to different sublattices. $\mathbf{S}$ is a quantum spin with
  magnitude $S=n/2$, where $n=$ positive integer. Examples of bipartite lattices include 1D spin chains, 2D square or
  honeycomb lattices, and 3D cubic lattices.

  \subsection{Two-spin problem}
   The semi-classical approach begins with the assumption that the quantum spins are ``close" to classical spins, and it is helpful to start by first
   analyzing the corresponding classical spin problem. For simplicity, we start by considering only two classical spins
   coupled by the Heisenberg interaction
   \[
      H=J\mathbf{S}_A\cdot\mathbf{S}_B. \\\\\ (J>0).   \]
    The classical spins obey Euler's equation of motion:
   \begin{equation}
   \label{eqm1}
   {\partial\mathbf{S}_{A(B)}\over\partial t}=J\mathbf{S}_{B(A)}\times\mathbf{S}_{A(B)}.
   \end{equation}

    This equation can be solved most easily by introducing the magnetization and
   staggered magnetism vectors $\mathbf{M}(\mathbf{N})=\mathbf{S}_A+(-)\mathbf{S}_B$,
   where it is easy to show from Eq.~\ (\ref{eqm1}) that
   \begin{eqnarray}
   \label{eqm3}
   {\partial \mathbf{M}\over\partial t} & = & 0,  \\ \nonumber
   {\partial\mathbf{N}\over\partial t} & = & J\mathbf{M}\times\mathbf{N},
   \end{eqnarray}
  indicating that classically, the staggered magnetization vector $\mathbf{N}$ rotates around the (constant) total magnetization vector $\mathbf{M}$. Let
   $\mathbf{S}_{A(B)}=S_{A(B)}\hat{r}_{A(B)}$, where $S_{A(B)}$ are the magnitudes of the spins $\mathbf{S}_{A(B)}$ and
  $\hat{r}_{A(B)}$ are unit vectors indicating the directions of $\mathbf{S}_{A(B)}$; then, the classical
  ground state has $\hat{r}_A=-\hat{r}_B$ with $\mathbf{M}=0$, i.e., the two spins are antiferromagnetically aligned. Note that the equation of
  motion given in Eq.~\eqref{eqm3} implies that ${\partial(\mathbf{N}^2)\over\partial t}=0$, i.e., the magnitude of $\mathbf{N}$ remains unchanged during its motion.
  Therefore, if we write $\mathbf{N}=N\hat{n}$, where $N$ is the magnitude of $\mathbf{N}$ and $\hat{n}$ is the unit vector denoting its
  direction, we find that only $\hat{n}$ changes under the equation of motion given in Eq.~\eqref{eqm3}.

   The effects of quantum mechanics can be seen most easily by observing that the equation of motion given in Eq.~\eqref{eqm3} describes the dynamics of a free rotor
   (a rigid rod with one end fixed such that the rod can rotate freely around the fixed end). A free rotor can be represented by a vector
   $\mathbf{r}=r_0\hat{r}$, where $r_0=$ constant is the length of the rod and $\hat{r}$ is
   the unit radial vector describing the orientation of the rod. The rod has an angular momentum of
   \begin{equation}
   \label{rotor1}
   \mathbf{L}=\mathbf{r}\times\mathbf{p}=r_0\hat{r}\times\mathbf{p},
   \end{equation}
   where $\mathbf{p}=mr_0\dot{\hat{r}}$ is the momentum and $m$ is the
   mass. Using Eq.~\eqref{rotor1}, we obtain
   \begin{subequations}
   \label{rotor2}
   \begin{eqnarray}
   \label{rotor2a}
   \hat{r}\times\mathbf{L}=-r_0\mathbf{p}=-mr_0^2\dot{\hat{r}}.
   \end{eqnarray}
    We also have
    \begin{equation}
    \label{rotor2b}
    \dot{\mathbf{L}}=0
    \end{equation}
    (conservation of angular momentum).
    \end{subequations}
    Comparing Eqs.\eqref{eqm3} and \eqref{rotor2}, we find that the equation of motion for two spins is equivalent to the equation of motion for a free rotor if we identify $\mathbf{L}\rightarrow\mathbf{M}$, $\hat{r}\rightarrow\hat{n}$ and $J=I^{-1}$, where
    $I=mr_0^2$ is the moment of inertia of the rotor.

    The quantum Hamiltonian of the free rotor is
    \[
       H_{rotor}={1\over2 I}\mathbf{L}^2, \]
    and its solution is well known. The eigenstates are the spherical
    harmonics $Y_{lm}(\theta,\phi)$ (where $\theta$ and $\phi$ specify the direction of the unit vector $\hat{r}$) with eigenvalues
    \[
     \mathbf{L}^2=l(l+1)\hbar^2,\\\\\\\\\\\  L_z=m\hbar,  \]
    and corresponding energies $E_l=l(l+1)\hbar^2/2I$, where $l$ and $m$ are integers such that $l\geq0$ and
    $l\geq|m|$. In particular, $\mathbf{L}(\mathbf{M})=0$ for the ground state of the quantum rotor, but the direction of the vector
    $\mathbf{r}(\mathbf{N})$ is completely uncertain ($Y_{00}(\theta,\phi)={1\over\sqrt{4\pi}}$) as a result of quantum fluctuations, indicating a breakdown of
    the classical solution, in which $\mathbf{n}$ is fixed in the ground state.
    (Alternatively, one can gain this understanding from the Heisenberg uncertainty principle, $\langle\delta\hat{r}\rangle\langle\delta\mathbf{L}\rangle>\hbar$.
     With $\mathbf{L}=0$ in the ground state, $\delta\mathbf{L}\equiv0$ and $\delta\hat{r}\rightarrow\infty$, the
    direction of the vector $\hat{r}$ becomes completely uncertain.)

     A moment of thought indicates that our mapping of the spin problem to the rotor problem cannot be totally correct. What happens
     if $\mathbf{S}_A$ is an integer spin and $\mathbf{S}_B$ is a half-odd-integer spin? Elementary
     quantum mechanics tells us that the ground state should carry half-odd-integer angular momentum. The possibility of such a scenario is missing in our
     rotor mapping, in which the spin magnitudes $S_{A(B)}$ do not appear.

    \subsection{Berry's phase}
      The missing piece in our mapping of the two-spin problem to the rotor model is the Berry's phase \cite{Berry84}, which is carried by spins but is
      absent in rotors. The correct spin-quantization rule is recovered only after this piece of physics is properly added
      into the rotor problem. First, let us review the Berry's phase carried by a single spin.

\begin{figure}[tbph]
    \includegraphics[width=4.0cm]{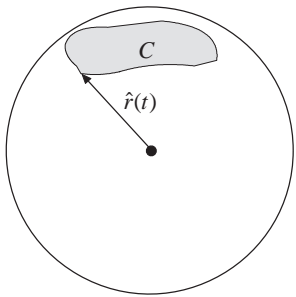}
    \caption{ Berry's phase with a magnetic monopole.}
    \label{fig:Berry}
    \end{figure}

      We recall that for a spin tracing out a closed path $\mathbf{C}$ on the surface of the unit sphere, the spin wavefunction acquires a Berry's phase
      $\gamma(\mathbf{C})=S\Omega(\mathbf{C})$, where $S$ is the spin magnitude and $\Omega(\mathbf{C})$ is the surface area under the closed
      path $\mathbf{C}$ on the unit sphere (see Fig. \ref{fig:Berry}). $S\Omega(\mathbf{C})$ can be represented more conveniently by imagining the spin trajectory as the trajectory of a particle carrying a unit charge moving on the surface of the unit sphere. In this case, the Berry's phase is simply the phase acquired by the charged particle if a magnetic monopole of strength $S$ (i.e., $\mathbf{B}(\mathbf{r})=(S/r^2)\hat{r}$) is placed at the center of the sphere. The Berry's phase acquired is the magnetic flux enclosed by the closed path $\mathbf{C}$.

      Let $S\mathbf{A}_M(\mathbf{r})$ be the vector potential associated with the monopole, i.e., $\nabla\times\mathbf{A}_M=\hat{r}/r^2$; then, in the
      ``charge + gauge field" representation, the effect of the Berry's phase can be described by a vector-potential term in the action:
      \begin{equation}
      \label{WZ1}
         S_{B}=\hbar S\Omega(C)=\hbar S\int dt\mathbf{A}_M(\hat{r})\cdot\dot{\hat{r}}.
      \end{equation}
         This is an example of a Wess-Zumino term for quantum particles. A more rigorous derivation of the Wess-Zumino action is given in Appendix A, where the action for a single spin in a magnetic field is derived {\em via} a path integral approach.

      We now revisit the two-spin problem. With the Berry's phases included, the Lagrangian of the corresponding rotor problem becomes
     \begin{equation}
     \label{La2}
      L={1\over2J}(\hat{n}\times\dot{\hat{n}})^2+\hbar S_A\mathbf{A}_{M}(\hat{r}_A)\cdot\dot{\hat{r}}_A+\hbar S_B\mathbf{A}_{M}(\hat{r}_B)\cdot\dot{\hat{r}}_B,
     \end{equation}
     where $\mathbf{N}=N\hat{n}=\mathbf{S}_A-\mathbf{S}_B$. To simplify the problem, we adopt the semi-classical approximation
     $\hat{r}_A=-\hat{r}_B$ in the Berry's phase terms, which is a reasonable approximation for states close to the classical ground
     state. With this approximation, we obtain
      \begin{equation}
     \label{La2a}
      L\rightarrow{1\over2J}(\hat{n}\times\dot{\hat{n}})^2+\hbar\Delta S\mathbf{A}_M(\hat{n})\cdot\dot{\hat{n}},
     \end{equation}
     where $\hat{n}=\hat{r}_A$ and $\Delta S=S_A-S_B$. The Hamiltonian of the system is
     \begin{equation}
      \label{h2}
        H_M={J\over2}\left(\mathbf{\Pi}-\hbar\Delta S\mathbf{A}_M(\hat{n})\right)^2,
      \end{equation}
      where
      $\mathbf{\Pi}=\dot{\hat{n}}/J$ is the canonical momentum of the rotor.

      $H_M$ is the Hamiltonian of a charged particle moving on the surface of a unit sphere with a magnetic monopole of strength $|\Delta S|$ located at the center of the sphere. The eigenstates of the Hamiltonian are well known and are called the monopole spherical
      harmonics \cite{Wu76}. The most interesting feature of the monopole spherical harmonics is that they allow half-odd-integer angular momentum states (which occur when $|\Delta S|$ is a half-odd-integer). The ground state carries an angular momentum of $L=|\Delta S|$ and is $(2|\Delta S|+1)$-fold degenerate, corresponding to the degeneracy of a quantum spin of magnitude $|\Delta S|$, in agreement with the exact result for the two-spin problem.

      \subsection{Non-linear-$\sigma$-model}
       The two-spin problem tells us that there are two important elements that we must keep track of when a classical spin problem is
       replaced with the corresponding quantum spin problem: a) quantum fluctuations, originating from the (non)-commutation relation between
       the canonical coordinates ($\mathbf{N}$) and momenta ($\mathbf{M}$), and b) Berry's phase, which dictates the quantization of the
       spins.  In the following, we generalize the rotor approach to the many-spin systems described by the antiferromagnetic (AFM) Heisenberg
       model, keeping in mind the above two elements.

       Following Haldane \cite{Haldane83PLA,Haldane83PRL}, we here consider Heisenberg antiferromagnets on a bipartite lattice described by the Hamiltonian given in
       Eq.~\eqref{hm1}. As in the two-spin problem, we introduce the magnetization vectors $\mathbf{M}(\mathbf{x}_i)$ and the staggered magnetization vectors
        $\mathbf{N}(\mathbf{x}_i)$ such that
        \begin{eqnarray}
        \label{sab}
        \mathbf{S}^A_i & = &  \mathbf{M}(\mathbf{x}_i)+\mathbf{N}(\mathbf{x}_i), \\  \nonumber
        \mathbf{S}^B_i & = &  \mathbf{M}(\mathbf{x}_i)-\mathbf{N}(\mathbf{x}_i),
        \end{eqnarray}
        where $\mathbf{S}^{A(B)}$ denote spins on the $A(B)$ sublattices of the bipartite lattice. We assume that the ground state of the
        quantum system is ``classical-like" with nearly anti-parallel spins on two nearest neighboring sites such that $\mathbf{M}(\mathbf{x}_i)\ll
        \mathbf{N}(\mathbf{x}_i)$, where both $\mathbf{M}(\mathbf{x})$ and
        $\mathbf{N}(\mathbf{x})$ are very slowly varying functions in space. (We show that this assumption can be justified in the
        following section.) The classical equation of motion for the spin at lattice site $i$ is
        \begin{equation}
        \label{eqma}
       {\partial\mathbf{S}^{A(B)}_i\over\partial t}=J\left(\sum_{j=\text{NN}(i)}\mathbf{S}^{B(A)}_j\right)\times\mathbf{S}^{A(B)}_i,
       \end{equation}
       where $j=\text{NN}(i)$ means that $j$ represents the nearest neighbor sites of $i$.

     Using Eq.~\eqref{sab}, after some straightforward algebra and taking the continuum limit, we obtain
    \begin{eqnarray}
    \label{eqmb}
     {\partial\mathbf{N}(\mathbf{x})\over\partial
     t} & \sim & Jz\mathbf{M}(\mathbf{x})\times\mathbf{N}(\mathbf{x}),  \\ \nonumber
     {\partial\mathbf{M}(\mathbf{x})\over\partial t} & \sim &
     -{Ja^2\over2}(\nabla^2\mathbf{N}(\mathbf{x}))\times\mathbf{N}(\mathbf{x}),
     \end{eqnarray}
     where $a$ is the lattice spacing and $z=2d$ is the coordination number. We have assumed a square (cubic)-type lattice and have adopted the slowly varying
     approximation
     \[
         \mathbf{M}(x_{i+1})+\mathbf{M}(x_{i-1})\sim2\mathbf{M}(x_i),
         \]
     \[
         \mathbf{N}(x_{i+1}))+\mathbf{N}(x_{i-1})\sim2\mathbf{N}(x_i)+a^2\partial^2_x\mathbf{N}(x_i),
         \]
         etc. in deriving the above result. We have also assumed $\mathbf{M}(\mathbf{x})$ to be small and have neglected all non-linear terms in
           $\mathbf{M}(\mathbf{x})$ in deriving Eq.~\eqref{eqmb}.

         To proceed further, we consider the situation in which all spins have the same magnitude  $S$. Then, it is easy to see from Eq.~\eqref{sab} that
         $\mathbf{N}(\mathbf{x})^2+\mathbf{M}(\mathbf{x})^2=S^2$ and $\mathbf{N}(\mathbf{x})\cdot\mathbf{M}(\mathbf{x})=0$.
         Assuming that $M=|\mathbf{M}(\mathbf{x})|\ll N=|\mathbf{N}(\mathbf{x})|\sim S$, we find from Eq.~\eqref{eqmb} that
         $M\sim\omega/(zJ)$ and $\omega\sim\sqrt{z}JaS|\mathbf{k}|$, where $\omega$ and $\mathbf{k}$ are the frequency and wavevector, respectively, of the
         fluctuations in $\mathbf{N}$. In particular, $M\ll N$ when $ak\ll \sqrt{z}$, i.e., when $\mathbf{N}(\mathbf{x})$ is slowly varying in space.

         In the following, we adopt the approximation $N\sim S$ and write $\mathbf{N}(\mathbf{x})=S\hat{n}(\mathbf{x})$, where
         $\hat{n}^2=1$. Eliminating $\mathbf{M}(\mathbf{x})$ from Eq.~\eqref{eqmb}, we obtain
         \begin{subequations}
     \begin{equation}
        \label{nlsm1}
       {\partial^2\hat{n}(\mathbf{x},t)\over\partial t^2}={z(SJa)^2\over2}\nabla^2\hat{n}(\mathbf{x},t),
    \end{equation}
       corresponding to the following classical action for the vector field $\hat{n}$:
     \begin{equation}
        \label{nlsm2}
      S_{\sigma}={1\over2}\int dt\int d^dx \left({1\over J}\left({\partial\hat{n}\over\partial t}\right)^2-{zJ(Sa)^2\over2}(\nabla\hat{n})^2\right),
    \end{equation}
    with the constraint $\hat{n}^2=1$.
    \end{subequations}
      $S_{\sigma}$ is the non-linear-$\sigma$ model (NL$\sigma$M) for the unit vector field $\hat{n}(\mathbf{x})$.

      Comparing Eqs.\eqref{nlsm2} and \eqref{La2a}, we see that the NL$\sigma$M can be viewed as a continuum model describing coupled rotors $\hat{n}(\mathbf{x})$. The first term in the action gives the
      kinetic energy for the rotors, which we have discussed in detail for the two-spin model. The second term represents the coupling between nearest neighboring
      rotors in the lattice spin model. We note that the term for the coupling between rotors has a magnitude of $\sim S^2$ and dominates over the kinetic energy in the limit of large $S$.

       A more systematic derivation of the NL$\sigma$M starting from Eq.~\eqref{sab} can be achieved by writing
      \begin{equation*}
         \mathbf{S}_i=\eta_i S \hat{n}(x_i)\sqrt{1-\left|\frac{\mathbf{M}(x_i)}{S}\right|^2}+\mathbf{M}(x_i),
      \end{equation*}
     where $\eta_i=e^{i\pi x}$ and we still have $\mathbf{N}(\mathbf{x})\cdot\mathbf{M}(\mathbf{x})=0$. Assuming that $\mathbf{M}(\mathbf{x})$ is small, we can integrate out $\mathbf{M}(\mathbf{x})$ in a power series expansion of $\mathbf{M}(\mathbf{x})$ in the path integral. The NL$\sigma$M for $\hat{n}(\mathbf{x})$ is thus obtained to the leading (Gaussian) order \cite{Auerbach94}.

    \subsubsection{Topological term}
      We next consider the Berry's phase contribution to the action. Following Appendix A, the total Berry's phase contribution is
    \begin{equation}
      \label{WZ2}
         S_{T}=\sum_iS_B(\hat{r}_i)\sim \hbar S\sum_i(-1)^i\Omega({\hat{n}_i}),
      \end{equation}
    where $S\Omega(\hat{r}_i)=S\int dt\mathbf{A}_M(\hat{r}_i)\cdot \dot{\hat{r}}_i$ is the Berry's phase for a single spin and $(-1)^i=1(-1)$ for sites on even (odd)
    sublattices. In the last step, we have assumed that the spins are almost anti-parallel. In the continuum limit, we obtain
     \begin{equation}
      \label{WZ3}
         S_{T}\sim {\hbar S\over2^d}\int d^dx\left({\partial\over\partial_x^1}\cdots{\partial\over\partial_x^d}\right) \Omega(\hat{n}(\mathbf{x})).
      \end{equation}
      $S_T$ is sensitive to the boundary conditions (see the discussion below), and we assume closed (periodic) boundary conditions in the
      following. The case of open boundary conditions is discussed afterward. For periodic boundary conditions, it is easy to see that $S_T$ is zero unless
      the integrand has a non-trivial topological structure.

       To evaluate $\partial_x\Omega$, we recall that $\Omega(\hat{n})$ measures the area on the surface of the sphere bounded by the trajectory
     $\hat{n}(t)$. Thus, the variation $\delta \Omega(\hat{n})$ due to a small variation in the trajectory $\delta\hat{n}$
     is simply
     \[
       \delta\Omega(\hat{n})=\int dt \delta\hat{n}\cdot(\hat{n}\times\partial_t\hat{n}),
       \]
     and
   \begin{equation}
      \label{WZ4}
         S_{T}={\hbar S\over2^d}\int d^dx\int dt\left[\left({\partial\over\partial_x^1}\cdots{\partial\over\partial_x^d}\right)\hat{n}\right]\cdot(\hat{n}\times\partial_t\hat{n}).
      \end{equation}
     The total effective action describing the quantum antiferromagnet is $S=S_{\sigma}+S_T$.

     The topological term is nonzero in one dimension and is usually written in the form
    \begin{subequations}
    \begin{equation}
      \label{topo1}
         {S_{T}\over\hbar}={\theta\over8\pi}\sum_{\mu,\nu=0,1}\int d^2x\varepsilon_{\mu\nu}\hat{n}\cdot(\partial_{\mu}\hat{n}\times\partial_{\nu}\hat{n}),
      \end{equation}
      where $x_0=t$, $x_1=x$, $\theta=2\pi S$ and $\varepsilon_{\mu\nu}$ is the rank-2 Levi-Civita antisymmetric tensor \cite{Haldane85,Affleck85}.
    The Pontryagin index
    \begin{equation}
      \label{topo2}
         Q={1\over8\pi}\sum_{\mu,\nu=0,1}\int d^2x\varepsilon_{\mu\nu}\hat{n}\cdot(\partial_{\mu}\hat{n}\times\partial_{\nu}\hat{n})= integer
      \end{equation}
   \end{subequations}
    measures how many times the 2[=1(space)+1(time)]-dimensional spin configuration $\hat{n}$ has wrapped around the unit sphere. In two dimensions,
  \[
    S_T\rightarrow{\hbar\theta\over2}\int dy {\partial Q(y)\over\partial y}=0,  \]
  where $Q(y)$ is the Pontryagin index that arises from summing over all spin configurations in the $y^{th}$ column of the two-dimensional lattice.
  The sum is zero for smooth spin configurations because $Q$ is an integer and thus cannot ``change smoothly" \cite{Haldane88-NLSM,Wen88,Fradkin88,Dombre88}.
  For the same reason, $S_T$ vanishes for any number of dimensions greater than one.
  However, one should be cautioned that this conclusion is valid only when we restrict ourselves to smooth spin configurations $\hat{n}(\mathbf{x},t)$
  when computing $S_T$. The Berry's phase may have a nonzero contribution if we also allow singular spin configurations in the theory. This is the case in
  $2+1$D, where monopole-like spin configurations are allowed in 3D space \cite{Haldane88-NLSM,ReadSachdev90}.

\subsection{Quantum spin chains and the Haldane conjecture}
  We now study the predictions of the effective action for quantum spin chains. In one dimension, the quantum spin chains are described by the path integral
 \[
   \int D[\hat{n}(x,t)]e^{{i\over\hbar}(S_{\sigma}(\hat{n})+S_T(\hat{n}))}.  \]

  We first consider the topological term. We note that $S_T=2\hbar\pi SQ$ and $e^{{i\over\hbar}S_T}=(-1)^{2SQ}$ ($Q=$ integer). In particular,
  $e^{{i\over\hbar}S_T}\equiv 1$ for integer spin chains, and the Berry's phase has no effect on the effective action. However,
 $e^{{i\over\hbar}S_T}=\pm1$ for half-odd-integer spin chains, depending on whether $Q$ is even or odd. There is no further distinction between spin chains
 with different spin values $S$ in $S_T$. This result leads to the first part of the Haldane conjecture, namely, that fundamental differences exist between
 integer and half-odd-integer spin chains \cite{Haldane88-NLSM}. To proceed further, we first consider integer spin chains, where $e^{{i\over\hbar}S_T}\equiv 1$ and the system
  is described by the ``pure" NL$\sigma$M $S_{\sigma}$.

 \subsubsection{Integer spin chains}
  We start by asking the following question: what are the plausible ground states described by $S_{\sigma}$? For this purpose, it is more convenient to consider a
 lattice version of $S_{\sigma}$:
  \begin{equation}
   \label{nlsm3}
    S_{\sigma}\rightarrow{1\over2}\int dt\sum_i \left({1\over J}\left({\partial\hat{n}_i\over\partial
 t}\right)^2+JS^2\hat{n}_i\cdot\hat{n}_{i+1}\right),
    \end{equation}
   with the corresponding Hamiltonian
  \begin{equation}
   \label{hrotor}
    H_{\sigma}={J\over2}\sum_i\left((\mathbf{L}_i)^2-S^2\hat{n}_i\cdot\hat{n}_{i+1}\right),
    \end{equation}
   where $\mathbf{L}_i$ is the angular momentum operator for the $i^{th}$ rotor. The Hamiltonian contains two competing terms, and we expect that it may describe
   two plausible phases, a strong coupling phase, in which the kinetic energy (first) term dominates, and a weak coupling phase, in which
  the potential energy (second) term dominates. A natural control parameter for this analysis is the spin magnitude $S$, which dictates the magnitude of the
  potential energy. In the first case (small $S$), in which the potential energy term is small, we expect that the ground state can be viewed, to a first
  approximation, as a product of local spin singlets, i.e.,
  $\mathbf{L}=0$ states,
  \[ |G\rangle=|0\rangle_1|0\rangle_2\cdots|0\rangle_N, \]
  where $|0\rangle_i$ represents the $\mathbf{L}=0$ state for the rotor on site $i$.
  The lowest-energy excitations are $\mathbf{L}=1$ states separated from the ground state by an excitation gap $\sim\hbar^2J$. This picture is believed to
  be correct as long as the magnitude of the potential energy term is much smaller than the excitation energy for the $\mathbf{L}=1$ state.
 In the second case, in which the potential energy term dominates (large $S$), we expect that the ground state is a magnetically ordered (N\'{e}el state) with
 $\hat{n}_i=\hat{n}_0$ at all sites $i$, where the excitations are Goldstone modes of the ordered state (spin waves).

  It turns out that this naive expectation is valid only in dimensions of $d>1$. In one dimension, the magnetically ordered state is not stable because of
  quantum fluctuations associated with the Goldstone mode (Mermin-Wigner-Hohenberg Theorem), and the ground state is always quantum disordered \cite{Mermin66,Hohenberg67}, i.e., a
  spin liquid state.  This result can be shown
  more rigorously through a renormalization group (RG) analysis of the NL$\sigma$M. We do not go through this analysis in detail in this article; instead, we simply assume
  that this is the case and examine its consequences. Readers interested in the RG analysis can consult, for example, references \cite{Polyakov75,Polyakov87,Brezin76}.

 Physically, this result means that after some renormalization, the ground state of integer spin chains can always be viewed as a product state of local spin
  singlets, irrespective of the spin magnitude $S$. The lowest-energy excitations are gapped spin triplet ($\mathbf{L}=1$) excitations. This is the Haldane
  conjecture for integer spin chains.

 \subsubsection{Half-odd-integer spin chains}
    The RG analysis cannot be straightforwardly applied to half-odd-integer spin chains because of the appearance of the topological term
 $S_T$. To understand why, let us again take the RG to the strong coupling limit and examine what happens in this case.

   To zeroth order, the Hamiltonian of the system consists only of the kinetic energy term. However, the rotors are moving under the influence of
   effective monopole potentials originating from $S_T$. In particular, all half-odd-integer spin chains have the same $S_T$
  with an effective magnetic monopole strength of $1/2$, corresponding to that of a spin-$1/2$ chain. In this case, the ground state of a single rotor has an angular momentum of $\mathbf{L}=1/2$ and is two-fold degenerate (see the discussion after Eq.~\eqref{h2}). The total degeneracy of the ground state is $2^N$, where $N$=number of lattice sites. This enormous degeneracy implies that the coupling between rotors cannot be neglected
  when we consider the rotor Hamiltonian given in Eq.~\eqref{hrotor}, and the strong coupling expansion simply tells us that the system behaves like a coupled-spin-$1/2$ chain \cite{Read90}.

  Fortunately, the antiferromagnetic spin-$1/2$ chain can be solved using the exact Bethe ansatz technique \cite{Giamarchi03}. The exact
  Bethe ansatz solution tells us that the antiferromagnetic spin-$1/2$ Heisenberg chain is critical, namely, the ground state has no long-range magnetic order but has a gapless excitation spectrum. Unlike integer spin chains, where the lowest-energy excitations carry spin $S=1$, the elementary excitation of this system
  has spin $S=1/2$. Combining this with the continuum theory leads to the Haldane conjecture for half-odd integer spin chains, namely, that they are all critical
  with elementary $S=1/2$ excitations.

 \subsubsection{Open spin chains and end states}
   The Haldane conjecture has been checked numerically for quantum spin chains with different spin magnitudes and has been found to be correct in all cases that
 have been studied thus far. One may wonder whether the difference in spin magnitudes may manifest at all in some low-energy properties of quantum spin chains. The answer is yes, when we consider open spin chains.

   Recall that we have always assumed periodic boundary conditions in deriving $S_T$. In fact, a periodic boundary condition is needed
  to define the Pontryagin index for the topological term $S_T$. For an open chain of length $L$, $S_T$ is replaced by \cite{Haldane83PLA,Affleck90,Ng94}
   \begin{eqnarray}
      \label{WZo}
         S_{T}^{(o)} & = &{\hbar\over2}\int^L_0 d^dx{\partial S_B(\hat{n}(x))\over \partial x}  \\ \nonumber
         & = & 2\pi\hbar SQ+{\hbar S\over2}\left(\Omega(\hat{n}(L))-\Omega(\hat{n}(0))\right),
      \end{eqnarray}
    where $2\pi SQ=\theta Q$ is the usual topological $\theta$ term that we obtain when $\Omega(\hat{n}(0))=\Omega(\hat{n}(L))$, i.e., when we consider periodic boundary
  conditions. An open chain differs from a closed chain in the existence of an additional boundary Berry's phase term with an effective spin magnitude of $S/2$.

  We now examine the effect of this additional Berry's phase term. First, we consider integer spin chains. Following the previous discussion, we expect the
 spin chain to be described by the strong coupling limit of the effective Hamiltonian given in Eq.~\eqref{hrotor}, except that the rotors at the two ends of the spin
 chain are subjected to monopole potentials of strength $S/2$, resulting in effective free spins of magnitude $S/2$ located at the ends of the spin chain.
 The two spins are coupled by a term $J_{eff}\sim JS^2e^{-L/\xi}$ when the coupling between rotors is considered, where $\xi\sim E^{-1}_g$ is the
 correlation length and $E_g$ is the spin gap. These end states can also be understood based on a wavefunction proposed by Affleck, Lieb, Kennedy and Tasaki (the AKLT state)
 for $S=1$ spin chains \cite{AKLT} (see section IV) and have been observed experimentally in $S=1$ spin chain materials \cite{s1end}. In modern terminology, the end states of
 integer spin chains are a manifestation of symmetry-protected topological (SPT) order \cite{Gu09,Chen12,Pollmann12},
 which manifests itself as a boundary action that is protected by rotational (SO(3)) symmetry.\footnote{For $S=1$ chains, the $S=1/2$ end states are protected by a weaker $Z_2\times Z_2$ symmetry \cite{Chen11,Chen11a}.}

  For half-odd-integer spin chains, the analysis is a bit more complicated. We start by rewriting Eq.~\eqref{WZo} for $S_T^{(o)}$ as follows \cite{Ng94}:
  \begin{eqnarray}
      \label{WZo1}
         S_{T}^{(o)} & = &  {\hbar\over2}\left(4\pi{1\over2}Q+S\left(\Omega(\hat{n}(L))-\Omega(\hat{n}(0))\right)\right)  \\ \nonumber
      & = & {\hbar\over2}\left(4\pi{1\over2}Q+{1\over2}\left(\Omega(\hat{n}(L))-\Omega(\hat{n}(0))\right)\right.  \\ \nonumber
        & & +\left.(S-{1\over2})\left(\Omega(\hat{n}(L))-\Omega(\hat{n}(0))\right)\right)
    \end{eqnarray}
   where we have replaced $S$ with $1/2$ in the usual topological (Pontryagin index) term and have divided the boundary Berry's phase term into two parts;
   the first part, when combined with the Pontryagin index term, is the total Berry's phase contribution for an open $S=1/2$ spin chain, and
   the second part is the additional contribution when $S>1/2$. Performing the strong coupling expansion as before, we find that the system behaves as an
   open spin-$1/2$ chain coupled to two end spins with a magnitude of ${1\over2}+{1\over2}(S-{1\over2})$. The problem of impurity end spins coupled to a spin-$1/2$
   chain has been analyzed using the bosonization technique, through which it was found that after the screening induced by the spin-$1/2$ chain (essentially a Kondo effect), a free
   spin with a magnitude of ${1\over2}(S-{1\over2})$ is left at each end of the spin chain \cite{Eggert92}. Note that the existence of end states in half-odd-integer spin chains
   is rather non-trivial because the bulk spin
   excitations are {\em gapless}. As a result, the end spins at the two ends of a half-odd-integer spin chain are coupled by a term
   $J_{eff}\sim JS^2/(L\ln L)$, where $L$ is the length of the spin chain. The excitation energy of the end state is logarithmically lower than the energy of the bulk
   spin excitations, which have an energy of $\sim J/L$ \cite{Ng94}. These predictions for open chains and end states based on the NL$\sigma$M plus topological $\theta$ term analysis have been verified numerically
   by means of density matrix renormalization group (DMRG) calculations \cite{Qin95}.

  \subsection{Higher dimensions and frustrated quantum antiferromagnets}
    The NL$\sigma$M approach to quantum antiferromagnets has been extended to higher dimensions and to frustrated quantum antiferromagnets. For simple
  antiferromagnets, $S_T$ vanishes in dimensions of $d>1$, and we need only consider the NL$\sigma$M, i.e., $S_{\sigma}$. As discussed before, $S_{\sigma}$ describes
 two plausible phases, the weak coupling phase, in which the ground state is antiferromagnetically ordered, and the strong coupling phase, in which the ground state
 is gapped. The weak coupling phase is favored for large spin magnitudes $S$. Various numerical and analytical studies have consistently demonstrated that the ground state is
 always N\'{e}el ordered for simple quantum antiferromagnets on a $2d$ square lattice, even for the smallest possible spin value of $S=1/2$ \cite{Manousakis91}.
 For this reason, physicists have turned to frustrated spin models to look for exotic spin liquid states.

   The NL$\sigma$M approach has generated interesting results when applied to weakly frustrated spin models, where the main effect of frustration is to
  reduce the effective coupling strength between rotors (for example, $J_1-J_2$ models, in which a next-nearest neighbor antiferromagnetic coupling is added to
  the Heisenberg model on a square lattice). In this case, it has been shown that spin-Peierls order can be obtained when discontinuous monopole-like spin configurations are included in the calculation of $S_T$ \cite{ReadSachdev90}.
  However, the method becomes questionable when applied to strongly frustrated spin systems, in which
   effective rotors become difficult to define locally, for example, the antiferromagnetic Heisenberg model on a kagome lattice.

  Generally speaking, a continuum theory is reliable only if the short-distance physics is captured correctly by the underlying classical or mean-field theory. A continuum theory becomes unreliable if the short-distance physics it assumes is not correct. This seems to be the case for the
  NL$\sigma$M approach when applied to strongly frustrated spin systems. In the following sections, we consider alternative methods of treating quantum spin systems, keeping in mind the physics that we have previously discussed.

\section{Resonant valence bond (RVB) states}

 The semi-classical approach, which is based on fluctuations around a presumed classical (N\'{e}el) order, is difficult to apply in frustrated
 lattice models. The difficulties arise from two main sources. First, different degenerate or quasi-degenerate classical ground states may exist in a frustrated spin system. It is difficult to include these quasi-degenerate classical ground states in the NL$\sigma $M description. Second, the effect of Berry's phases becomes intractable because of the complicated (classical) spin trajectory.

 The term geometric frustration (or frustration for short) was introduced by Gerard Toulouse in the context of frustrated magnetic systems \cite{Toulouse77,Toulouse77a}. Indeed, frustrated magnetic systems had long been studied prior to that time.
 Early work included a study conducted by G. H. Wannier \cite{Wannier50} on the classical Ising model on a triangular lattice with antiferromagnetically coupled nearest neighbor spins, which serves as the simplest example of geometric frustration \cite{BookDiep}.
 Because of the AFM coupling, two nearest neighboring spins $A$ and $B$ tend to be anti-parallel. Then, a third spin $C$ that is a neighbor of both $A$ and $B$ is \emph{frustrated} because its two possible orientations, up and down,
 both have the same energy. The classical ground state has a high level of degeneracy. As a result, we cannot choose a classical spin order as the starting point for constructing the NL$\sigma $M for the quantum $S=1/2$ $XXZ$ model
\[
  H=J_z\sum_{\langle i,j\rangle}S_i^{(z)}S_j^{(z)}+J_{\perp}\sum_{\langle i,j\rangle}\left(S_i^{(x)}S_j^{(x)}+S_i^{(y)}S_j^{(y)}\right)
\]
 with $J_z>>J_{\perp}$ because there exist infinite spin configurations with the same classical energy. We note that the spin-spin correlation has been found to decay following a power law at zero temperature
 in the exact solution for the classical Ising model \cite{Stephenson70}.

\begin{figure}[tbph]
\includegraphics[width=3.6cm]{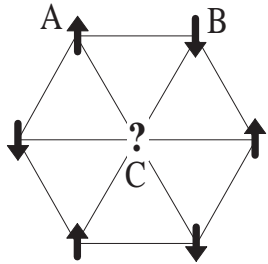}
\caption{ Geometric frustration. The spin $C$ is frustrated because either the up or down orientation will give rise to the same energy in the AFM Ising limit.} \label{fig:frustration}
\end{figure}

 In this case, an alternative approach is a variational wavefunction, in which we essentially must \emph{guess} the ground state wavefunction based on
  experience or physical intuition. A very important idea related to this approach
 is the resonating valence bond (RVB) concept for spin-$1/2$ systems suggested by Anderson. The term RVB was first coined by Pauling \cite{Pauling} in the
 context of metallic materials. Anderson revived interest in this concept in 1973 when he constructed a non-degenerate quantum ground state
 for an $S=1/2$ AFM system on a triangular lattice \cite{Anderson73}.  A valence bond is a spin singlet state constructed from
 two $S=1/2$ spins at sites $i$ and $j$, given by
\begin{equation}
(i,j)=\frac{1}{\sqrt{2}}(\left\vert \uparrow _{i}\downarrow
_{j}\right\rangle -\left\vert \downarrow _{i}\uparrow
_{j}\right\rangle ), \label{valencebond}
\end{equation}%
 and an RVB state is a tensor product of valence bond states, whose wavefunction is given by%
\begin{equation}
\left\vert \Psi _{RVB}\right\rangle =\sum_{i_{1}j_{1}\cdots
i_{n}j_{n}}a_{(i_{1}j_{1}\cdots i_{n}j_{n})}\left\vert
(i_{1},j_{1})\cdots (i_{n},j_{n})\right\rangle ,  \label{RVB1}
\end{equation}%
 where $(i_{1},j_{1})\cdots (i_{n},j_{n})$ are dimer configurations covering the entire lattice. The wavefunction is summed over all possible ways in which
 the lattice can be divided into pairs of lattice sites (i.e., dimers). The quantities $a_{(i_{1}j_{1}\cdots i_{n}j_{n})}$ are variational parameters determined by minimizing the ground-state energy of a given Hamiltonian. For a quantum disordered antiferromagnet, it has been proposed that the valence bond pairs in the RVB construction are
 dominated by short-range pairs, resulting in liquid-like states with no long-range spin order. The corresponding spin correlation function
 $\langle\mathbf{S}_i.\mathbf{S}_j\rangle$ in the RVB state may be short in range, with a finite correlation length (usually called short-range RVB (sRVB)), or may decay with distance following a power law (algebraic spin liquid states). The state is called a valence-bond solid (VBS) state if a single dimer configuration dominates in the ground state. An algebraic spin liquid state is usually invariant under all symmetry operations allowed by the lattice, whereas a VBS state usually breaks the translational or rotational lattice symmetry.

\begin{figure}[tbph]
\includegraphics[width=6.4cm]{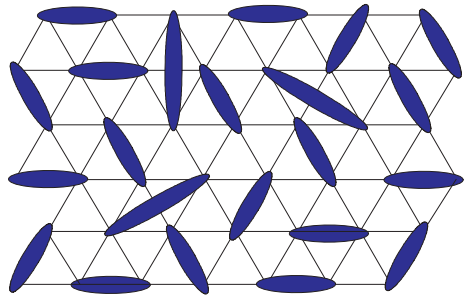}
\caption{ A spin-singlet dimer configuration covering a lattice. An RVB state is a superposition of such configurations.} \label{fig:RVB}
\end{figure}

  The wavefunction given in Eq.~\eqref{RVB1}, which is parameterized by $a_{(i_{1}j_{1}\cdots i_{n}j_{n})}$, has too many variational degrees of freedom even after the translational and rotational symmetries of the wavefunction are considered and must be simplified for practical purposes. A solution has been proposed by
  Baskaran, Zou and Anderson \cite{BZA}, who noted that the Bardeen-Cooper-Schrieffer (BCS) states for superconductors are direct product states of spin-singlet Cooper pairs and suggested that
  good RVB wavefunctions can be constructed from BCS wavefunctions {\em via} Gutzwiller projection, denoted by $P_{G}$:
\begin{eqnarray}
\left\vert \Psi _{RVB}\right\rangle &=&P_{G}\left\vert \Psi
_{BCS}\right\rangle ,  \label{RVBPG} \\
\left\vert \Psi _{BCS}\right\rangle &=&\prod\limits_{\mathbf{k}}(u_{\mathbf{k%
}}+v_{\mathbf{k}}c_{\mathbf{k}\uparrow }^{\dag
}c_{-\mathbf{k}\downarrow }^{\dag })\left\vert 0\right\rangle ,
\notag
\end{eqnarray}
 where $c_{\mathbf{k}\uparrow }^{\dag }$ and $c_{-\mathbf{k}\uparrow }^{\dag } $ are electron creation operators and the numerical coefficients $u_{\mathbf{k}}$ and $v_{\mathbf{k}}$ are determined from a trial BCS mean-field Hamiltonian $H_{BCS}$ through the Bogoliubov-de Gennes equations, i.e., the RVB wavefunction is
 fixed by the parameters determining $H_{BCS}$. The number of electrons at each lattice site may take a value of 0, 1 or 2 in the original BCS wavefunctions.
 The Gutzwiller projection $P_{G}$ removes all wavefunction components with doubly occupied sites from the BCS state and freezes the charge
 degrees of freedom. A half-filled Mott insulator state is obtained if the total number of electrons is equal to the number of lattice
  sites. We note that the technique of Gutzwiller projection is currently being widely applied to other mean-field wavefunctions $\left\vert \Psi _{MF}\right\rangle $
  to study Mott insulating states in diverse physical systems. Interesting and energetically favorable wavefunctions are often obtained when
 $\left\vert \Psi _{MF}\right\rangle $ is chosen properly.

 In addition to representing spins by electrons or fermions, one may also use Schwinger bosons to represent spins to construct RVB wavefunctions (see also the discussion after
 Eq.~\eqref{slaveparticle}). It is easy to recognize that in general, almost any mean-field wavefunction $\left\vert \Psi _{MF}\right\rangle $ can be employed to
 construct a corresponding spin state as follows:
\begin{equation}
\left\vert \Psi _{Spin}\right\rangle =P_{G}\left\vert \Psi
_{MF}\right\rangle ,  \label{RVBPG2}
\end{equation}%
 where $\left\vert \Psi _{MF}\right\rangle $ is the ground state of a trial mean-field Hamiltonian $H_{trial}(c,c^{\dag};a_1,...,a_N)$,
 where $c_{i\sigma }^{\dag }(c_{i\sigma })$ can represent either fermions or bosons and $a_1,...,a_N$ are
 variational parameters determined by minimizing the energy of the parent spin Hamiltonian.\footnote{ For historical reasons, the fermion representation is also
 called the slave-boson representation, and the Schwinger boson representation is also called the slave-fermion representation. In the context of doped Mott insulators, one can decompose the electron annihilation operator
 as $c_{i\sigma}=h_i^{\dagger} f_{i\sigma}$, where $f_{i\sigma}$ carries a charge-neutral spin and $h_i^{\dagger}$ is the (spinless) hole creation operator. If the spinon operator $f_{i\sigma}$ is fermionic,
 then the charge carrier ($h_i^{\dagger}$) is a ``slave boson", whereas if the spinon operator is bosonic, then the charge carrier is a ``slave fermion".}
 The invention of Gutzwiller projection techniques enables us to construct a large variety of variational spin wavefunctions, of which the best is the one with the lowest energy.

 The most important difference between the fermion and boson constructions is that they lead to very different sign structures in the spin wavefunction
 $\left\vert \Psi _{RVB}\right\rangle $. In a bosonic wavefunction, when two spins (note that only spin degrees of freedom remain
 after Gutzwiller projection) at different sites are interchanged, the wavefunction does not change, whereas the wavefunction does change sign when two spins
 are interchanged in a fermionic wavefunction. These different sign structures represent very different quantum entanglement structures in the corresponding
 RVB wavefunctions. A famous example is Marshall's sign rule \cite{Marshall55} for the AFM Heisenberg model on a bipartite lattice, where the Heisenberg exchange exists only
 between bonds linking sites in different sublattices. Marshall's theorem tells us that the ground state for such an AFM system is a spin-singlet state
 with positive-definite coefficients in the Ising basis $\left\{(-1)^{N_{A\downarrow }}\left\vert \sigma _{1}\cdots \sigma_{N}
 \right\rangle \right\} $, where $N_{A\downarrow }$ is the number of down spins in sublattice $A$ and $N$ is the number of lattice
 sites. Using this result, Liang, Doucot and Anderson \cite{Liang88} proposed the use of the following trial ground-state RVB\ wavefunction
 for spin-$1/2$ Heisenberg antiferromagnets on a square lattice:
 \begin{eqnarray}
 \left\vert \Psi _{LDA}\right\rangle &=&\sum_{i_{\alpha }\in A,j_{\beta }\in B}h(i_{1}-j_{1})\cdots h(i_{n}-j_{n})  \notag \\
 &&\times (-1)^{N_{A\downarrow }}\left\vert (i_{1},j_{1})\cdots (i_{n},j_{n})\right\rangle ,  \label{LDA}
 \end{eqnarray}
 where $h(r)$ represents a positive-definite function of the bond length $r$. This particular wavefunction can be conveniently represented as a Gutzwiller-projected wavefunction in the Schwinger boson representation, whereas the representation of the same wavefunction in terms of fermions is far from straightforward \cite{Read89a}. However, it has been shown that the projected BCS wavefunction given in Eq.~\ (\ref{RVBPG}) will satisfy Marshall's sign rule provided that the spatial Fourier transformation of $u_{\mathbf{k}}$ and $v_{\mathbf{k}}$ (=
 $u_{ij}$ and $v_{ij}$) connects only sites in different sublattices in a bipartite lattice \cite{LiTao07,Yunoki06}

 It has been noted by Ma \cite{Ma88} that the sum of states $\left\vert(i_{1},j_{1})\cdots
 (i_{n},j_{n})\right\rangle $, with $i_{\alpha }\in A$ and $j_{\beta }\in B$, forms an overcomplete set for spin-singlet states in a bipartite
 lattice. Because $h$ is a positive function, it can be interpreted as a weight factor in a Monte Carlo simulation based on
 loop gas statistics. Such a calculation has been performed for large lattices by Liang \textit{et al.} \cite{Liang88}, and a very accurate ground-state wavefunction for the AFM
 Heisenberg model on a square lattice was obtained. The wavefunction can give rise to either long-range or short-range spin correlations depending on the choice of $h(r)$.

\begin{figure}[tbph]
\includegraphics[width=6.4cm]{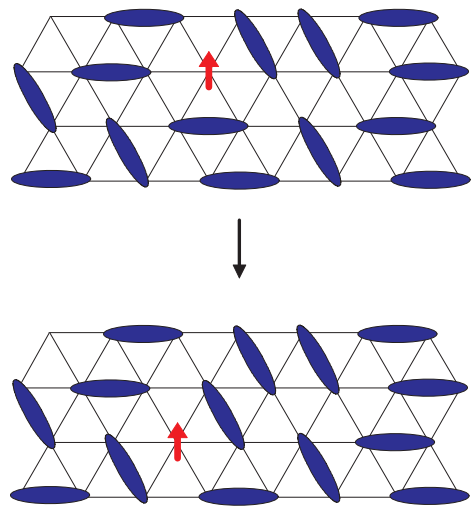}
\caption{ A spinon excitation on top of an RVB ground state.} \label{RVB-spinon}
\end{figure}

 Once a proper RVB ground-state wavefunction has been constructed, the next natural question is what are the low-energy dynamics, or the elementary excitations on top of
 the ground states? A natural candidate for excitation is to break a spin-singlet pair in the ground state to form a spin-triplet excited state with two
 unpaired spins. For a long-range magnetically ordered state, it has been found that the two unpaired spins will bind together closely in space and
 that the resulting elementary excitations will be localized spin-triplet excitations with well-defined energy and momentum.
 This is nothing but a spin wave or magnon excitation, as guaranteed by the Goldstone theorem. By contrast, for a QSL state with short-range spin
 correlation, it has been proposed that the two unpaired spins may interact only weakly with each other and can be regarded as independent spin-$1/2$ elementary
  excitations called spinons. The existence of $S=1/2$ spinon excitations is one of the most important predictions in QSLs and is crucial to the experimental
 verification of QSLs. The process through which a spin-$1$ magnon turns into two independent spin-$1/2$ spinons is an example of \emph{fractionalization}. Whether fractionalization of spin excitations actually occurs in a particular spin system is a highly non-trivial question. A systematic way to examine whether fractionalization may occur in a spin model was first proposed by X.G. Wen \cite{Wen89-Confinment,Wen91} based on the concept of confinement/deconfinement in lattice gauge
 theory.\footnote{This criterion for fractionalization works only in dimensions $d>1$. In one dimension, gauge fields are always confining, while spinons appear in energy spectrum as the gapless spin-$1/2$ excitations of the quantum antiferromagnet Heisenberg model \cite{Mudry94a,Mudry94b}.}
 This approach is explained in the following subsection, where the gauge theory for QSLs is introduced.

\subsection{RVB theory and gauge Theory}

 This subsection presents a brief survey of how RVB theory is implemented in practice, especially how low-energy effective field theories for QSL
 states are constructed, which is crucial for characterizing QSLs. We discuss a few common examples of QSLs and define the $SU(2)$, $U(1)$ and
 $Z_2$ spin liquid states. The nature of the $U(1)$ QSL state is then further illuminated by relating it to a Fermi liquid state through a
 Mott metal-insulator transition. We shall see that analytical approaches have strong limitations and should be complemented by numerical approaches in practice.

 One complication associated with the RVB construction is that there exist, in general, different mean-field states $\left\vert \Psi_{MF}\right\rangle $ that
 correspond to the same RVB spin wavefunction after Gutzwiller projection. This redundancy originates from the enlarged Hilbert space in the boson/fermion representation
 for spins and is called \emph{gauge redundancy} or \emph{gauge symmetry}. Gutzwiller projection removes this redundancy, resulting in a unique
 state in spin Hilbert space. To see how this occurs, we consider the fermion representation of $S=1/2$ spin operators \cite{Abrikosov65,BZA,Baskaran88}:
 \begin{subequations}
 \label{slaveparticle}
 \begin{equation}
 \vec{S}_{i}=\frac{1}{2}\sum_{\alpha \beta }f_{i\alpha }^{\dag }\vec{\sigma}_{\alpha \beta }f_{i\beta },  \label{spin-fermion}
 \end{equation}%
 where $\alpha ,\beta =\uparrow ,\downarrow $ are spin indices, $f_{i\alpha }^{\dag }(f_{i\alpha })$ is the fermion creation (annihilation) operator, and $\vec{\sigma}=(\sigma ^{1},\sigma^{2},\sigma ^{3})$ represents the Pauli matrices. It is easy to confirm that the three
 components of $\vec{S}_{i}$ satisfy the $SU(2)$ Lie algebra relation, $[S_{i}^{\lambda },S_{j}^{\mu }]=i\epsilon _{\lambda \mu \nu
 }S_{i}^{\nu }\delta _{ij}$, where $\lambda ,\mu ,\nu =1,2,3$ and $\epsilon _{\lambda \mu \nu }$ is the antisymmetric tensor. Hence,
 Eq.~\eqref{spin-fermion} is a representation of $SU(2)$ spins. However, the local Hilbert space for two fermions contains four Fock states,
 $\left\{ \left\vert 0\right\rangle,f_{\uparrow }^{\dag }\left\vert 0\right\rangle =\left\vert\uparrow \right\rangle ,f_{\downarrow }^{\dag }\left\vert
 0\right\rangle =\left\vert \downarrow \right\rangle ,f_{\uparrow}^{\dag }f_{\downarrow }^{\dag }\left\vert 0\right\rangle
 =\left\vert \uparrow \downarrow \right\rangle \right\} $; this is larger than the physical spin Hilbert space for spin-$1/2=\left\{ \left\vert
 \uparrow \right\rangle ,\left\vert \downarrow \right\rangle\right\} $, and we need to impose the
 single-occupancy constraint
 \begin{equation}
 \sum_{\alpha }f_{i\alpha }^{\dag }f_{i\alpha }=1
 \label{n-constraint}
 \end{equation}
 \end{subequations}
 to remove the unphysical states to obtain a proper spin representation. This is what the Gutzwiller projection does.
 The construction presented in Eq.~\eqref{slaveparticle} is equally applicable for bosons (the Schwinger boson representation) because the $SU(2)$ Lie algebra is independent of the statistics of the represented particles. In the following, we focus on the fermion representation approach because it has been found to be a more fruitful approach for constructing QSLs. Readers who are interested in the Schwinger boson
 approach may refer to reference \cite{Arovas88} for details.

 There are multiple choices of $\left\{ f_{i\alpha }\right\} $ available to represent spin operators even once the single-occupancy
 constraint is satisfied and the statistics of the particles have been chosen. For example, a new set of $\left\{f_{i\alpha }\right\} $ can be obtained through an
 $U(1)$ gauge transformation:
 \begin{equation*}
 f_{i\alpha }\rightarrow f_{i\alpha }^{\prime }=e^{i\theta(i)}f_{i\alpha }.
 \end{equation*}
 It is easy to verify that $\left\{ f_{i\alpha }^{\prime }\right\}$ forms another representation of spin operators by replacing $f_{i\alpha }$ with $%
 f_{i\alpha }^{\prime }$ in Eq.~\eqref{slaveparticle}, independent of whether the $f$s are fermions or bosons. This multiplicity is
 called gauge redundancy or gauge symmetry in the literature. We call it gauge redundancy here because symmetry usually refers to situations in which there are
 multiple physically distinct states with the same properties, e.g., there is a degeneracy in energy. However, the gauge degree of freedom we discuss here is not
 a ``real" symmetry among different physical states. Here, two gauge-equivalent states are the {\em same} state in the spin
 Hilbert space. They just ``look" different when they are represented by particles that live in an enlarged Hilbert space. There is
 no way to distinguish them physically \cite{Wen02-PSG}.

 The gauge redundancy in the fermion representation of $S=1/2$ spins extends beyond $U(1)$. There exists an additional $SU(2)$ gauge structure
 because of the particle-hole symmetry in the fermion representation, which is absent in the Schwinger boson representation. An elegant way of showing this
 $SU(2)$ gauge structure was suggested by Affleck, Zou, Tsu and Anderson \cite{AZHA88}, who introduced the following $2\times 2$ matrix operator:%
 \begin{equation}
 \Psi =\left(
 \begin{array}{cc}
 f_{\uparrow } & f_{\downarrow }^{\dag } \\
 f_{\downarrow } & -f_{\uparrow }^{\dag }%
 \end{array}\right) .  \label{matrixpsi}
 \end{equation}%
 It is straightforward to show that the spin operator can be re-expressed in terms of $\Psi $ as%
 \begin{equation}
 \vec{S}_{i}=\text{tr}\left( \Psi _{i}^{\dagger }\vec{\sigma}\Psi_{i}\right) \text{.}  \label{SSU2}
\end{equation}%
 The single-occupancy condition given in Eq.~\eqref{n-constraint} also leads to the identities
 \begin{subequations}
\begin{equation}
 \label{v-constraint}
f_{i\uparrow }f_{i\downarrow }=f_{i\uparrow }^{\dag
}f_{i\downarrow }^{\dag }=0.
\end{equation}%
  Together with Eq.~\eqref{v-constraint}, Eq.~\eqref{n-constraint} can be rewritten in the following compact vector form:
\begin{equation}
\text{tr}\left( \Psi _{i}\vec{\sigma}\Psi _{i}^{\dagger }\right)
=0. \label{n-constraintSU2}
\end{equation}%
\end{subequations}

 We now consider the following $SU(2)$ gauge transformation of $\Psi $:%
 \begin{equation}
 \Psi _{i}\rightarrow \Psi _{i}^{^{\prime }}=\Psi_{i}W_{i},W_{i}\in SU(2). \label{SU2transform}
 \end{equation}%
 The spin operator $\vec{S}_{i}$ in Eq.~\eqref{SSU2} remains invariant under this transformation because $W_{i}W_{i}^{\dag }=1$. The single-occupancy constraint given in Eq.~\eqref{n-constraintSU2} is also invariant because $W_{i}\vec{\sigma}W_{i}^{\dag }$ represents a rotation of vector $\vec{\sigma}$ but all components of tr$\left( \Psi _{i}\vec{\sigma}\Psi _{i}^{\dagger}\right) $ are zero, i.e., $\Psi _{i}\rightarrow \Psi_{i}^{^{\prime}}=\Psi_{i}W_{i}$ is also a valid representation for $S=1/2$ spins.

We show now how RVB theory is implemented in an analytical fermionic approach. For concreteness, we consider an
 AFM Heisenberg model on a lattice:
\begin{equation}
H=J\sum_{\left\langle ij\right\rangle }\vec{S}_{i}\cdot
\vec{S}_{j}, \label{HAFM}
\end{equation}
 where $\left\langle ij\right\rangle $ denotes a nearest neighbor bond and $ J>0$. The spin exchange $\vec{S}_{i}\cdot \vec{S}_{j}$ can be
 written in terms of fermionic (spinon) operators:
\begin{equation}
\vec{S}_{i}\cdot \vec{S}_{j}=\frac{1}{4}\sum_{\alpha \beta }\left(
2f_{i\alpha }^{\dag }f_{i\beta }f_{j\beta }^{\dag }f_{j\alpha
}-f_{i\alpha }^{\dag }f_{i\alpha }f_{j\beta }^{\dag }f_{j\beta
}\right) ,  \label{SiSj1}
\end{equation}
  where we have used the relation $\vec{\sigma}_{\alpha \beta }\cdot\vec{\sigma}_{\alpha ^{\prime }\beta ^{\prime }}=2\delta _{\alpha
 \beta ^{\prime }}\delta _{\alpha ^{\prime }\beta }-\delta _{\alpha\beta }\delta _{\alpha ^{\prime }\beta ^{\prime }}$. The
 constraint given in Eq.~\eqref{n-constraint} or \eqref{n-constraintSU2} can be imposed by inserting delta functions into the imaginary-time path integral.
 The corresponding partition function is
\begin{eqnarray}
Z &=&\int D[f,\bar{f}]\exp [-S(f,\bar{f})]\prod_{i}\delta \left(
\sum\nolimits_{\alpha }\bar{f}_{i\alpha }f_{i\alpha }-1\right)  \notag \\
&&\times \delta \left( \sum\nolimits_{\alpha \beta }\epsilon
_{\alpha \beta }f_{i\alpha }f_{i\beta }\right) \delta \left(
\sum\nolimits_{\alpha \beta }\epsilon _{\alpha \beta
}\bar{f}_{i\alpha }\bar{f}_{i\beta }\right) , \label{Z1}
\end{eqnarray}
where the action $S(f,\bar{f})$ is given by
\begin{equation}
S(f,\bar{f})=\int_{0}^{\beta }d\tau \left( \sum_{i\alpha
}\bar{f}_{i\alpha }\partial _{\tau }f_{i\alpha }-H\right) .
\label{S1}
\end{equation}
 The delta functions can be represented by the integration over real auxiliary fields $a_{0}^{l}(i)$ on all sites $i$,
 $l=1,2,3$. Using the relation $\delta \left( x\right) =\int \frac{dk}{2\pi }e^{ikx}$, we obtain%
\begin{equation}
Z=\int D[f,\bar{f};a]\exp [-S(f,\bar{f};a)],  \label{Z2}
\end{equation}%
with%
\begin{eqnarray}
S(f,\bar{f};a) &=&S(f,\bar{f})-i\left\{ \sum_{i}a_{0}^{3}\left(
\sum\nolimits_{\alpha }\bar{f}_{i\alpha }f_{i\alpha }-1\right)
\right.
\notag \\
&&\left. +\left[ (a_{0}^{1}+ia_{0}^{2})\sum\nolimits_{\alpha \beta
}\epsilon _{\alpha \beta }f_{i\alpha }f_{i\beta }+h.c.\right]
\right\} .  \label{S2}
\end{eqnarray}%
 It is generally believed (but has not been proven) that the partition function $Z$ will remain invariant under a Wick rotation of the fields $%
 a_{0}^{l}$ in the path integral, namely, we can replace $ia_{0}^{l}$ with $a_{0}^{l}$. Then, the action becomes
\begin{eqnarray}
S(f,\bar{f};a) &=&S(f,\bar{f})-\left\{ \sum_{i}a_{0}^{3}\left(
\sum\nolimits_{\alpha }\bar{f}_{i\alpha }f_{i\alpha }-1\right)
\right.
\notag \\
&&\left. +\left[ (a_{0}^{1}+ia_{0}^{2})\sum\nolimits_{\alpha \beta
}\epsilon _{\alpha \beta }f_{i\alpha }f_{i\beta }+h.c.\right]
\right\} .  \label{S3}
\end{eqnarray}

 The action given in Eq.~\eqref{S3} serves as the starting point for theoretical analysis. The path integral is difficult to solve, and approximate methods are generally needed. We start with a mean-field theory in which we assume that the path integral is dominated by saddle
 points characterized by equal-time expectation values of the operators $\sum\nolimits_{\alpha }f_{i\alpha }^{\dag }f_{i\alpha}$,
 $\sum\nolimits_{\alpha \beta }\epsilon _{\alpha \beta }f_{i\alpha}f_{i\beta}$ and $a_0^{l}(i)$:
\begin{eqnarray}
\chi _{ij} &=&\sum\nolimits_{\alpha }\left\langle f_{i\alpha
}^{\dag
}f_{j\alpha }\right\rangle ,  \notag \\
\Delta _{ij} &=&\sum\nolimits_{\alpha \beta }\epsilon _{\alpha
\beta }\left\langle f_{i\alpha }f_{j\beta }\right\rangle ,  \notag
\\
 a_0^{l} & = & \langle a_0^{l}(i)\rangle, \label{MFparameter}
\end{eqnarray}%
 where $\epsilon_{\alpha\beta}$ is the totally antisymmetric tensor ($\epsilon_{\uparrow\downarrow}$=1), $l=1,2,3$. It is easy to verify that $\chi _{ij}$ and $\Delta _{ij}$ satisfy the
  relations $\chi _{ij}=\chi _{ji}^{\ast }$ and $\Delta_{ij}=\Delta _{ji}$. Note that any time-dependent fluctuations in $\Delta_{ij}, \chi_{ij}$ and
 $a_{0}^{l}(i)$ are ignored in mean-field theory. With these approximations, we arrive at the following mean-field Hamiltonian:%
\begin{eqnarray}
H_{MF} &=&\sum_{\left\langle ij\right\rangle }-\frac{3}{8}J\left[
(\chi _{ji}\sum\nolimits_{\alpha }f_{i\alpha }^{\dag }f_{j\alpha
}\right.  \notag
\\
&&+\left. \Delta _{ij}\sum\nolimits_{\alpha \beta }\epsilon
_{\alpha \beta }f_{i\alpha }^{\dag }f_{j\beta }^{\dag }+h.c)-|\chi
_{ij}|^{2}-|\Delta
_{ij}|^{2}\right]  \notag \\
&&+\sum_{i}\left\{ a_{0}^{3}\left( \sum\nolimits_{\alpha
}f_{i\alpha }^{\dag
}f_{i\alpha }-1\right) \right.  \notag \\
&&\left. +\left[ (a_{0}^{1}+ia_{0}^{2})\sum\nolimits_{\alpha \beta
}\epsilon _{\alpha \beta }f_{i\alpha }f_{i\beta }+h.c.\right]
\right\} ,  \label{HMF}
\end{eqnarray}%
 where $\chi _{ij}$, $\Delta _{ij}$ and $a_{0}^{l}$ are determined by minimizing the ground-state energy with the exact constraint condition\
(\ref{n-constraint}) replaced with the average constraint
 \begin{equation}
 \sum_{\alpha }\langle f_{i\alpha }^{\dag }f_{i\alpha }\rangle=1.
 \label{a-constraint}
 \end{equation}
 The spin exchange term $\vec{S}_{i}\cdot \vec{S}_{j}$ in Eq.~\eqref{SiSj1} can be evaluated within the mean-field assumption using the Wick theorem.
 Maintaining spin rotation invariance in the calculation, we obtain
\begin{equation}
 \label{mfenergy}
\left\langle \vec{S}_{i}\cdot \vec{S}_{j}\right\rangle
=-\frac{3}{8}\left( \chi _{ij}^{\ast }\chi _{ij}+\Delta
_{ij}^{\ast }\Delta _{ij}\right) .
\end{equation}
in mean-field theory.

 Physically, the mean-field theory outlined above is equivalent to assuming that the ground state of the spin system is given by
 a mean-field wavefunction $\left\vert \Psi_{MF}\right\rangle$ without Gutzwiller projection. The spin exchange
 energy \eqref{mfenergy} evaluated in this way is usually not a good estimate of the energy of the ``real" spin wavefunction. In practice,
 this mean-field theory provides an effective way to obtain a BCS Hamiltonian to construct a Gutzwiller-projected wavefunction.
 Whether the spin wavefunction obtained through Gutzwiller projection is a good wavefunction for the spin Hamiltonian can only be tested by evaluating the energy of the wavefunction numerically (see section \ref{GutzwillerP}).

  In the following section, we assume that the Gutzwiller-projected wavefunction $P_G\left\vert \Psi_{MF}\right\rangle$ is a sufficiently good starting point
 to locate the true ground state of the spin Hamiltonian. In this case, we expect that the ground and low-energy states constructed
 from $H_{MF}$ are adiabatically connected to the corresponding Gutzwiller-projected wavefunctions and that we may construct an effective low-energy
 Hamiltonian/Lagrangian of the spin system from fluctuations around $H_{MF}$ through the usual path integral technique. The fluctuations in
 $\Delta_{ij},\chi_{ij}$ and $a^l_0(i)$ describe spin-singlet excitations and are usually called \emph{ gauge fluctuations}. Before discussing gauge
 fluctuations, we first discuss the effect of gauge redundancy on the mean-field states.

 To illustrate, we consider two mean-field QSL states with different structures of
 the mean-field parameters $\left\{ \chi _{ij},\Delta _{ij},a_{0}^{l}(i)\right\} $. We place the states on a simple square lattice. The first state
 is the uniform RVB state with%
 \begin{subequations}
 \label{examples}
\begin{eqnarray}
 \label{urvb}
\chi _{ij} &=&0, \\  \nonumber
\Delta _{ij} &=&\left\{
\begin{array}{cc}
\Delta , & \text{NN bonds,} \\  \nonumber
0, & \text{others,}%
\end{array}%
\right. \\  \nonumber
 a_{0}^{l} & = &0 \ (l=1,2,3).
\end{eqnarray}%
The second example considered is the zero-flux state given by%
\begin{eqnarray}
\label{zeroflux}
 \chi _{ij} &=&\left\{
\begin{array}{cc}
\chi , & \text{NN bonds,} \\  \nonumber
 0, & \text{others,}%
\end{array}%
\right. \\
\Delta _{ij} &=&0, \\  \nonumber
 a_{0}^{l} &=& 0 \ (l=1,2,3).
\end{eqnarray}%
\end{subequations}
 $\Delta$ and $\chi $ are real numbers. We show that irrespective of their very different appearances, these two mean-field ansatze actually give rise to the
 same spin state after Gutzwiller projection. The two states are gauge equivalent because they can be
transformed into each other through a proper gauge transformation.

 The Hamiltonian given in Eq.~\eqref{HMF} retains a local $SU(2)$ structure, which originates from the gauge redundancy in the fermion representation of spin. This local $%
 SU(2)$ symmetry becomes explicit if we introduce a doublet field $\psi =\left( f_{\uparrow },f_{\downarrow }^{\dag }\right) ^{T}$
 and a $2\times 2$ matrix%
\begin{equation*}
u_{ij}=\left(
\begin{array}{cc}
\chi _{ij} & \Delta _{ji}^{\ast } \\
\Delta _{ij} &  -\chi_{ji}%
\end{array}\right) .
\end{equation*}%
The mean-field Hamiltonian \eqref{HMF} can be written in a compact manner as%
\begin{eqnarray}
H_{MF} &=&\sum_{\left\langle ij\right\rangle }\frac{3}{8}J\left[ \frac{1%
}{2}\text{Tr}(u_{ij}^{\dag }u_{ij})-(\psi _{i}^{\dag }u_{ij}\psi _{j}+h.c.)%
\right]  \notag \\
&&+\sum_{i,l}a_{0}^{l}\psi _{i}^{\dag }\tau ^{l}\psi _{i},
\label{HMFSU2}
\end{eqnarray}%
where the $\tau ^{l},\ l=1,2,3$, are the Pauli matrices. From
Eq.~\eqref{HMFSU2}
we can clearly see that the Hamiltonian $H_{MF}$ is invariant under a local $%
SU(2)$ transformation $W_{i}$:
\begin{eqnarray}
\psi _{i} &\rightarrow &W_{i}\psi _{i},  \notag \\
u_{ij} &\rightarrow &W_{i}u_{ij}W_{j}^{\dag }.  \label{SU2MF}
\end{eqnarray}%
This $SU(2)$ gauge transformation is the same as that in
(\ref{SU2transform}), where $\Psi =\left( \psi ,i\sigma _{2}\psi
^{\dag }\right) ^{T}$.

 Because of this $SU(2)$ gauge structure, if we regard the ansatz $\left( u_{ij},a_{0}^{l}\tau ^{l}\right)$ as labeling a physical spin wavefunction
 $|\Psi _{spin}^{\left(u_{ij},a_{0}^{l}\tau ^{l}\right)}\rangle=P_{G}|\Psi_{MF}^{(u_{ij},a_{0}^{l}
 \tau ^{l})}\rangle$, then such a label is not a one-to-one label. Two ansatze related by an $SU(2)$ gauge transformation, $\left(u_{ij},a_{0}^{l}\tau ^{l}\right) $ and
 $\left(u'_{ij},a_{0}^{'l}\tau ^{l}\right)=\left(W(u_{ij}),W(a_{0}^{l}\tau ^{l})\right) $, label the same
 physical spin wavefunction:%
\begin{eqnarray}
|\Psi _{spin}(\{\alpha _{i}\})\rangle &=&P_{G}|\Psi_{MF}^{(W(u_{ij}),W(a_{0}^{l}\tau ^{l}))}\rangle  \notag \\
 &=&P_{G}|\Psi _{MF}^{(u_{ij},a_{0}^{l}\tau ^{l})}\rangle
\label{Psi-PG2}
\end{eqnarray}%
 where $W(u_{ij})=W_{i}u_{ij}W_{j}^{\dagger }$ and $W(a_{0}^{l}(i)\tau^{l})=W_{i}a_{0}^{l}(i)\tau ^{l}W_{i}^{\dagger }$, $W_{i}\in
 SU(2)$. The uniform RVB state and the zero-flux state discussed above
 denote the same physical spin state because they are related by a gauge transformation,
 \[ W_i=\exp\left(i{\pi\over4}\tau^2\right). \]

  More generally, the existence of gauge redundancy implies that the low-energy fluctuations in spin systems have a similar
 redundancy. To measure gauge fluctuations, we introduce the loop variables
 \[ P(C_i)=u_{ij}u_{jk}\cdots u_{li}, \]
 where $i,j,k,\cdots,l$ denote a loop of lattice sites that passes through site $i$. $P(C_i)$ measures gauge fluxes and has the general form
  \[
    P(C_i)=A(C_i) \tau^0 +\mathbf{B}(C_i)\cdot\vec{\tau},  \]
  where $\tau^0$ is the identity matrix and $\vec{\tau}=\{\tau^1,\tau^2,\tau^3\}$ represents the Pauli matrices, $A(C_i)$ and $\mathbf{B}(C_i)$ measure the $U(1)$ and $SU(2)$
  components, respectively, of the gauge flux. For a translationally invariant mean-field state, we can find a gauge with
  $\mathbf{B}(C_i)=\hat{n}B(C_i)$, where $A(C_i)$ and $B(C_i)$ are proportional to the area of the loop. Under a gauge transformation,
 \[ P(C_i)\rightarrow W_iP(C_i)W_i^{\dag},  \]
  and the ``direction" of $\hat{n}$ changes. The presence of gauge redundancy means that we may perform gauge transformations to change
 the ``local" directions of $\hat{n}$, but the physical spin state remains unchanged.

  For a given mean-field state, it is useful to distinguish between two kinds of gauge transformations: those that change the
 mean-field ansatz $\left\{ u_{ij},a_{0}^{l}(i)\right\} $ and those that do not. The latter constitute a subgroup of the original $SU(2)$
 symmetry called an invariant gauge group (IGG) \cite{Wen02-PSG}:
 \begin{equation}
 IGG\equiv \left\{ W_{i}|W_{i}u_{ij}W_{j}^{\dagger
 }=u_{ij},W_{i}\in SU(2)\right\} .  \label{IGG}
 \end{equation}%
 It can be shown rather generally that for a stable QSL state, physical gapless gauge excitations exist only for those
 fluctuations belonging to the IGG of the corresponding mean-field ansatz. Therefore, it is important to understand the structure of
 the IGGs in spin liquid states. Within the fermionic $SU(2)$ formalism, there are only three plausible kinds of IGG: $SU(2)$, $U(1)$ and $Z_{2}$.
 We call the corresponding spin liquids $SU(2)$, $U(1)$ and $Z_{2}$ spin liquids. $SU(2)$ spin liquids have $\mathbf{B}(C_i)=0$ with $IGG=SU(2)$.
 They are rather unstable because of the existence of a large amount of gapless $SU(2)$ gauge field fluctuations. $U(1)$
 spin liquids have $\mathbf{B}(C_i)$ pointing in only one direction for all loops $C_i$.  The condensation of fluxes in one ``direction" provides an
 Anderson-Higgs mechanism for $SU(2)$ fluxes in ``directions" perpendicular to $\mathbf{B}(C)$ and turns the IGG into $U(1)$. The low-energy fluctuations
 are $U(1)$ gauge field fluctuations. $Z_2$ spin liquids have $\mathbf{B}(C_i)$ pointing in different directions for different loops that pass
 through the same site $i$. The gauge fluctuations are all gapped because the Anderson-Higgs mechanism now applies to fluxes in all
 directions. A few examples of mean-field ansatze for these three types of spin liquid states are presented in the following subsections.

 \subsection{U(1) gauge fluctuations}
   We briefly discuss the $U(1)$ gauge theory in regard to two examples of spin liquids that are believed to exist in nature (see section V).
   The first example is the zero-flux state given in Eq.~\eqref{zeroflux}, for which $\Delta_{ij}=a_{0}^{l}=0$ and $\chi_{ij}=\chi$ in the mean-field ansatz.

 It is easy to see that $\mathbf{B}(C_i)\equiv\mathbf{0}$ and that the IGG of such a QSL is $SU(2)$, i.e., the zero-flux state describes a $SU(2)$ spin liquid. The low-energy fluctuations are $SU(2)$ gauge fluctuations. Here, we do not consider the full $SU(2)$ gauge fluctuations; we consider only the phase fluctuations of $\chi _{ij}$, i.e., $U(1)$ gauge fluctuations. The consideration of only $U(1)$ gauge fluctuations for the zero-flux state can be justified in a slave-rotor theory for the Hubbard model \cite{Lee05} or in a phenomenological Landau Fermi-liquid-type approach for spin liquid states {\em near the metal-insulator transition} (see the next subsection).

 Upon writing $\chi_{ij}=\chi e^{ia_{ij}}$, where $a_{ij}$ denotes phase fluctuations, it is straightforward to see that
   \[
    P(C_i)\propto\exp\left(i\Phi(C_i)\mathbf{\tau}^3\right),   \]
  where $\Phi(C_i)=\left(a_{ij}+a_{jk}+\cdots+a_{li}\right)$ is the total $U(1)$ gauge flux enclosed by the loop, i.e., the phase fluctuations of $\chi_{ij}$
  represent one component of the $SU(2)$ gauge fluctuations.

 The effective Lagrangian describing these low-energy phase fluctuations is
\begin{eqnarray}
L^{(0)} &=&\sum_{i\alpha }\bar{f}_{i\alpha }(\partial _{\tau
}-a_{0})f_{i\alpha }
\notag \\
 &&+\frac{3}{8}\sum_{\left\langle ij\right\rangle }\left( J\chi
 e^{ia_{ji}}\sum\nolimits_{\alpha }\bar{f}_{i\alpha }f_{j\alpha}+h.c.\right),   \label{Leff1}
\end{eqnarray}%
 and the corresponding Lagrangian in the continuum limit is
\begin{eqnarray}
L^{(0)} &=&\int d\vec{r}\sum_{\alpha }\bar{f}_{\alpha
}(\vec{r})(\partial _{\tau
}-a_{0})f_{\alpha }(\vec{r})  \notag \\
&&+\frac{1}{2m^{\ast }}\bar{f}_{\alpha }(\vec{r})(-i\triangledown +\vec{a}%
)^{2}f_{\alpha }(\vec{r}),  \label{Leff2}
\end{eqnarray}%
 where $m^{\ast }$ is the effective mass for the spinon energy dispersion determined by $J\chi$ and the vector field $\vec{a}(\vec{r})$ is
given by the lattice gauge field $a_{ij}$ through%
\begin{equation}
a_{ij}=(\vec{r}_{i}-\vec{r}_{j})\cdot \vec{a}\left( \frac{\vec{r}_{i}+\vec{r}%
_{j}}{2}\right) .
\end{equation}%
 Thus, the low-energy effective field theory describes non-relativistic spin-$1/2$ fermions (spinons) coupled to the $U(1)$ gauge field
 $\left(a_0(\vec{r}), \vec{a}(\vec{r})\right)$ in the continuum limit.

  The other spin liquid state we introduce here is the $\pi$-flux state \cite{Affleck-Marston88,Kotliar88} on a square lattice given by
  $\Delta_{ij}=a_0^{l}=0$ and
 \begin{eqnarray}
 \label{piflux}
 \chi _{i,i+\hat{\mu}} &=&\left\{
 \begin{array}{cc}
 \chi , & \mu=x, \\
 i\chi(-1)^{i_x}, & \mu=y.
 \end{array}\right.
 \end{eqnarray}
 It is easy to see that $P(C_i)\propto\exp\left(i\pi\mathbf{\tau}^3\right)$ per square plaquette in the mean-field ansatz, i.e., the $\pi$-flux state
 has $IGG=U(1)$ and is a $U(1)$ spin liquid.

 The zero-flux and $\pi$-flux states are physically distinct states because of their different IGGs. Their mean-field spinon dispersions are also
 qualitatively different. The zero-flux state has a
 mean-field dispersion of $E_0(\vec{k})=-J\chi(\cos k_x+\cos k_y)$, whereas the $\pi$-flux state has $E_{\pi}(\vec{k})=\pm J\chi\sqrt{\cos^2k_x+\cos^2k_y}$
 with a reduced Brillouin zone. The continuum theory describes non-relativistic fermions with a large Fermi surface in the zero-flux state and describes
 Dirac fermions with four Fermi points ($\mathbf{k}=(\pm\pi/2,\pm\pi/2)$) in the $\pi$-flux state \cite{Affleck-Marston88}. The effective continuum theory for the $\pi$-flux state has the form
\begin{equation}
 \label{l22}
  L^{(\pi)}=\sum_{\mu\sigma}\left(\bar{\psi}_{+\sigma}(\partial_{\mu}-ia_{\mu})\tau_{\mu}\psi_{+\sigma}+
    \bar{\psi}_{-\sigma}(\partial_{\mu}-ia_{\mu})\tau_{\mu}\psi_{-\sigma}\right),
 \end{equation}
  where $\mu=0,1,2$. The two-component Dirac spinor fields $\psi_{\pm\sigma}$ describe two inequivalent Dirac nodes in
  the spinon spectrum \cite{Affleck-Marston88}. The two effective low-energy Lagrangians $L^{(0)}$ and $L^{(\pi)}$ describe two different types of spin liquid
 states that are believed to exist in nature. We discuss these states again in section V.

  The continuum action $L$ serves as the starting point for studying the stability and low-energy properties of spin liquid states.
 Integrating out the fermion fields (at each momentum shell) gives rise to a Maxwellian potential energy term in the gauge field:
 \begin{equation*}
 \frac{1}{2g^{2}(\Lambda)}(\triangledown \times \vec{a})^{2},
 \end{equation*}%
 where $g(\Lambda)$ is a running gauge coupling constant in the sense of renormalization group theory, which depends on the energy or momentum scale $\Lambda$.
 If $g(\Lambda)\rightarrow 0$ in the low-energy and long-wavelength limit of $\Lambda\rightarrow0$, then the gauge fluctuations become increasingly weak.
 The corresponding interaction between two fermions becomes too weak to bind them together, and the elementary excitations in the spin system are spin-$1/2$
 fermionic excitations called spinons. This phenomenon is called deconfinement, and the ground state is a filled Fermi sea of spinons. By contrast, if
 $g(\Lambda)\rightarrow \infty$ as $\Lambda\rightarrow0$, then two spinons will always be confined together to form a magnon. This phenomenon is called
 confinement. In this case, the mean-field QSL ground state breaks down into a spin-ordered state because of the strong gauge fluctuations, and magnon excitations are recovered in this ordered state.

 It is not exactly clear which kinds of mean-field QSL states are stable against gauge fluctuation. It is generally believed that $Z_{2}$ QSL states are stable because $Z_{2}$ (Ising) gauge theories are deconfining \cite{Fradkin79}, whereas $SU(2)$ QSL states are unstable because of the presence of large gauge fluctuations.
 The case of $U(1)$ QSL states is more nontrivial. The $SU(2)$ gauge group and the corresponding gauge fields are compact in spin liquid states. To reflect the compactness of the $U(1)$ gauge group, one must replace the electromagnetic field tensor $F_{\mu\nu}^2$
 with $2(1-\cos F_{\mu\nu})$. This periodic version of $U(1)$ gauge theory is called compact $U(1)$ gauge theory.
 A pure compact $U(1)$ gauge theory always gives rise to confinement in two dimensions \cite{Polyakov75, Polyakov77}, but whether deconfinement is possible in the presence of a matter field is an open question. Herbut \textit{et al.} have argued that the theory is always confining in the presence of a Fermi surface \cite{Herbut03HSSM} or nodal fermions \cite{Herbut03HS}. Their conclusion depends
 on an approximate effective action for the gauge field obtained by integrating out the fermions to the lowest order. However, this approximation is questionable for gapless fermions. Indeed, Hermele \textit{et al.} \cite{Hermele04} proved that when the spin index is generalized to $N$
 flavors, deconfinement arises in the case of $2N$ 2-component Dirac fermions coupled to complex $U(1)$ gauge fields for sufficiently large
 $N$, thus providing a counter example to confinement. Further renormalization group analysis for compact quantum electrodynamics in $2+1$D
 shows that deconfinement occurs when $N>N_c=36/\pi^3\simeq 1.161$, where $N$ is the number of fermion replicas. This implies
 that a $U(1)$ spin liquid is stable at the physical value of $N=2$ \cite{Nogueira05}. Moreover, by mapping the spinon Fermi surface in $2+1$D
 to an infinite set of (1+1)-dimensional chiral fermions, Lee \cite{LeeSS08} argued that an instanton has an infinite scaling dimension
 for any $N>0$. Therefore, the QSL phase is stable against instantons, and the noncompact $U(1)$ gauge theory
 is a good low-energy description.

   We note that mechanisms other than confinement arising from gauge fluctuations may also lead to the instability of $U(1)$ QSLs, such as Amperean pairing \cite{LeeLee07} and spin-triplet pairing \cite{Galitski07} between spinons.

  A non-trivial prediction of the $U(1)$ gauge theory for spin liquids is that it leads to charge excitations with a soft gap \cite{NgLee07}, which can be detected by means of their AC conductivities $\sigma(\omega)$. It has been predicted that $\sigma(\omega)\sim\omega^{\alpha}$
 in these spin liquid states, with $\alpha\sim3.33$ in a non-relativistic spin liquid and $\alpha=2$ in a Dirac fermion spin liquid \cite{Potter2013}. It is expected that this soft gap and the related charge fluctuations will manifest themselves most clearly when the system is near the metal-insulator transition (see the next subsection).

 Because charge fluctuations will manifest themselves near the metal-insulator transition, spin liquids
 in ``weak" Mott insulators become an interesting topic \cite{Senthil08,Podolsky09,Grover10} for investigation. To study the effect of charge fluctuations near the metal-insulator transition,
 Lee and Lee \cite{Lee05} began with the Hubbard model and developed a $U(1)$ gauge theory with the help of the slave-rotor representation \cite{Florens04}. A number of physical phenomena, including transport properties \cite{Nave07} and Kondo effect \cite{Ribeiro11}, have been studied using this framework.
 Charge fluctuations correspond to higher-order spin ring-exchange terms in terms of the spin Hamiltonian \cite{Misguich98,Yang10}.

 \subsubsection{Mott transition: relation between Fermi and spin liquids}

  Zhou and Ng \cite{ZhouNg13} proposed a different way to understand $U(1)$ spin liquids near the Mott transition. They proposed that spin liquids near the Mott transition can be regarded as ``Fermi liquids" with a constraint imposed on the current operator. For isotropic systems, they observed that the charge current carried by quasi-particles,
\begin{subequations}
\begin{equation}
\mathbf{J}=\frac{m}{m^{\ast
}}(1+{\frac{F_{1}^{s}}{d}})\mathbf{J}^{(0)}, \label{Je}
\end{equation}
is renormalized by the Landau parameter $F_1^s$ in Fermi liquid theory, but the thermal current,
\begin{equation}
\mathbf{J_Q}=\frac{m}{m^{\ast}}\mathbf{J_Q}^{(0)}, \label{Jq}
\end{equation}
\end{subequations}
 is not, where $\mathbf{J}^{(0)}$ and $\mathbf{J_q}^{(0)}$ are the charge and thermal currents, respectively, carried by the corresponding non-interacting fermions and $d$ is the number of dimensions of the system. For systems with Galilean invariance, the charge current carried by quasi-particles is not renormalized, and $\frac{m^{\ast}}{m}=1+{\frac{F_{1}^{s}}{d}}$ \cite{Baym}. However, this is not valid in general for electrons in crystals, where Galilean invariance is lost. In this case, $\frac{m^{\ast }}{m}\neq1+{\frac{F_{1}^{s}}{d}}$, and the charge
 current carried by quasi-particles is renormalized through quasi-particle interaction. In the special case in which $1+F_{1}^{s}/d\rightarrow 0$ while
 $\frac{m^{\ast }}{m}$ remains finite, $\mathbf{J}\rightarrow 0$, suggesting that the fermionic system is in a special state wherein spin-$1/2$ quasi-particles
 do not carry charge due to interaction but still carry entropy. This is exactly what one expects for spinons in QSL states.

  These authors noted that the limit of $1+F_{1}^{s}/d\rightarrow 0$ is a singular point in Fermi liquid theory and that higher-order $\mathrm{q}$- and $\omega$-dependent terms should be included
 in the Landau interaction to ensure that finite results are obtained when calculating physical response functions. Expanding at small $\mathrm{q}$ and $\omega$, they obtained
\begin{equation}
 \frac{1+F_{1}^{s}(\mathrm{q},\omega )/d}{N(0)}\sim
\alpha-\beta \omega ^{2}+\gamma_t q^{2}_t+\gamma_l q^2_l,
\label{f1}
\end{equation}
 where $q_t\sim\nabla\times$ and $q_l\sim\nabla$ are associated with the transverse (curl) and longitudinal (gradient) parts, respectively, of the small-$\vec{q}$ expansion. In a QSL state, $\alpha=0$. They found that to ensure that the system is in an incompressible (insulator) state, it is necessary to have $\gamma_l=0$.

 To show that this phenomenology actually describes fermionic spin liquids with $U(1)$ gauge fluctuations, Zhou and Ng \cite{ZhouNg13} considered a Landau Fermi liquid with
 interaction parameters of $F_{0}^{s}(q)$ and
 $F_{1}^{s}(q)$ only. The long-wavelength and low-energy dynamics of the Fermi liquid are described by the following effective Lagrangian:
 \begin{equation}
 L_{\text{eff}}=\sum_{\mathrm{k},\sigma }\left[ c_{\mathrm{k}\sigma}^{\dagger }(i\frac{\partial }{\partial t}-\xi _{\mathrm{k}})c_{\mathrm{k}%
 \sigma }-H^{\prime }(c^{\dag },c)\right] ,  \label{leff}
 \end{equation}%
 where $c_{\mathrm{k}\sigma }^{\dagger }(c_{\mathrm{k}\sigma })$ is the spin-$\sigma $ fermion creation (annihilation) operator with momentum $\mathrm{k}$ and
 \begin{equation}
 H^{\prime }(c^{\dag },c)=\frac{1}{2N(0)}\sum_{q}\left[ \frac{F_{1}^{s}(q)}{v_{F}^{2}}\mathbf{j}(q)\cdot \mathbf{j}(-q)+F_{0}^{s}(q)n(q)n(-q)\right] \label{h'eff}
\end{equation}
 describes the current-current and density-density interactions between
 quasi-particles \cite{Leggett65,Larkin64}, where $q=(\mathrm{q},\omega )$ and $v_{F}=\hbar k_{F}/m^{\ast }$ is the Fermi velocity.

 The current and density interactions can be decoupled by introducing fictitious gauge potentials $\mathbf{a}$ and $\varphi
$ (Hubbard-Stratonovich transformation) as follows:
 \begin{equation}
 H^{\prime }(c^{\dag },c)\rightarrow \sum_{q}\left[ \mathbf{j}\cdot \mathbf{a}+n\varphi -{\frac{1}{2}}\left( \frac{n}{m^{\ast }}\frac{d}{F_{1}^{s}(q)}\mathbf{%
a}^{2}+\frac{N(0)}{F_{0}^{s}(q)}\varphi ^{2}\right) \right] ,
\label{h'eff2}
\end{equation}%
 where $n$ is the fermion density. The equality $d(n/m^{\ast })=N(0)v_{F}^{2}$ was used in formulating Eq.~\eqref{h'eff2}.

 The Lagrangian presented in  Eq.~\eqref{leff} and \eqref{h'eff2} can be rewritten in the standard form of $U(1)$ gauge theory by noting that in this representation, the
 fermion current is given by
 \begin{equation*}
 \mathbf{j}=\frac{-i}{2m^{\ast }}\sum_{\sigma }\left[ \psi _{\sigma}^{\dag
}\nabla \psi _{\sigma }-(\nabla \psi _{\sigma }^{\dagger })\psi _{\sigma }%
\right] -\frac{n}{m^{\ast }}\mathbf{a},
\end{equation*}%
 where $\psi _{\sigma }(\mathrm{r})=\int e^{-i\mathrm{k\cdot r}}c_{\mathrm{k}\sigma }$ is the Fourier transform of $c_{\mathrm{k}\sigma }$. The
 Lagrangian can be written as
 \begin{subequations}
 \label{lspin}
 \begin{equation}
 L=\sum_{\sigma }\int d^{d}\mathrm{r}\left[ \psi _{\sigma }^{\dagger }(i\frac{\partial }{\partial t}-\varphi )\psi _{\sigma }-H(\psi _{\sigma }^{\dagger
},\psi _{\sigma })\right] +L(\varphi ,\mathbf{a}),  \label{leff2}
\end{equation}%
where
\begin{equation}
H(\psi _{\sigma }^{\dagger },\psi _{\sigma })=\frac{1}{2m^{\ast }}|(\nabla -i%
\mathbf{a})\psi _{\sigma }|^{2}  \label{heff}
\end{equation}%
and
\begin{equation}
L(\varphi ,\mathbf{a})={\frac{1}{2}}\int d^{d}\mathrm{r}\left[ \frac{n}{%
m^{\ast }}(1+\frac{d}{F_{1}^{s}})\mathbf{a}^{2}+\frac{N(0)}{F_{0}^{s}}%
\varphi ^{2}\right] .
\end{equation}%

Using Eq.~\eqref{f1}, they find that in the small-$q$ limit, the
transverse part of $L(\varphi ,\mathbf{a})$ in the spin
liquid state is given by
\end{subequations}
\begin{equation}
L_{t}(\varphi ,\mathbf{a})=-\frac{n}{2m^{\ast }}\int
d^{d}\mathrm{r}\left[ \beta (\frac{\partial \mathbf{a}}{\partial
t})^{2}-\gamma _{t}(\nabla \times \mathbf{a})^{2}\right] .
\label{lt}
\end{equation}%
 The Lagrangian as expressed in Eq.~\eqref{lspin} together with Eq.~\eqref{lt} is the standard Lagrangian used to describe QSLs with $U(1)$ gauge fluctuations. The analysis can be rather straightforwardly
 generalized to a $U(1)$ spin liquid with Dirac fermion dispersion. The appearance of a soft charge gap in $U(1)$ spin liquids can be
 understood from the phenomenological form of $F_{1}^s(q,\omega)$ \ as expressed in Eq.~\eqref{f1}, which suggests that the quasi-particles carry vanishing charges
 only in the limit of $q,\omega\rightarrow0$. The appearance of a non-vanishing $\beta$ in \eqref{f1} leads to an AC conductivity $\sigma(\omega)$ with a power-law form. This picture is very different from theories of spin liquid states that start from simple spin models in which charge fluctuations are absent at all energy scales and suggests that charge fluctuations are important in regions near the Mott transition. We note that charge fluctuations can be (partially) incorporated into the spin models through ring-exchange terms.

 The close relationship between Fermi liquids and spin liquid states suggests an alternative picture of the Mott metal-insulator transition with respect to that put forward by Brinkman and Rice \cite{BrinkmanRice}, who proposed that a metal-insulator (Mott) transition is characterized by a diverging effective mass ${\frac{m^{\ast}}{m}} \to \infty $ and an inverse compressibility $\kappa \to 0$ at the Mott transition point, with a correspondingly vanishing quasi-particle renormalization weight $Z\sim {\frac{m}{m^{\ast}}}\to 0$. The diverging effective mass and vanishing quasi-particle weight imply that the Fermi liquid state is destroyed at the Mott transition and that the Mott insulator state is distinct from the Fermi liquid state on the metal side.

\begin{figure}[tbph]
\includegraphics[width=6.4cm]{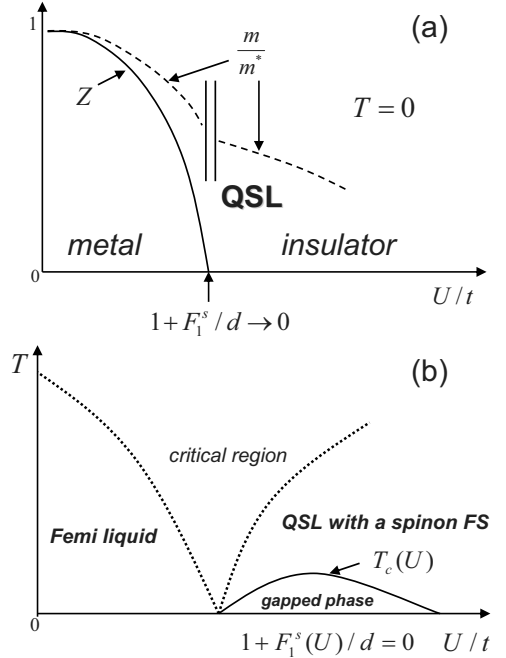}
\caption{ \cite{ZhouNg13} (a) Schematic zero-temperature phase diagram for the Mott transition. $%
U$ is the strength of the Hubbard interaction, and $t$ is the hopping
integral. The electron quasi-particle weight and the quasi-particle
charge current $\sim 1+F_1^s/d$ vanish at the critical point,
whereas the effective mass remains finite. (b) Schematic phase
diagram showing finite-temperature crossovers and possible
instability toward gapped phases at lower temperatures. There
exists a (finite-temperature) critical region around $U_c$ where the phenomenology is not applicable.} \label{phases}
\end{figure}

 The phenomenology described here suggests an alternative picture in which the Fermi surface is not destroyed, but the Landau quasi-particles are converted into spinons $(1+{\frac{F_{1}^{s}}{d}})\rightarrow0)$ at the Mott transition. In particular, the effective mass $m^*/m$ may not diverge at the metal-insulator transition, although $Z\rightarrow0$. A schematic phase diagram for the Mott (metal-QSL) transition is presented in Fig. \ref{phases} for a generic Hubbard-type Hamiltonian with a hopping integral $t$ and an on-site Coulomb repulsion $U$. The system is driven into a Mott insulator state at zero temperature at $U=U_{c}$, where $1+F_{1}^{s}(U>U_{c})/d=0$. This picture suggests that a $U(1)$ spin liquid state is likely to exist in an insulator close to the Mott transition.

  The point $1+F_{1}^{s}/d=0$ is a critical point in Fermi liquid theory called the Pomeranchuk point. The Fermi surface is unstable with respect to
 deformation when $1+F_{1}^{s}/d<0$. The criticality of this point implies that the QSLs obtained in this way are marginally stable because of large critical fluctuations. A similar conclusion can be drawn from $U(1)$ gauge theory by analyzing the $U(1)$ gauge fluctuations. As a result, QSLs with large Fermi surfaces are, in general, susceptible to the formation of other, more stable QSLs at lower temperatures, such as $Z_{2}$ QSLs or valence-bond solid (VBS) states that gap out part of
 or the entire Fermi surface. This is indicated schematically in the phase diagram shown in Fig. \ref{phases}(b), where the
 system is driven into a gapped QSL at low temperatures of $T<T_{c}(U)$ on the insulating side. The nature of the low-temperature QSLs depends on the
 microscopic details of the system and cannot be determined based on the above phenomenological considerations.

\subsection{$Z_2$ spin liquid states}
    An example of a $Z_2$ spin liquid state was first constructed by Wen \cite{Wen91} for a $J_1-J_2$ Heisenberg model on a square lattice, where $J_1$ and $J_2$ are the nearest
  neighbor and next nearest neighbor Heisenberg interactions,
  respectively. Wen considered the mean-field ansatz
  \begin{subequations}
  \begin{equation}
 u_{i,i+\hat{\mu}}=\left(
 \begin{array}{cc}
 \chi & 0 \\
 0 &  -\chi %
 \end{array}\right)
 \end{equation}
 where $\hat{\mu}=\hat{x},\hat{y}$, and
 \begin{equation}
 u_{i,i\pm\hat{x}+\hat{y}}=u_{i,i\mp\hat{x}-\hat{y}}=\left(
 \begin{array}{cc}
 0 & \Delta_0\pm i\Delta_1 \\
 \Delta_0\mp i\Delta_1 &  0 %
 \end{array}\right) ,
 \end{equation}
 where $\chi$, $\Delta_0$ and $\Delta_1$ are nonzero real numbers; $a_0^{2,3}=0$; and $a_0^1\neq0$.
\end{subequations}
   It is easy to check $P(C)$ for two loops: $C_1=i\rightarrow i+\hat{x}\rightarrow i+\hat{x}+\hat{y}\rightarrow i$ and
   $C_2=i\rightarrow i+\hat{y}\rightarrow i+\hat{y}-\hat{x}\rightarrow i$. We obtain
\begin{subequations}
 \begin{equation}
  P(C_1)=\chi^2\left(\Delta_0\tau^1+\Delta_1\tau^2\right)
 \end{equation}
 and
 \begin{equation}
  P(C_2)=-\chi^2\left(\Delta_0\tau^1-\Delta_1\tau^2\right),
 \end{equation}
 \end{subequations}
  which clearly demonstrates that $\mathbf{B}(C_1)\neq\mathbf{B}(C_2)$ and that the spin liquid state described above is a $Z_2$ spin liquid
  state. The mean-field ground state describes a half-filled spinon band with a band dispersion given by
  $E_{\pm}(\mathbf{k})=\pm\sqrt{\varepsilon_1(\vec{k})^2+\varepsilon_2(\vec{k})^2+\varepsilon_3(\vec{k})^2}$, where
  \begin{eqnarray*}
    \varepsilon_1(\vec{k}) & = & 2J_1\chi(\cos(k_x)+\cos(k_y)),
    \\ \nonumber
     \varepsilon_2(\vec{k}) & = & 2J_2\Delta_0(\cos(k_x+k_y)+\cos(k_x-k_y))+a_0^1,
    \\ \nonumber
   \varepsilon_3(\vec{k}) & = &
   2J_2\Delta_1(\cos(k_x+k_y)-\cos(k_x-k_y)).
  \end{eqnarray*}
  Note that the spinon spectrum is fully gapped.

   Many other examples of $Z_2$ spin liquid states have been constructed in the literature. For instance, a nodal gapped $Z_2$ spin liquid state was
   proposed by Balents, Fisher and Nayak \cite{Balents98} and by Senthil and Fisher \cite{Senthil00}. The corresponding mean-field ansatz
   includes nearest neighbor and next nearest neighbor hopping as well as d-wave pairing on nearest neighbor bonds on the square lattice:
  \begin{subequations}
  \begin{equation}
  u_{i,i+\hat{x}}=\left(
   \begin{array}{cc}
    \chi_1 & \Delta \\
    \Delta &  -\chi_1 %
    \end{array}\right),
  \end{equation}
  \begin{equation}
  u_{i,i+\hat{y}}=\left(
   \begin{array}{cc}
    \chi_1 & -\Delta \\
    -\Delta &  -\chi_1 %
   \end{array}\right),
  \end{equation}
 and
 \begin{equation}
 u_{i,i\pm\hat{x}\pm\hat{y}}=\left(
   \begin{array}{cc}
    \chi_2 & 0 \\
    0 &  -\chi_2 %
   \end{array}\right) ,
 \end{equation}
 where $\chi_1$, $\chi_2$, and $\Delta$ are nonzero real numbers; $a_0^{1,2}=0$; and $a_0^3\neq0$.
 \end{subequations}
 The spinon dispersion is given by $E_{\pm}(\mathbf{k})=\pm\sqrt{\varepsilon(\vec{k})^2+\Delta(\vec{k})^2}$, where
  \begin{eqnarray*}
    \varepsilon(\vec{k}) & = & 2J_1\chi_1(\cos(k_x)+\cos(k_y)) \\ \nonumber
                         &   & +2J_2\chi_2(\cos(k_x+k_y)+\cos(k_x-k_y))+a_0^3, \\ \nonumber
    \Delta(\vec{k}) & = &  2J_1\Delta(\cos(k_x)-\cos(k_y))+a_0^3,
  \end{eqnarray*}
  and is found to be gapless at four $\vec{k}$ points with a linear dispersion. Thus, this spin liquid is a $Z_2$ nodal spin liquid.
  We reiterate that $Z_2$ spin liquid states are expected to be the most stable because the $SU(2)$ gauge fields are gapped and the fermionic spins are interacting only through short-range interactions.

\begin{figure}[tbph]
\includegraphics[width=5.4cm]{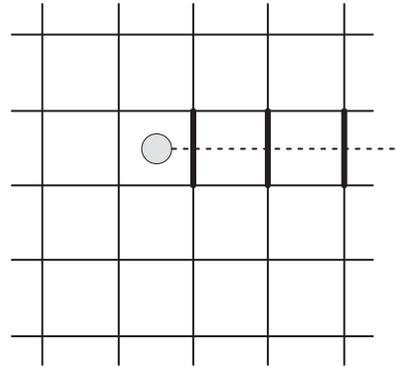}
\caption{ A $Z_2$ vortex created by flipping the signs of the $u_{ij}$ on the bonds cut by the dashed line (indicated by thick lines).}
\label{Z2Vortex}
\end{figure}

  It has been observed by Wen \cite{Wen91} that in addition to spinons, a soliton-type excitation exists in a $Z_2$ spin liquid.
  This excitation is nothing but a $\pi$ flux in the $Z_2$ gauge field, called a ``$Z_2$ vortex". This $Z_2$ vortex can be described
  by a new mean-field ansatz,
  $$
  \tilde{u}_{ij}=u_{ij}\Theta_{ij},
  $$
  where $\Theta_{ij}=\pm 1$ generates a $\pi$ flux on a lattice. One possible choice of $\Theta_{ij}$ is illustrated in Fig. \ref{Z2Vortex},
  where $\Theta_{ij}=-1$ on the bonds cut by the dashed line and $\Theta_{ij}=1$ on the other bonds. An interesting consequence of such a
  $Z_2$ vortex is that the statistics of a spinon can be changed from bosonic to fermionic and vice versa if it is bound to a vortex. Therefore,
  $Z_2$ spin liquids may contain charge-neutral spin-$1/2$ spinons with both bosonic and fermionic statistics \cite{Ng99}.
  The dynamics of $Z_2$ vortices can give rise to interesting physical consequences \cite{Ng99,Qi09}.

It is worth noting that the $J_1-J_2$ model on a square lattice has been well studied. The lowest-energy $Z_2$ spin liquid state is a nodal spin liquid with four Dirac points \cite{Capriotti2001,Hu2013}, labeled as Z2Azz13 in the projected group symmetry classification scheme
\cite{Wen02-PSG}, which we discuss in section \ref{sec:PSG}. This nodal $Z_2$ spin liquid is energetically competitive with calculations performed using the DMRG \cite{Jiang2012,Gong2014} and PEPS \cite{Wang13} approaches.

 {\em Relation to superconductivity:} RVB theory were developed not only for QSLs but also for high-T$_c$ superconductivity \cite{Anderson1987}. It is generally believed that $Z_2$ spin liquid states may become superconductors upon doping \cite{LeeRMP06}.
The superconducting state inherits novel properties from its QSL parent, and new phenomena may also emerge. For instance, it has been proposed that doping a kagome system can give rise to an exotic superconductor with an $hc/4e$-quantized flux (as opposed to the usual $hc/2e$ quantization) \cite{Ko09}.

\subsection{Numerical realization of Gutzwiller projection: variational Monte
Carlo method and some results}\label{GutzwillerP}

 The theories of QSL states rely heavily on the reliability of Gutzwiller-projected wavefunctions. In this subsection, we discuss how
 Gutzwiller projection is performed numerically in practice and how the physical observables can be evaluated using a Monte Carlo method for a given
 projected wavefunction $\left\vert \Psi _{RVB}\right\rangle =P_{G}\left\vert \Psi _{MF}\right\rangle $.

 Two types of mean-field ansatz are frequently used in constructing QSL states. The first one contains only (fermionic) spinon
 hopping terms $\chi$, and the mean-field ground state is a half-filled Fermi sea. The second one includes both hopping terms and pairing terms $\Delta$, and the
 mean-field ground state is a BCS-type state with a fermion energy gap. These two types of wavefunctions describe $U(1)$ and $Z_{2}$ spin liquid states, respectively, with the proper choice of hopping and pairing parameters. For a given spin Hamiltonian, we can determine these hopping and pairing parameters by optimizing the ground-state energy. Therefore, this approach is called the variational Monte Carlo (VMC)\ method.

 For a projected Fermi sea state, the mean-field ground-state wavefunction on a lattice with $N$ sites can be constructed by filling the $N$ lowest states
 in the mean-field band:%
\begin{equation*}
\left\vert \Psi _{FS}\right\rangle =\prod_{\sigma }\prod_{k=1}^{N/2}\psi
_{k\sigma }^{\dagger }\left\vert 0\right\rangle ,
\end{equation*}%
 where $\sigma =\uparrow ,\downarrow $ is the spin index and the states are sorted in order of ascending energy, $E_{1}\leq \cdots \leq E_{N/2}<E_{F}$. $\psi_{k\sigma }^{\dagger }$ creates an eigenstate in the mean-field band and can be expressed as%
\begin{equation*}
\psi _{k\sigma }^{\dagger }=\sum_{i}a_{k}\left( i\right) c_{i\sigma
}^{\dagger },
\end{equation*}%
 where each value of $i$ denotes a site and $c_{i\sigma }^{\dagger }$ is a local fermion creation operator. The eigenstate wavefunction $a_{k}\left( i\right) $
 does not depend on the spin index $\sigma $ for spin-singlet states because of the spin rotational symmetry. More explicitly,%
\begin{equation}
 \left\vert \Psi _{FS}\right\rangle =\prod_{\sigma}\prod_{i=1}^{N/2}\left( \sum_{j=1}^{N}a_{i}\left( j\right)
 c_{j\sigma }^{\dag }\right) \left\vert 0\right\rangle ,
\label{PsiFock}
\end{equation}%
 and the Gutzwiller-projected wavefunction can be written in terms of the product of three factors:%
\begin{eqnarray}
P_{G}\left\vert \Psi _{FS}\right\rangle &=&\sum_{\left\{ \sigma _{i}\right\}
}\text{sgn}\left\{ i_{1},\cdots ,i_{N/2},j_{1},\cdots ,j_{N/2}\right\}
\notag \\
&&\times \det \left[ A\left( i_{1},\cdots ,i_{N/2}\right) \right]  \notag \\
&&\times \det \left[ A\left( j_{1},\cdots ,j_{N/2}\right) \right] \left\vert
\sigma _{1},\cdots ,\sigma _{N}\right\rangle ,  \label{PgFock}
\end{eqnarray}%
where $\left\vert \sigma _{1},\cdots ,\sigma _{N}\right\rangle $ is a state
in the Ising basis with $N/2$ up spins located at sites $i_{1},\cdots
,i_{N/2}$ and $N/2$ down spins located at sites $j_{1},\cdots ,j_{N/2}$; sgn%
$\left\{ i_{1},\cdots ,i_{N/2},j_{1},\cdots ,j_{N/2}\right\} $ is the sign
of the permutation $P=\left\{ i_{1},\cdots ,i_{N/2},j_{1},\cdots
,j_{N/2}\right\} $; and $A\left( i_{1},\cdots ,i_{N/2}\right) $ is an $N/2\times
N/2$ matrix given by%
\begin{equation}
A\left( i_{1},\cdots ,i_{N/2}\right) =\left(
\begin{array}{ccc}
a_{1}\left( i_{1}\right) & \cdots & a_{1}\left( i_{N/2}\right) \\
\cdots & \ddots & \cdots \\
a_{N/2}\left( i_{1}\right) & \cdots & a_{N/2}\left( i_{N/2}\right)%
\end{array}%
\right) .  \label{matrixA}
\end{equation}

A BCS-type mean-field ground state with spin-singlet pairing can be written
as%
\begin{equation}
\left\vert \Psi _{BCS}\right\rangle =e^{\frac{1}{2}\sum_{i,j}W_{ij}(c_{i%
\uparrow }^{\dag }c_{j\downarrow }^{\dag }-c_{i\downarrow }^{\dag
}c_{j\uparrow }^{\dag })}\left\vert 0\right\rangle,  \label{Psipair}
\end{equation}%
 where $i$ and $j$ are site indices and $W_{ij}=W_{ji}$ for fermionic spin-singlet pairing. For a system with lattice
 translational symmetry, $W_{ij}$ can be written explicitly as%
\begin{equation*}
W_{ij}=-\sum_{\mathbf{k}}\frac{v_{\mathbf{k}}}{u_{\mathbf{k}}}e^{-i\mathbf{k}%
\cdot (\mathbf{R}_{i}-\mathbf{R}_{j})},
\end{equation*}%
 where $u_{\mathbf{k}}$ and $v_{\mathbf{k}}$ are given in the BCS form. In the more general situation in which lattice translational symmetry is lost, the $W_{ij}$s are determined from the Bogoliubov-de Gennes equations. Gutzwiller projection retains only states with a number of electrons equal to the number of lattice sites and removes all terms with more than one electron per site, i.e.,
\begin{equation}
\left\vert \Psi _{RVB}\right\rangle =P_{G}\left(
\sum\nolimits_{i<j}W_{ij}c_{i\uparrow }^{\dag }c_{j\downarrow }^{\dag
}\right) ^{N/2}\left\vert 0\right\rangle .  \label{Pgpair0}
\end{equation}%
In the spin representation, the projected BCS state can be written as%
\begin{eqnarray}
P_{G}\left\vert \Psi _{BCS}\right\rangle &=&\sum_{\left\{ \sigma
_{i}\right\} }\text{sgn}\left( i_{1},\cdots ,i_{N/2},j_{1},\cdots
,j_{N/2}\right)  \notag \\
&&\times \det \left[ w\left( i_{1},\cdots ,i_{N/2},j_{1},\cdots
,j_{N/2}\right) \right]  \notag \\
&&\times \left\vert \sigma _{1},\cdots ,\sigma _{N}\right\rangle ,
\label{Pgpair}
\end{eqnarray}%
where $\left\vert \sigma _{1},\cdots ,\sigma _{N}\right\rangle $ is a state
in the Ising basis with $N/2$ up spins located at sites $i_{1},\cdots
,i_{N/2}$ and $N/2$ down spins located at sites $j_{1},\cdots ,j_{N/2}$ and $%
w\left( i_{1},\cdots ,i_{N/2},j_{1},\cdots ,j_{N/2}\right) $ is an $%
N/2\times N/2$ matrix given by%
\begin{equation}
w\left( i_{1},\cdots ,i_{N/2},j_{1},\cdots ,j_{N/2}\right) =\left(
\begin{array}{ccc}
W_{i_{1}j_{1}} & \cdots & W_{i_{1}j_{N/2}} \\
\cdots & \ddots & \cdots \\
W_{i_{N/2}j_{1}} & \cdots & W_{i_{N/2}j_{N/2}}%
\end{array}%
\right) .  \label{matrixw}
\end{equation}

 A key observation regarding these two projected wavefunctions, Eqs.\eqref{PgFock} and \eqref{Pgpair}, is that both of them can be written as a determinant
 or as a product of two determinants. This allows us to evaluate a projected wavefunction numerically. For a large system, the number of degrees
 of freedom increases exponentially with the system size. In this case, the Monte Carlo method is applied to evaluate the energy,
 magnetization and spin correlation for these projected wavefunctions \cite{Horsch83,Gros89}. Below, we briefly
 describe how the MC method works. Those who are interested in the details may refer to Gros\cite{Gros89}.

 The expectation value of an operator $\Theta $ in a system with the spin wavefunction $|\Psi\rangle $ can be written as
\begin{equation}
\langle \Theta \rangle =\frac{\langle \Psi |\Theta |\Psi \rangle }{\langle
\Psi |\Psi \rangle }=\sum_{\alpha ,\beta }\langle \alpha |\Theta |\beta
\rangle \frac{\langle \Psi |\alpha \rangle \langle \beta |\Psi \rangle }{%
\langle \Psi |\Psi \rangle },  \label{MC1}
\end{equation}%
 where the spin configurations $|\alpha \rangle $ and $|\beta\rangle $ are states in the Ising basis with $N/2$ up spins and
 $N/2$ down spins. This sort of expectation value is recognized to be amenable to a Monte Carlo (MC) evaluation \cite{Horsch83}.
The expectation value expression given in Eq. (\ref{MC1}) can be rewritten as
\begin{eqnarray}
\langle \Theta \rangle &=&\sum_{\alpha }\left( \sum_{\beta }\frac{\langle
\alpha |\Theta |\beta \rangle \langle \beta |\Psi \rangle }{\langle \alpha
|\Psi \rangle }\right) \frac{|\langle \alpha |\Psi \rangle |^{2}}{\langle
\Psi |\Psi \rangle }  \notag \\
&=&\sum_{\alpha }f(\alpha )\rho (\alpha ),  \label{MC2}
\end{eqnarray}%
with
\begin{eqnarray*}
f(\alpha ) &=&\sum_{\beta }\frac{\langle \alpha |\Theta |\beta \rangle
\langle \beta |\Psi \rangle }{\langle \alpha |\Psi \rangle }, \\
\rho (\alpha ) &=&\frac{|\langle \alpha |\Psi \rangle |^{2}}{\langle \Psi
|\Psi \rangle }.
\end{eqnarray*}%
It follows that
\begin{equation*}
\rho (\alpha )\geqslant 0,\sum_{\alpha }\rho (\alpha )=1.
\end{equation*}%
Note that for a ``local operator" $\Theta$ (e.g., $\Theta=\vec{S}_i\cdot\vec{S}_j$) and a given spin configuration $|\alpha \rangle$,
only a limited number of ``neighbor" configurations $|\beta\rangle$ give rise to a nonvanishing $\langle \alpha |\Theta|\beta \rangle$. As noted by Horsch and Kaplan \cite{Horsch83},
the computation time for the ratio $\frac{\langle \beta |\Psi \rangle }{\langle \alpha |\Psi \rangle }$ is of $O(N^2)$. Therefore, $\langle \Theta \rangle $ can be evaluated by means of a random walk
through spin configuration space with weight $\rho (\alpha )$. As in
the standard MC method, the probability $T(\alpha \rightarrow \alpha
^{\prime })$ of transitioning from one configuration $\alpha $ to another configuration $%
\alpha ^{\prime }$ can be chosen as follows:%
\begin{equation*}
T(\alpha \rightarrow \alpha ^{\prime })=\left\{
\begin{array}{cc}
1, & \rho (\alpha ^{\prime })\ge \rho (\alpha ), \\
\frac{\rho (\alpha ^{\prime })}{\rho (\alpha )}, & \rho (\alpha ^{\prime
})<\rho (\alpha ).%
\end{array}%
\right.
\end{equation*}%
The new configuration $\alpha ^{\prime }$ is accepted with probability $%
T(\alpha \rightarrow \alpha ^{\prime })$.

 Because $\langle \alpha |\Psi \rangle $ is either a determinant or a product of two determinants, the computation time for $\langle\alpha |\Psi \rangle $
 is of $O(N^{3})$. The computational resource consumption for the MC weight factor $T(\alpha \rightarrow \alpha ^{\prime })$ is not too high, and consequently, this MC
 method is feasible for Gutzwiller projection. Moreover, the computation time of the ratio $T(\alpha \rightarrow \alpha ^{\prime })$ can be reduced to
 $O(N^{2})$ if the corresponding matrix $A(\alpha ^{\prime })$ or $w(\alpha ^{\prime })$ in Eq.~\ (\ref{matrixA}) or (\ref{matrixw}) differs from
 $A(\alpha )$ or $w(\alpha )$ by only one row or column. This can be achieved by properly choosing the spin update procedure, e.g., the interchange of two
 opposite spins. This algorithm was first introduced by Ceperley \textit{et al.} for the MC\ evaluation of a fermionic trial
 wavefunction \cite{Ceperley77}.

 As a variational method, the VMC\ method not only yields an upper bound on the ground-state energy for a spin Hamiltonian but also provides
 detailed information on the trial ground state. This information is useful for understanding the nature of the ground-state wavefunction. In the remainder
 of this subsection, we discuss some numerical results regarding Gutzwiller-projected wavefunctions on one- and two-dimensional frustrated lattices.

\subsubsection{One-dimensional lattice}

 One-dimensional systems usually  serve as benchmarks for comparison because exact solutions are often available. It turns out that
 $P_{G}\left\vert \Psi _{FS}\right\rangle $, which is gauge equivalent to $P_{G}\left\vert \Psi _{BCS}\right\rangle $ in one
 dimension, is an excellent trial wavefunction for the ground state of the one-dimensional Heisenberg model. The energy for
 $P_{G}\left\vert \Psi _{FS}\right\rangle $ is higher than that of the exact ground state by only $0.2\%$ \cite{Yokoyama87,Gros87,Gebhard87}. The spin-spin correlation
 decays following a power law at large distances, $\langle \vec{S}_{i}\cdot \vec{S}_{i+r}\rangle \sim \frac{(-1)^{r}}{|r|}$, consistent with the results obtained through
 bosonization \cite{Luther75}. Indeed, it has been shown that this Gutzwiller-projected wavefunction is the exact ground state
 of the Haldane-Shastry model \cite{Haldane88,Shastry88},
\begin{equation*}
H_{H-S}=\frac{J}{2}\sum_{i=1}^{N}\sum_{r=1}^{N-1}\frac{1}{\sin ^{2}(\pi r/N)}%
\vec{S}_{i}\cdot \vec{S}_{i+r},
\end{equation*}%
 which describes an AFM Heisenberg chain with long-range coupling (a periodic version of $1/r^{2}$ exchange).

 Excited states with $S_{z}=m=(N_{\uparrow }-N_{\downarrow })/2$ can also be constructed, where $N_{\uparrow }$ and $N_{\downarrow}$ are the numbers of up
 and down spins, respectively, in the wavefunction. The lowest-energy state in the subspace with $S_{z}=m$ is given by%
\begin{equation}
P_{G}\left\vert \Psi _{m}\right\rangle =P_{G}\prod_{|k|\leq k_{F\uparrow
}}\psi _{k\uparrow }^{\dagger }\prod_{|k|\leq k_{F\downarrow }}\psi
_{k\uparrow }^{\dagger }\left\vert 0\right\rangle ,  \label{PGPsim}
\end{equation}%
where $k_{F\sigma }=\pi (N_{\sigma }-1)/N=\pi (N_{\sigma }-1)/(N_{\uparrow
}+N_{\downarrow })$. With the help of this trial wavefunction, the spin
susceptibility $\chi $ can be calculated \cite{Gros87}. It is found that $%
\chi $ is close to the value obtained from the exact
solution \cite{Griffiths64}. The numerical results are summarized
in Table~\ref{tab:Gutz1D}.

\begin{widetext}

\begin{table}[htpb]
\caption{\cite{Gros89} Comparison of ground-state energy and spin susceptibility in one dimension. The
first row shows the results for the projected Fermi sea. The second row
shows the results for the exact ground state of the Heisenberg model.}
\label{tab:Gutz1D}%
\begin{ruledtabular}
\begin{tabular}{ccc}
           & $\langle \vec{S}_{i}\cdot \vec{S}_{i+1}\rangle $   &   $\chi $               \\
\hline
Gutzwiller & $-0.442118$ \cite{Gebhard87}                        & $0.058\pm 0.008$ \cite{Gros87} \\
Exact      & $-0.443147$ \cite{Lieb68}                           & $0.0506$ \cite{Griffiths64}    \\
\end{tabular}
\end{ruledtabular}
\end{table}

\end{widetext}

\subsubsection{Triangular lattice}
 Historically, the AFM spin-$1/2$ Heisenberg Hamiltonian on a triangular lattice was the first model to be proposed for the microscopic realization of a spin
 liquid ground state \cite{Fazekas74}. However, the minimum-energy configuration for the classical Heisenberg model on a triangular lattice
 is well known to be the 120$^{\circ }$ N\'{e}el state. There has been a long-standing debate regarding whether the frustration together with quantum fluctuations could
 destroy the long-range 120$^{\circ }$ N\'{e}el order, leading to a spin liquid state. Many trial wavefunctions have been proposed as the ground state of the
 nearest neighbor Heisenberg model on a triangular lattice, including a chiral spin liquid state \cite{Kalmeyer87} and 120$^{\circ }$-N\'{e}el-order states
 with quantum mechanical corrections \cite{Huse88,Sindzingre94}. In 1999, Capriotti \textit{et al.} \cite{Capriotti99} utilized the Green's function Monte Carlo
 (GFMC) method with the stochastic reconfiguration technique to obtain the state of the model with the lowest energy (to our knowledge; the ground state energy per site
 is $0.5458\pm 0.0001 $), which exhibits 120$^{\circ }$ long-range N\'{e}el order. More recently, the three-sublattice 120$^{\circ }$-N\'{e}el-order has been further confirmed by DMRG \cite{White07}.

 It thus seemed that for a triangular lattice, the possibility of a spin liquid state had been ruled out. However, the story continues. It was found that a
 four-spin ring exchange stabilizes the projected Fermi sea state against a long-range AFM state \cite{Motrunich05}. Because multi-spin ring exchange reflects
 the charge fluctuations in the vicinity of the Mott transition, this result provides theoretical support for the search for spin liquid
 states in a Mott insulating state close to the metal-insulator transition.

 The model Hamiltonian that contains both nearest neighbor Heisenberg exchange and four-spin ring
 exchange is
\begin{equation}
{H}_{\mathrm{ring}}=J\sum_{\begin{picture}(17,10)(-2,-2) \put (0,0) {\line
(1,0) {12}} \put (0,0) {\circle*{5}} \put (12,0) {\circle*{5}} \end{picture}%
}P_{12}+J_{ring}\sum_{\begin{picture}(26,15)(-2,-2) \put (0,0) {\line (1,0)
{12}} \put (6,10) {\line (1,0) {12}} \put (0,0) {\line (3,5) {6}} \put
(12,0) {\line (3,5) {6}} \put (6,10) {\circle*{5}} \put (18,10)
{\circle*{5}} \put (0,0) {\circle*{5}} \put (12,0) {\circle*{5}}
\end{picture}}\left( P_{1234}+P_{1234}^{\dagger }\right) ~,  \label{Hring}
\end{equation}%
 where $P_{12}=2\vec{S}_{1}\cdot \vec{S}_{2}+\frac{1}{2}$ interchanges the two spins at site $1$ and site $2$ and the four-spin
 exchange operators satisfy the following relations: $P_{1234}^{\dagger }=P_{4321}$ and $%
 P_{1234}+P_{4321}=P_{12}P_{34}+P_{14}P_{23}-P_{13}P_{24}+P_{13}+P_{24}-1$.

\begin{figure}[tbph]
\includegraphics[width=8.4cm]{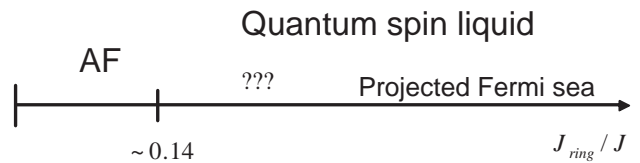}
\caption{ \cite{Motrunich05} Variational phase diagram for the Hamiltonian presented in (\protect\ref{Hring}).}
\label{ring-exchange-phases}
\end{figure}

 By comparing the trial energies of the AF-ordered states proposed by Huse and Elser \cite{Huse88} with those of various fermionic spin liquid states, Motrunich
 found that the ring-exchange term favors a spin liquid ground state over the AFM-ordered state \cite{Motrunich05}. The results are
 summarized in Fig. \ref{ring-exchange-phases}. For small ring exchange, i.e., $J_{ring}/J\lesssim 0.14$, the ordered states are of lower energy. However, for $J_{ring}/J\gtrsim 0.14$, spin liquid states are energetically favored. For larger values of $J_{ring}/J\gtrsim 0.3-0.35$, the optimal spin liquid state is the projected Fermi sea state. In the intermediate regime, optimized wavefunctions with extended anisotropic $s$-wave, $d_{x^{2}-y^{2}}$,\ and $
 d_{x^{2}-y^{2}}+id_{xy}$ spinon pairings have similar energies.

 Recently, a novel $Z_2$ spin liquid state on a triangular lattice was proposed, where the paired fermionic spinons preserve all symmetries of the system
 and the system has a gapless excitation spectrum with quadratic bands that touch at $q=0$. It was shown through the VMC method that
 this $Z_2$ spin liquid state has a highly competitive energy when $J_{ring}/J$ is realistically large \cite{Mishmash13}.

\subsubsection{Kagome lattice}

\begin{figure}[tbph]
\includegraphics[width=6.4cm]{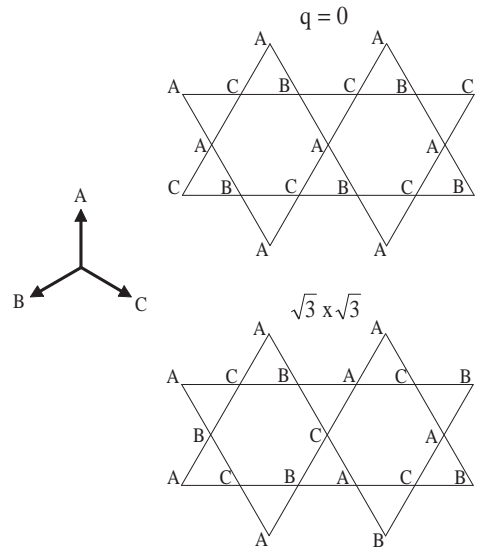}
\caption{ Two classical planar N\'{e}el states ($\mathrm{{q}=0}$ and $%
\protect\sqrt{3}\times \protect\sqrt{3}$) on a kagome lattice. A, B and C
specify three coplanar spin orientations with intersection angles of 120$^{\circ }$.}
\label{Kagome-classical-GS}
\end{figure}

 Unlike the case of a triangular lattice, the classical Heisenberg model on a kagome lattice has an infinite number of degenerate ground states that are connected to
 one another by continuous ``local" distortions of the spin configuration \cite{Villain80}. This property holds
 on any lattice with corner-sharing units, such as checkerboard, kagome, and pyrochlore lattices \cite{Moessner98}. For instance, on a kagome lattice
 formed by corner-sharing triangles, the nearest neighbor Heisenberg Hamiltonian can be written as the sum of the squares of the total spins $\vec{S%
 }_{\bigtriangleup }=\vec{S}_{1}+\vec{S}_{2}+\vec{S}_{3}$ of individual triangles that share only one vertex:%
 \begin{equation*}
 H=J\sum_{\bigtriangleup }(\vec{S}_{\bigtriangleup })^{2}.
 \end{equation*}%
 Classical ground states are obtained whenever $\vec{S}_{\bigtriangleup }=0$. This triangle rule fixes the relative orientations of the three classical
 spins of a triangle at 120$^{\circ }$ from each other in a plane, but it does not fix the relative orientation of the plane of one
 triad with respect to the planes of the triads on neighboring triangles. These degrees of freedom lead to a continuous local
 degeneracy of the ground states. Note that this degeneracy exists even if we restrict ourselves to coplanar spin states. Two of the simplest examples \cite{Sachdev92} are the three sublattice planar states shown in Fig. \ref{Kagome-classical-GS} for the $\mathrm{q}=0$ and $\sqrt{3\times }%
\sqrt{3}$ ordered states.

 The large classical ground-state degeneracy must be lifted by quantum fluctuations. The nature of the ground state for the quantum model
 is highly speculative because of the enormous degeneracy in the classical model. Many arguments have been presented in the literature regarding what kind of
 ground state is favored, and this issue is still under debate \cite{BookDiep}. In the
 following, we discuss the $U(1)$ QSL state, which is one of the promising candidates for the ground state of a spin-$1/2$ Heisenberg antiferromagnet on a kagome lattice.

 Inspired by neutron scattering experiments on herbertsmithite, ZnCu$_{3}$(OH)$_{6}$Cl$_{2}$, Ran \textit{et al.} constructed a series of variational
 wavefunctions of $U(1)$ spin liquids on a kagome lattice \cite{Ran07}. The corresponding mean-field ansatz involves only fermionic spinon hopping on
nearest neighbor bonds:%
\begin{equation*}
H_{MF}=J\sum_{\langle ij\rangle \sigma }(\chi _{ij}f_{j\sigma }^{\dagger
}f_{i\sigma }+h.c.),
\end{equation*}%
 where the complex field $\chi _{ij}$ lives on the links between two neighboring sites. For a kagome lattice, the mean-field states are characterized by
 the $U(1)$ gauge fluxes through the triangles and hexagons. Large-$N$ expansion suggests several candidate mean-field states \cite{Marston91,Hastings00}:
 (i) VBS states, which break translation symmetry; (ii) a spin liquid state (SL-$[\frac{\pi }{2},0]$) with a flux of $+\pi /2$ through each triangle on the kagome
 lattice and zero flux through the hexagons, which is a chiral spin liquid state that breaks time-reversal symmetry; (iii) a spin liquid state (SL-$[\pm
 \frac{\pi }{2},0]$) with staggered $\pi /2$ fluxes through the triangles ($+\pi /2$ through up triangles and $-\pi /2$ through down triangles) and zero
 flux through the hexagons; (iv) a spin liquid state (SL-$[ \frac{\pi }{2}%
 ,\pi ]$) with a flux of $+\pi /2$ flux through each triangle and a flux of $\pi $ through each hexagon; (v) a uniform RVB spin liquid state (SL-$[0,0]$) with zero
 flux through both triangles and hexagons, which has a spinon Fermi surface; and (vi) a $U(1)$-Dirac spin liquid state (SL-$[0,\pi ]$) with zero
 flux through the triangles and a flux of $\pi$ through each hexagon, which has four flavors of two-component Dirac fermions.

 By performing VMC calculations on $8\times 8\times 3$ and $12\times 12\times 3$ lattices, Ran \textit{et al.} \cite{Ran07} found that
 the $U(1)$-Dirac spin liquid state (SL-$[0,\pi ]$) has the lowest energy among states (i)-(vi) listed above after Gutzwiller
 projection, with a ground-state energy of $-0.429J$ per site. Note that there is no tunable parameter in this $U(1)$-Dirac spin liquid state. This energy is remarkably favorable because the value is very close to the exact diagonalization result when extrapolated to the thermodynamic limit. A comparison among the ground-state energies determined using this VMC method and other numerical methods is presented in Table \ref{tab:Kagome}. The authors also found that the $U(1)$-Dirac spin
 liquid state is stable against VBS ordering and chiral spin liquid states with fluxes of $\theta $ through the triangles and $(\pi -2\theta )$
 through the hexagons. The spin correlation functions exhibit algebraic decay with distance because of the Dirac nodes in the spinon spectrum.

\begin{table}[htpb]
\caption{Comparison of the ground-state energies (in units of $J$) determined using different methods for the nearest neighbor Heisenberg model on a kagome lattice.
In the VMC method, the $U(1)$-Dirac spin liquid state (SL-$[0,\pi ]$) is used.}
\label{tab:Kagome}
\begin{ruledtabular}
\begin{tabular}{cc}
Method                               & Energy per site                \\
\hline
Exact diagonalization                & $-0.43$ \cite{Waldtmann98}      \\
Coupled cluster method               & $-0.4252$ \cite{Farnell01}      \\
Spin-wave variational method         & $-0.419$ \cite{Arrachea04}      \\
VMC method                           & $-0.429$ \cite{Ran07}           \\
\end{tabular}
\end{ruledtabular}
\end{table}

 We note that exact diagonalization \cite{Leung93,Lecheminant97,Waldtmann98,Mila98} and DMRG calculations \cite{Jiang08,Yan11,Jiang12,Depenbrock12} strongly
 indicate the existence of a spin gap and seem to rule out the $U(1)$-Dirac spin liquid scenario. However, this disagreement may be a finite-size
 effect. The applicability of exact diagonalization is limited to very small lattices of up to 36 sites, and the maximum cylinder circumference used in the DMRG approach is only
 17 lattice spacings.
 Very recently, through the combination of the Lanczos algorithm for projected fermionic wavefunctions with the Green's function Monte Carlo technique,
 Iqbal, Becca, Sorella, and Poilblanc \cite{Iqbal13,Iqbal14} found that the gapless $U(1)$-Dirac spin liquid is competitive with gapped $Z_2$ spin
 liquids. By performing a finite-size extrapolation of the ground-state energy, these authors obtained an energy per site of $E/J = -0.4365(2)$, which is
 within three error bars of the estimates obtained using the DMRG method.
 In summary, the $U(1)$-Dirac spin liquid state has proven to be a good candidate for describing a critical phase on a kagome lattice.

\subsection{Classification of spin liquid states: quantum orders and projective symmetry groups}\label{sec:PSG}

 The use of Gutzwiller-projected wavefunctions can be made more systematic by using a powerful approach based on classifying spin liquid states according to their symmetry properties. For classical systems, it was observed by Landau that symmetry is a universal property shared by all macroscopic states within the same phase, irrespective of microscopic details. Consequently, the symmetry (or broken symmetry) associated with classical order parameters serves as a powerful tool for characterizing different classical phases. This approach can be generalized to quantum spin systems described by Gutzwiller-projected wavefunctions, with additional constraints.

  For spin liquid states described by Gutzwiller-projected wavefunctions, one might expect that the quantum phases could be classified according to the symmetry properties of the mean-field ansatz $\left( u_{ij},a_{0}^{l}\tau ^{l}\right)$. However, the usual classical symmetry group (SG) is insufficient for classifying these states for two reasons: (i) Because of the gauge redundancy, different mean-field descriptions exist for the same QSL state. For instance, the uniform RVB state and the zero-flux state correspond to the same spin state, and the $d$-wave RVB state on a square lattice is also the $\pi$-flux state.  (ii) QSL states may have inherent (phase) structures contained in the mean-field ansatz $\left( u_{ij},a_{0}^{l}\tau ^{l}\right) $ that cannot be fully distinguished based on the SG constructed for classical systems. To address this issue, X.G. Wen proposed a new mathematical object called a projective symmetry group (PSG) \cite{Wen02-PSG}, which generalizes Landau's approach and has now become an important tool in studying QSLs and the quantum phase transitions between different QSL states.

 Wen proposed that the symmetry of the mean-field ansatz $\left(u_{ij},a_{0}^{l}\tau ^{l}\right) $ is a universal property and
 serves as a kind of ``quantum number" that can be used to characterize quantum orders in QSLs. The macroscopic properties of the ansatz are characterized by its
 projective symmetry group (PSG). An element of a PSG is a combined operation consisting of a symmetry transformation $U$ followed by a local
 gauge transformation $G_{U}(i)$. The PSG of a given mean-field ansatz consists of all combined operations that leave the ansatz
unchanged, i.e.,
\begin{equation}
PSG\equiv \{G_{U}|G_{U}U(u_{ij})=u_{ij},G_{U}(i)\in SU(2)\},
\label{PSG0}
\end{equation}%
 where $U(u_{ij})=\tilde{u}_{ij}\equiv u_{U(i),U(j)}$, $G_{U}U(u_{ij})\equiv G_{U}(i)\tilde{u}_{ij}G_{U}^{\dagger }(j)$, $U$ generates the
 symmetry transformation (SG), and $G_{U}$ is the associated gauge transformation. From this definition, it is easy to see that
 \[ SG\equiv {PSG\over IGG}.  \]

 The PSGs of two mean-field ansatze related by a gauge transformation $W$ are obviously also related. From $WG_{U}U(u_{ij})=W(u_{ij})$,
 where $W(u_{ij})\equiv W_{i}u_{ij}W_{j}^{\dagger } $, we obtain $WG_{U}UW^{-1}W(u_{ij})=W(u_{ij})$. Therefore, if $G_{U}U$ belongs to
 the PSG of the mean-field ansatz $u_{ij}$, then $WG_{U}UW^{-1}$ belongs to the PSG of the gauge-transformed ansatz $W(u_{ij})$. We see that the
 gauge transformation $G_{U}$ associated with the transformation $U$ changes in the following way under an $SU(2)$ gauge transformation $W$:
\begin{equation}
G_{U}(i)\rightarrow W(i)G_{U}(i)W(U(i))^{\dagger } . \label{E-PSG}
\end{equation}%
Wen proposed that mean-field ansatze with different PSGs belong to different classes of QSL
 states, just as classical states with different SGs belong to different classical phases.

  As examples, we consider the PSGs of the zero-flux state given in Eq.~\eqref{zeroflux} and the $\pi$-flux state given in Eq.~\eqref{piflux} on a square lattice.
  For illustration, let us consider the PSG associated with translational symmetry. First, we consider the zero-flux state.
 The mean-field ansatz given in Eq.~\eqref{zeroflux} is invariant under the translation transformations $T_{x}(i\rightarrow i+\hat{x})$ and $T_{y}(i\rightarrow i+\hat{y})$ and the gauge
 transformation $G(\theta)=e^{i\theta\tau^3}$. The elements of the PSG have the form $G_UU$; $G_U=\pm G(\theta)$, and $U=(T_x)^n(T_y)^m$, where $n$ and $m$ are arbitrary
 integers. The $\pi$-flux state is different. The mean-field ansatz given in Eq.~\eqref{piflux} breaks translational symmetry in
the $x$ direction because of the odd number of lattice sites. Thus, we  naively expect that the PSG should consist of elements $G_UU$
 with $G_U=\pm G(\theta)$ and $U=(T_x)^{2n}(T_y)^m$. However, this is incorrect because the two mean-field ansatze
\begin{eqnarray*}
 \chi _{i,i+\hat{\mu}} &=&\left\{
 \begin{array}{cc}
 \chi , & \mu=x \\
 i\chi(-1)^{i_x}, & \mu=y
 \end{array}\right.
 \end{eqnarray*}
 and
 \begin{eqnarray*}
 \chi _{i,i+\hat{\mu}} &=&\left\{
 \begin{array}{cc}
 \chi , & \mu=x \\
 i\chi(-1)^{i_x+1}, & \mu=y
 \end{array}\right.
 \end{eqnarray*}
 are actually related by a gauge transformation $W_i=(-1)^{i_y}\mathbf{\tau}^0$ and correspond to the same physical spin state. As a result, the
 transformations $G_{U'}U'$ with $G_{U'}=\pm G(\theta)(-1)^{i_y}\mathbf{\tau}^0$ and $U'=(T_x)^{2n+1}(T_y)^m$ are also elements of the PSG for the $\pi$-flux state.
 The zero-flux state and the $\pi$-flux state have different PSGs and therefore belong to different classes of $U(1)$ QSL states.

 More generally, other lattice symmetry operations (reflections and rotations),
 such as the parity transformations $P_{xy}\left( (i_{x},i_{y})\rightarrow (i_{y},i_{x})\right) $ and
 $P_{x\bar{y}}\left( (i_{x},i_{y})\rightarrow(-i_{y},-i_{x})\right) $ on a square lattice,
 the spin rotation transformation and the time-reversal transformation, are
 also considered when constructing PSGs, in addition to translations. The spin rotational symmetry of spin liquid states requires the mean-field ansatz to take the form:
\begin{eqnarray}
u_{ij} &=&i\rho _{ij}W_{ij},  \notag \\
\rho _{ij} &=&\text{real number},  \notag \\
W_{ij} &\in &SU(2).  \label{spin-rotation}
\end{eqnarray}

 We end with a brief discussion of an issue related to techniques for the classification of PSGs. For any two given symmetry transformations,
 their corresponding PSG elements must satisfy certain algebraic relations determined by the symmetry transformations. Solving these equations allows us to construct a PSG of a type called an \emph{algebraic PSG}. The name algebraic PSG is introduced to distinguish such PSGs from the invariant PSGs defined above. Any invariant PSG is an algebraic PSG; however, an algebraic PSG is not necessarily an invariant PSG unless there exists an ansatz such that the algebraic PSG is the total symmetry group of that ansatz.

 To provide an example, we again consider translations. The two translation elements $T_{x}$ and $T_{y}$ satisfy the following relation:
\begin{equation}
T_{x}T_{y}T_{x}^{-1}T_{y}^{-1}=1.
\end{equation}%
 From the definition of a PSG, we find that the two PSG elements $G_{x}T_{x} $ and $G_{y}T_{y}$ must satisfy the algebraic relation
\begin{eqnarray}
&&G_{x}T_{x}G_{y}T_{y}(G_{x}T_{x})^{-1}(G_{y}T_{y})^{-1}  \notag \\
&=&G_{x}T_{x}G_{y}T_{y}T_{x}^{-1}G_{x}^{-1}T_{y}^{-1}G_{y}^{-1}  \notag \\
&=&G_{x}\left( i\right) G_{y}\left( i-\hat{x}\right) G_{x}^{-1}\left( i-\hat{%
y}\right) G_{y}^{-1}\left( i\right) \in \mathcal{G},  \label{TxTy}
\end{eqnarray}%
 where we denote the IGG by $\mathcal{G}$. Each solution ($G_xT_x, G_yT_y$) of equation (\ref{TxTy}) is an algebraic PSG for $T_{x}$ and $T_{y}$.
 By adding other symmetry transformations, we can find and classify all algebraic PSGs associated with a given symmetry group.
 Because an invariant PSG is always an algebraic PSG, we can check whether an algebraic PSG is an invariant PSG by constructing
 an explicit ansatz $u_{ij}$. If an algebraic PSG supports an ansatz $u_{ij}$ with no additional symmetries, then it is an invariant
 PSG. Through this method, we can classify symmetric spin liquids in terms of PSGs.

 In reference \cite{Wen02-PSG}, Wen utilized PSGs to classify QSL states with spin rotational symmetry, time-reversal symmetry
 and all lattice symmetries on a square lattice. Later, the PSG classification approach for symmetric QSLs was applied to
 triangular \cite{ZhouWen02}, star \cite{Choy09}, and kagome \cite{Lu11} lattices. The PSG classification scheme can also be
 generalized to bosonic QSL states \cite{Wang06,WangPSG10} and to QSL states that break spin rotational symmetry and/or time-reversal symmetry \cite{KouWen09,Bieri16}.

\section{Beyond RVB approaches}

  There are many reasons to go beyond the simple RVB approach for $S=1/2$ spin systems, for example, the discovery of a plausible spin liquid state in a spin $S=1$ system \cite{HDZhou2011} and the rise in interest in Mott insulators in systems with strong spin-orbit coupling where rotational symmetry is broken and the ground state cannot be a pure spin singlet\cite{Jackeli09}. What is the nature of the spin liquid states in these systems? More importantly, we are interested in the possibility of exotic spin liquid states beyond the RVB description, where the elementary excitations may possess exotic properties beyond the simple spinon picture.

   We introduce some of these developments in this section. We start by introducing the generalization of the RVB approach to spin systems with strong spin-orbit coupling and to $S>1/2$ spin systems in sections IV.1 and IV.2, followed by the introduction of matrix product states and projected entangled pair states in section IV.3, which are completely different ways of constructing spin wavefunctions compared with the RVB approach. We end this section with an introduction to the Kitaev honeycomb model, which represents yet another different approach to constructing spin wavefunctions in a system with strong spin-orbit coupling with exotic properties beyond the simple spinon picture.

\subsection{RVB and its generalization to spin systems with strong spin-orbit coupling}

 Strong spin-orbit coupling may cause interesting experimental consequences that are absent in systems with spin rotational symmetry.
An example suggested by Zhou et al. \cite{Zhou08} is presented here, in which strong spin-orbit coupling in Ir atoms is used to explain the anomalous behavior of the Wilson ratio observed in Na$_4$Ir$_3$O$_8$, which was experimentally proposed \cite{Na4Ir3O8} as the first candidate for a 3D QSL on a hyperkagome lattice with fermionic spinons.

Although the Curie-Weiss constant is estimated to be as large as $\theta _{W}\sim 650$ K in Na$_4$Ir$_3$O$_8$, indicating strong AFM coupling,
 there is no observed thermodynamic and magnetic anomaly indicative of long-range spin ordering down to $2$ K. The specific heat ratio $\gamma=C_V/T$ shows a rather sharp peak at a temperature of $T_c\sim20$ K, indicating the existence of a phase transition or crossover at $T_c$. By contrast, the spin susceptibility $\chi\left( T\right) $ is nearly independent of temperature for all temperatures $T\ll \theta_W$.
Using the experimental values of the spin susceptibility $\chi$ and the specific heat ratio $\gamma $ at the specific heat peak at $\sim 20$ K,
for $T>T_c$, the Wilson ratio $R_{W}=\pi^{2}k_{B}^{2}\chi /3\mu _{B}^{2}\gamma $ of the material is 0.88, which is
 very close to that of a Fermi gas where $R_{W}$ is unity. Therefore, for a wide range of temperatures $T_{c}<T<\theta_W$, the system seems to behave as a Fermi liquid of spinons. Below $T_{c}$, the specific heat decreases to zero as $C_V\sim T^{2}$, suggesting a line nodal gap in the low-lying quasi-particle spectrum. However, this picture needs to be reconciled with the observation that the spin susceptibility $\chi $ remains almost constant, resulting in an anomalously large Wilson ratio of $R_W\gg 1$ at temperatures of $T<T_c$.

 The spins in Na$_4$Ir$_3$O$_8$ originate from the low-spin $5d^{5}$ Ir$^{4+}$ ions, which form a 3D network in the form of a corner-sharing hyperkagome lattice.
Chen and Balents \cite{ChenBalents08} suggested that because of the large atomic number, the spin-orbit coupling in Ir atoms is expected to be strong. In the following section, we explain the anomalous Wilson ratio based on a modified RVB spin liquid picture in which both spin-singlet and spin-triplet pairings exist in the spin-pairing wavefunction.

Based on the experimental observations discussed above, Zhou et al. \cite{Zhou08} proposed that a simple spinon hopping Hamiltonian $H_0$ determines the physics of the spin liquid state at $T>T_c$, where there exists a finite spinon Fermi surface, and that a spinon pairing gap characterized by $H_{pair}$
opens up at $T<T_c$. The power-law behavior $C_{V}\propto T^{2}$ that is observed at low temperatures of $T<T_{c}$ indicates that the gap has line nodes
on the Fermi surfaces. To determine the pairing symmetry, Zhou et al. noted that a group theoretical analysis indicates that a spin-triplet pairing state on a cubic lattice can create only full or point nodal gaps \cite{Sigrist91}, which seems to imply singlet pairing.  However, because of the broken
inversion symmetry on a hyperkagome lattice \cite{crystallography}, the spin-singlet and spin-triplet pairing states are, in general, mixed together
in the presence of spin-orbit coupling \cite{Gorkov01,Frigeri03}.
\footnote{ In general, for a many-spin system in which spin rotational symmetry is broken, the spin $S=0$ state(s) will mix with spin $S\geq1$ states even in the presence of spatial inversion symmetry. The only exception is the two-spin system, in which inversion symmetry provides a good quantum number that separates the spin-singlet state from the spin-triplet states. Because the RVB approach begins from mean-field spin wavefunctions that are superpositions of two-spin pairing states, broken inversion symmetry is needed for the construction of mixed spin-singlet and spin-triplet states.}

In terms of the $d$-vector, the gap function
$\Delta _{\alpha \beta }(\mathbf{k})$ ($\alpha,\beta=\uparrow,\downarrow$) has the general matrix form \cite{Leggett75},
\begin{equation}
\Delta (\mathbf{k})=i\left( d_{0}\left( \mathbf{k}\right) \sigma _{0}+\mathbf{d}\left( \mathbf{k}\right) \cdot \mathbf{\sigma }\right) \sigma _{y},
\label{delta}
\end{equation}
 and the spinon pairing must be singlet or a singlet-with-triplet admixture because of spin-orbit coupling in order to have line nodes \cite{Zhou08}.

We now consider the spin susceptibility of such mixed states. Zhou et al. showed that if both singlet and triplet pairings are present and the spin-orbit scattering is much weaker than the pairing gap $\Delta$, then the $k$-dependent electronic contribution to the spin susceptibility is given by
$${\chi _{ii}(\mathbf{k})\over \chi_{N}(\mathbf{k})}= 1-\frac{d_{0}d_{0}^{\ast }+d_{i}^{\ast }d_{i}}{d_{0}d_{0}^{\ast }
+\mathbf{d\cdot d}^{\ast }}+ \frac{d_{0}d_{0}^{\ast}+d_{i}^{\ast }d_{i}}{d_{0}d_{0}^{\ast }+\mathbf{d\cdot d}^{\ast }}Y(\mathbf{k};T),$$
where $i=x,y,z$; $\chi_{N}$ is the normal state contribution at $\Delta=0$; and $Y(\mathbf{k};T)$ is the $k$-dependent Yosida function \cite{Leggett75}.
Under the assumption that the $d$-vector is pinned by the lattice, for a polycrystalline sample, one must average over
all spatial directions, resulting in
\begin{equation}  \label{chip}
{\frac{\chi _{s}}{\chi _{N}}}=\frac{2}{3}-\frac{2}{3}\frac{\left\vert
d_{0}\right\vert ^{2}} {\left\vert d_{0}\right\vert^{2}+\left\vert \mathbf{d}\right\vert ^{2}}+(\frac{1}{3}+\frac{2}{3}\frac{ \left\vert d_{0}\right\vert
^{2}}{\left\vert d_{0}\right\vert^{2}+\left\vert \mathbf{d}\right\vert ^{2}})Y(T),
\end{equation}
where $Y\left( T\right)$ is the (spatially averaged) Yosida function, which vanishes at zero temperature; $\chi _{s}$ is the spin
susceptibility below $T_{c}$; and $\chi _{N}$ is the Pauli spin susceptibility in the normal state. Therefore, $\chi _{s}/\chi _{N}$ reduces to
$\frac{2}{3}-\frac{2}{3}\frac{\left\vert d_{0}\right\vert ^{2}}{\left\vert d_{0}\right\vert^{2}+ \left\vert \mathbf{d}\right\vert ^{2}}$
at zero temperature. If the spin-triplet pairing dominates, then $\chi _{s}/\chi_{N}\rightarrow \frac{2}{3}$,
whereas if the spin-singlet pairing dominates, then $\chi _{s}/\chi _{N}\rightarrow 0$.
 However, neither of these cases is observed in experiments; instead, $\chi$ changes only negligibly below $T_c$ \cite{Na4Ir3O8}.
This suggests that strong spin-orbit coupling is needed to explain the absence of a marked change in $\chi$ below $T_c\sim 20$ K.

It is well known that in conventional BCS singlet superconductors, the Knight shift, which is proportional to the Pauli paramagnetic susceptibility,
changes very little below $T_c$ for heavy elements such as Sn and Hg \cite{Adroes59}.
It is understood that this is caused by the destruction of spin conservation due to the spin-orbit coupling.
A clear explanation was presented by Anderson \cite{Anderson59} using the notion of time-reversed pairing states.
 We first consider the imaginary part of the spin response function $\chi^{\prime\prime} (q, \omega)$.
If the total spin is conserved, then the dynamics are diffusive and $\chi^{\prime\prime} (q, \omega)$ will have a central peak in $\omega$ space
with a width of $Dq^2$, which goes to zero as $q \rightarrow 0$.  Superconductivity gaps out all low-frequency excitations,
thus removing this central peak.  By the Kramers-Kronig relation, the real part $\chi^\prime (q=0, \omega=0)$ vanishes in the superconducting ground state.
In the presence of spin-orbit coupling, the total spin is not conserved but rather decays with a lifetime $\tau_s$.
In this case, $\chi^{\prime\prime}(q=0,\omega)$ has a central peak with a width of ${1\over \tau_s}$.
The superconducting gap (formed by a pair of time-reversal states) $\Delta$ cuts a hole in $\chi^{\prime\prime}(\omega)$ for $\omega < \Delta$
but leaves the $\omega \gg \Delta$ region intact, consistent with the physical expectation that the high-frequency region should
be unaffected by pairing.  By the Kramers-Kronig relation, $\chi^\prime$ will be reduced, but if the spin-orbit coupling is sufficiently strong that
\begin{equation}
{1\over \tau_s} \gg \Delta, \label{delta-tau}
\end{equation}
then the reduction will be small, i.e.,
$$
{\chi_s \over \chi_N} = 1-{\cal O} (\Delta \tau_s) .
$$
Eq. (\ref{delta-tau}) is the strong spin-orbit coupling condition that is required to have very little change in the spin susceptibility below $T_c$.
We emphasize that the criterion for discriminating strong from weak spin-orbit coupling that is given by Eq.~\ (\ref{delta-tau}) is completely different from the usual criterion,
which compares the spin-orbit energy, $\lambda$, with the splitting of the $t_{2g}$ levels, $E_3$ \cite{ChenBalents08}.
Another way to explain the large Wilson ratio observed in Na$_4$Ir$_3$O$_8$ was provided by Chen and Kim \cite{ChenKim13}, in which strong spin-orbit coupling is still essential.

 From a theoretical perspective, the PSG classification scheme has been applied to classify the spin liquid states on a kagome lattice with the Dzyaloshinskii-Moriya (DM) interaction \cite{Dodds13}. More recently, to test the validity of the RVB approach in constructing wavefunctions for spin systems with strong spin-orbit coupling, Sze, Zhou and Ng \cite{Sze2016} applied the Gutzwiller-projected wavefunction of fermion pairing states to study the $S=1/2$ anisotropic Heisenberg (XXZ) chain
  \begin{equation}
  \label{xxz}
  H=J_z\sum_{i}S^z_iS^z_{i+1}+J_{\perp}\sum_i\left(S^x_iS^x_{i+1}+S^y_iS^y_{i+1}\right),
  \end{equation}
   where $J_{\perp},J_z>0$. This model can be mapped to the isotropic (XXX) Heisenberg model with the Dzyaloshinskii-Moriya (DM) interaction,
   \[
   \sum_i\mathbf{D}\cdot(\mathbf{S}_i\times\mathbf{S}_{i+1}),  \]
    in one dimension with open boundary conditions through the transformation $U=\exp(-i\sum_{n}{n\theta\over 2}S_n^z)$ with $\cos\theta=J_z/J_{\perp}$ and $D = J_{\perp}\sin\theta$,
    where $U^{\dagger}H_{XXZ}U=H_{J+DM}$, with $H_J$ denoting the isotropic Heisenberg model with interaction $J$.

    Trial mean-field wavefunctions with the general pairing
\[
\Delta (\mathbf{k})=i\left( d_{0}\left( \mathbf{k}\right) \sigma _{0}+\mathbf{d}\left( \mathbf{k}\right) \cdot \mathbf{\sigma }\right) \sigma _{y}
\]
are being considered for the construction of the corresponding Gutzwiller-projected wavefunctions. The trial ground-state wavefunctions have the best energy when the $d$-vector has the form $d_0=0$ and $\mathbf{d}(k)=d_z\hat{z}=i\Delta\sin k$ for $J_z>J_{\perp}$ (Ising regime), whereas the preferred form is $d_0=0$ and $\mathbf{d}(k)=d_y\hat{y}=\Delta\sin k$ for $J_z<J_{\perp}$ (planar regime). The overlap between the trial ground-state wavefunction and the exact ground-state wavefunction obtained through exact diagonalization is better than  $95\%$  in all cases that have been considered. Notably, the pairing state with $\mathbf{d}(k)=d_y\hat{y}=\Delta\sin k$ does not conserve $S_z^{tot}$ and is not considered in the classification scheme used in reference \cite{Dodds13}.

\subsection{RVB approach to $S>1/2$ systems}
 Historically, the search for spin liquid states has been focused on spin $1/2$ systems because such systems have the strongest quantum mechanical fluctuation effects (see section II) when the unfrustrated Heisenberg model is considered. The situation is different when we consider spin systems with frustrated interactions \cite{Chandra88}. In this case, it is not obvious whether a spin liquid state is more likely to exist in systems of lower spin. In fact, it has recently been found that gapless spin liquid states may exist in a two-dimensional spin-$1$ compound Ba$_3$NiSb$_2$O$_9$ under high pressure \cite{Cheng12}. In this subsection, we examine how we can construct spin liquid states for $S>1/2$ systems by generalizing the RVB approach developed for $S=1/2$ systems. It should be noted that
 there are multiple possible methods of generalization. For example, Greiter and Thomale \cite{Greiter09} constructed a chiral spin liquid state using
 a fractional quantum Hall wavefunction, whereas Xu {\it et al.} \cite{Xu12} constructed a spin liquid state
 for an $S=1$ system by representing a spin of $1$ as the sum of two $S=1/2$ spins. Liu, Zhou, and Ng \cite{LZN10a,LZN10b} have developed an alternative approach in which a spin $S$ is represented by $2S+1$ fermions. In the following section, we consider this last approach, and we demonstrate the existence of fundamental differences between half-odd-integer spin and integer spin systems in this approach.

 We begin with the fermion representation of general spins. To generalize the fermion representation of $S=1/2$ spins to an arbitrary spin $S$,
 Liu, Zhou and Ng \cite{LZN10a,LZN10b} introduce $2S+1$ species of fermionic operators $c_m$ that satisfy anti-commutation relations,
 \begin{eqnarray}\label{Fermi}
 \{c_m,c^\dagger_n\}=\delta_{mn},
 \end{eqnarray}
 where $m,n=S,S-1,\cdots,-S$. The spin operator can be expressed in terms of these operators as follows:
 \begin{eqnarray*}
 \hat {\mathbf S}=C^\dagger {\mathbf I}C,
 \end{eqnarray*}
 where $C=(c_S, c_{S-1},\cdots,c_{-S})^T$ and $I^a$ $(a=x,y,z)$ is a $(2S+1)\times(2S+1)$ matrix whose matrix elements are given by
 \begin{equation*}
 I^a_{mn}=\langle S,m|S^a|S,n\rangle.
 \end{equation*}
 It is straightforward to show that the resulting spin operator $\hat{\mathbf{S}}$  satisfies the $SU(2)$ angular momentum algebra.
 Under a rotational operation, $C$
 is a spin-$S$ ``spinor" transforming as $C_m\to D^{S}_{mn}C_{n}$ and $\hat{\mathbf S}$ is a vector transforming as $S^a\to
 R_{ab}S^{b}$; here, $D^S$ is the $2S+1$-dimensional irreducible representation of the $SU(2)$ group generated by $\mathbf I$, and $R$
 is the adjoint representation.

 As in the $S=1/2$ case, a constraint that there must be only one fermion per site is needed to project the fermionic system into
 the proper Hilbert space representing spins, i.e.,
 \begin{equation}
 \label{contr1} (\hat N_i-N_f)|\mathrm{phy}\rangle=0,
 \end{equation}
 where $i$ is the site index and $N_f=1$ (the particle picture, one fermion per site). Alternatively, it is straightforward to show
 that the constraint $N_f=2S$ (the hole picture, one hole per site) equivalently represents a spin. The $N_f=1$ representation can be
 mapped to the $N_f=2S$ representation {\em via} a {\em particle-hole transformation}. For $S=1/2$, the particle picture
 and the hole picture are identical, reflecting an intrinsic particle-hole symmetry of the underlying Hilbert space, which is
 absent for $S\ge1$.

 Following Affleck, Zou, Hsu and Anderson \cite{AZHA88}, Liu, Zhou and Ng \cite{LZN10b} introduce another ``spinor" $\bar C=(c^\dagger_{-S}, -c^\dagger_{-S+1},
 c^\dagger_{-S+2},\cdots,(-1)^{2S}c^\dagger_{S})^T$, whose components can be written as $\bar C_m=(-1)^{S-m}c^\dagger_{-m}$,
 where the index $m$ runs from $S$ to $-S$ as for $C$. Upon combining $C$ and $\bar C$ into a $(2S+1)\times2$ matrix $\psi=(C,\bar C)$,
 it is straightforward to see that the spin operators can be re-expressed as
\begin{eqnarray}\label{spin}
\hat {\mathbf S}=\frac{1}{2}\mathrm{Tr}(\psi^\dagger {\mathbf
I}\psi)
\end{eqnarray}
and that the constraint can be expressed as
\begin{eqnarray}\label{contr2}
\label{newcontr1}\mathrm{Tr}(\psi\sigma_z\psi^\dagger)=2S+1-2N_f=\pm(2S-1),
\end{eqnarray}
where the $+$ sign implies $N_f=1$ and the $-$ sign implies $N_f=2S$.

 We now examine the internal symmetry group associated with the redundancy in the fermion representation. The internal symmetry group is different for integer and
 half-odd-integer spins; it is $U(1)\bar\otimes Z_2=\{e^{i\sigma_z\theta}, \sigma_x e^{i\sigma_z\theta}=e^{-i\sigma_z\theta}\sigma_x;\theta\in \mathbb{R}\}$ \
 for the former and $SU(2)$ for the latter. The reason for this difference can be qualitatively understood as follows: Note that $C$ and $\bar C$ are not independent. The
 operators in the internal symmetry group ``mix" the two fermion operators in the same row of $C$ and $\bar C$, i.e., $c_{S}$ and $c_{-S}^\dagger$. For integer spins, $c_0$ and $(-1)^Sc_0^\dagger$ will be ``mixed". For the relation $\{c_0,c_0^\dagger\}=1$
 to remain invariant, there are only two possible methods of ``mixing": one is a $U(1)$ transformation, and the other is interchanging the two operators.
 These operations form the $U(1)\bar\otimes Z_2$ group. For half-odd-integer spins, the pair $(c_0,(-1)^Sc_0^\dagger)$ does not exist, and the symmetry group
 is the maximum $SU(2)$ group. Thus, the difference between integer and half-odd-integer spins is a fundamental property of the fermion representation.

 Now let us see how the constraint expressed in Eq.~\eqref{newcontr1} transforms under the symmetry groups. For $S=1/2$, constraint
 given in Eq.~\eqref{newcontr1} is invariant under the transformation $\psi\to\psi W$ because the right-hand side vanishes (as a result of the
 particle-hole symmetry of the Hilbert space). For integer spins, if $W=e^{i\sigma_z\theta}$, then $W\sigma_z W^\dagger=\sigma_z$,
 and Eq.~\eqref{newcontr1} is invariant. If $W=\sigma_xe^{i\sigma_z\theta}$, then $W\sigma_z W^\dagger=-\sigma_z$, meaning that the ``particle" picture ($+$
 sign in Eq.~\eqref{contr2}) and the ``hole" picture ($-$ sign in Eq.~\eqref{contr2}) are transformed into each other.

 For a half-odd-integer spin with $S\ge 3/2$, $W\in SU(2)$ is a rotation, and we may extend the constraint into a vector form in a manner similar to the $S=1/2$
 case, such that Eq.~\eqref{contr2} becomes
 \begin{eqnarray}\label{veccontr}
 \mathrm{Tr}(\psi\vec\sigma\psi^\dagger)=(0,0,\pm(2S-1))^T.
 \end{eqnarray}
 Under the group transformation $\psi\to\psi W$,
 \begin{eqnarray}\label{veccontr2}
 \mathrm{Tr}(\psi\vec\sigma\psi^\dagger)\rightarrow(R^{-1})(0,0,\pm(2S-1))^T,
 \end{eqnarray}
 where $W\sigma^aW^\dagger=R_{ab}\sigma^b$, $a,b=x,y,z$, i.e., $R$ is a 3 by 3 matrix representing a 3D rotation. The transformed
 constraint represents a new Hilbert subspace, which is still a $(2N+1)$-dimensional irreducible representation of the spin
 $SU(2)$ algebra. Any measurable physical quantity, such as the spin $\mathbf S$, remains unchanged in this new Hilbert space. Therefore,
 for half-odd-integer spins ($S\ge 3/2$), there exist infinitely many ways of imposing the constraint that gives rise to a Hilbert
 subspace representing a spin. However, for integer spins, there exist only two possible constraint representations.

 The fermion representation can be used to construct mean-field Hamiltonians for spin models with arbitrary spins after the spin-spin interaction is
 written down in terms of fermion operators. For the spin-$1/2$ case, the Heisenberg interaction can be written as (see section III)
 \begin{eqnarray}\label{S1/2}
 \hat{\mathbf{S}}_i\cdot\hat {\mathbf{S}}_j&=&-\frac{1}{8}
 \mathrm{Tr}:(\psi_i^\dagger\psi_j \psi_j^\dagger\psi_i):\nonumber
 \\&=&-\frac{1}{4}:(\chi_{ij}^\dagger\chi_{ij}+\Delta_{ij}^\dagger\Delta_{ij}):,
 \end{eqnarray}
 where
 \begin{equation}
 \chi_{ij}=C_i^\dagger C_j, ~\quad \Delta_{ij}=\bar C_i^\dagger C_j.
 \end{equation}
 The definitions of $\chi_{ij}$ and $\Delta_{ij}$ in the above form can be extended to arbitrary spins. The only difference is that for an integer spin,
 $\chi_{ji}=\chi_{ij}^\dagger$ and $\Delta_{ji}=-\Delta_{ij}$, whereas for a half-odd-integer spin, $\chi_{ji}=\chi_{ij}^\dagger$ and
 $\Delta_{ji}=\Delta_{ij}$. The parity of the pairing term $\Delta_{ij}$ differs for integer and half-odd-integer spins \cite{LZN10b}. For
 $S=1$, it can be shown, after some straightforward algebra, that the Hamiltonian can be written as \cite{LZN10b}
\begin{eqnarray}\label{H1}
  \hat{\mathbf{S}}_i\cdot\hat{\mathbf{S}}_j&=&-\frac{1}{2}\mathrm{Tr}:(\psi_i^\dagger\psi_j\psi_j^\dagger\psi_i):\nonumber\\
&=&-:(\chi_{ij}^\dagger\chi_{ij} +\Delta_{ij}^\dagger\Delta_{ij}):.
\end{eqnarray}

  However, for $S>1$, we cannot write the spin-spin interaction $\hat{\mathbf{S}}_i\cdot\hat{\mathbf{S}}_j$ in terms of
 $\chi_{ij}$ and $\Delta_{ij}$ alone. In the case of $S=3/2$, triplet hopping and pairing terms must be introduced to represent the Heisenberg interaction.
 Generally speaking, quintet and higher multipolar hopping and pairing operators are needed to represent the Heisenberg
 Hamiltonian when $S$ becomes larger \cite{LZN10b}. In the following, we restrict ourselves to $S=1$ systems.

 In this case, the mean-field Hamiltonians are BCS-type Hamiltonians, as in the case of $S=1/2$ spins. The physical spin wavefunction can be
 obtained by applying Gutzwiller projection to the mean-field ground state. There are two major differences between $S=1$ and $S=1/2$ spin systems: (1) Because
 of the different internal symmetry group ($U(1)\bar\otimes Z_2$), $S=1$ spin liquid states are of either the $U(1)$ or $Z_2$ type. There are no $SU(2)$ spin liquid states for integer spin
 systems in the fermionic construction. Therefore, we expect that in general, spin liquid states for integer spin systems, if they exist, are more stable against
 gauge fluctuations. (2) The difference in parity of the pairing terms leads to interesting possibilities for obtaining topological spin liquid states in
 $S=1$ systems that are not easy to realize in $S=1/2$ systems \cite{LZN10a,Bieri12}. This difference leads to
 the existence of a Haldane phase in the bilinear-biquadratic Heisenberg spin chain in the fermionic description \cite{LZN12}.

 Finally, we note the existence of a fundamental difference in the excitation spectrum of an $S=1$ spin system compared with that of an $S=1/2$ system,
 under the assumption that the ground states are spin singlets. For an integer spin system, we can form spin-singlet states in a lattice
 with either an even or an odd number of lattice sites $N$, as long as $N>1$, whereas for a half-odd-integer spin system, spin-singlet states can be formed only in a lattice with an even
 number of sites. In the RVB approach, angular momentum $L=1$ excitations of the system are formed by
 Gutzwiller projecting the excited states in BCS theory, i.e., by breaking a pair of spin singlets in the BCS ground state. The
 resulting excited state consists of two excited spinons, which are $S=1/2$ objects for spin $1/2$ systems but are $S=1$ objects for
 spin $1$ systems. In an $S=1$ spin liquid, these two $S=1$ spinons together form an $L=1$ excitation.

  There is, however another method of forming an $L=1$ excitation in a spin-$1$ spin liquid. Beginning from a lattice system with $N$ sites, we may
 form an $L=1$ excitation by rearranging the spins such that the system is a product of spin-singlet ground states for $N-1$ of the sites plus a single
 spin-$1$ spinon. This excitation is a non-perturbative, topological excitation that cannot be achieved by simply Gutzwiller projecting a BCS excited state
 in the RVB construction. It has been demonstrated in reference \cite{LZN14} that the construction of these two kinds of excitations gives rise to the so-called
 one-magnon and two-magnon excitation spectra in the Haldane phase of the $S=1$ bilinear-biquadratic Heisenberg model.
 Similar construction approaches are not possible for $S=1/2$ systems.

 \subsection{Matrix product state (MPS) and projected entangled pair state (PEPS)}

   In this subsection, we discuss two approaches to spin liquid states that have completely different starting points from those of the RVB, or Gutzwiller-projected mean-field theory, approach we discussed in section III. We begin with matrix product states (MPSs) and projected entangled pair states (PEPSs), which represent another popular class of variational wavefunctions that are currently being applied to spin systems.
  Translationally invariant MPSs in spin chains were first constructed and studied by Fannes, Nachtergaele and Werner \cite{Fannes92}
 as an extension of the AKLT state \cite{AKLT}; in this context, the authors called them {\em finitely correlated states}.
 The term MPS was coined by Kl\"{u}mper, Schadschneider and Zittartz \cite{Klumper93}, who extended the AKLT state in a different way.
 Later, \"{O}stlund and Rommer \cite{Ostlund95} realized that the state resulting from DMRG \cite{White92}
 can be written as an MPS.
 This approach is very successful for one-dimensional systems and can be generalized to systems of two (or more) dimensions.

  First, let us consider the quantum wavefunction of a one-dimensional spin system that is translationally invariant with a local Hamiltonian  $H$. The wavefunction can be generally expressed as
  \begin{equation}
  \label{mps1}
  |\Psi\rangle=\sum_{s_1,s_2,\cdots,S_N}\phi(s_1,s_2,\cdots,s_N)|s_1,s_2,\cdots,s_N\rangle,
  \end{equation}
  where $|s_1,s_2,\cdots,s_N\rangle$ represents a spin configuration with spins $s_i$ on sites $i=1,2,\cdots,N$ and $\phi(s_1,s_2,\cdots,s_N)$ is the
  amplitude of the spin configuration in the quantum state $|\Psi\rangle$. Because of the spin-spin interaction, spin configurations at far away sites are generally
  correlated, and we cannot write $\phi(s_1,s_2,\cdots,s_N)=\phi_0(s_1)\phi_0(s_2)\cdots\phi_0(s_N)$ in general. The MPS approach is a powerful method
  of constructing wavefunctions with non-local quantum correlations. The trick is to extend the direct-product wavefunction $\phi(s_1,s_2,\cdots,s_N)=\phi_0(s_1)\phi_0(s_2)\cdots\phi_0(s_N)$ to matrix products.

  More explicitly, we associate a matrix $A^{s}$ with each spin state $s$; then, the wavefunction amplitude $\phi(s_1,s_2,\cdots,s_N)$ can be written as
  \begin{equation}
  \label{mps2}
  \phi(s_1,s_2,\cdots,s_N)={\rm Tr}\{A^{s_1}[1]A^{s_2}[2]\cdots A^{s_N}[N]\},
  \end{equation}
  where the trace is used to impose the periodic boundary condition.
  As an example, we consider an $S=1/2$ two-spin system and choose $A^{\uparrow}=\sigma_z$ and $A^{\downarrow}=\sigma_x$, where the $\sigma$s are Pauli matrices. It
 is easy to see that in this case, $\phi(\uparrow,\uparrow)=\phi(\downarrow,\downarrow)\neq0$ and
 $\phi(\uparrow,\downarrow)=\phi(\downarrow,\uparrow)=0$. A different choice of $A^{\uparrow}=\sigma_+$ and $A^{\downarrow}=\sigma_-$ yields
 $\phi(\uparrow,\downarrow)=\phi(\downarrow,\uparrow)\neq0$ and $\phi(\uparrow,\uparrow)=\phi(\downarrow,\downarrow)=0$. The
 correlation between the different spin states on the two sites is determined by the matrix $A^s$ that is chosen to link the sites. Extending the construction
 to more than two sites, one sees that the choice of the matrices $A^{\sigma}$ determines the quantum entanglement structure of the wavefunction.

  When the MPSs are treated as variational wavefunctions, one may determine the number of variational parameters in the wavefunctions by means of a simple counting argument.
  The number of parameters $P$ appearing in an MPS wavefunction in the form of Eq.~\eqref{mps2} depends on the size of the matrix $A$ and the number of available states
  $S$ per site. In  general,  $P\sim S\times M^2$ for an $M\times M$ matrix as long as $P<S^N$, where $N$ is the number of sites in the system. Thus, MPS
  wavefunctions are generally variational wavefunctions with a large number of built-in variational parameters.
  As the dimension $M\rightarrow\infty$, MPSs can represent any quantum state of the many-body Hilbert space with arbitrary accuracy.
  In practice, the low-energy states of gapped local Hamiltonians in one dimension can be efficiently represented by MPSs with a finite value of $M$ \cite{Hastings07, Verstraete06b}.
  The DMRG method \cite{White92} and its generalizations \cite{Schollwock05} can be viewed as systematic approaches for constructing MPS variational wavefunctions
  as the size of the system gradually increases.

  The MPS construction can be extended in several ways. First, it can be extended to higher dimensions by replacing
  the matrices $A$ (= rank 2 tensors) with higher-rank tensors $T$. These wavefunctions are presently known as projected entangled pair states (PEPSs)
  \cite{VerstraeteCirac04a,VerstraeteCirac04b}.
  Second, the local correlation or entanglement between a pair of sites in a PEPS can be generalized
  to a cluster (or simplex), resulting in states called projected entangled simplex states (PESSs) \cite{Xie14}.
  A representative example of a PESS is the simplex solid state proposed by Arovas \cite{Arovas08}.

\subsubsection{Valence-bond solids and MPSs in one dimension}
   The physics of an MPS or PEPS wavefunction is encoded in the tensors linking neighboring spin states.
   In general, these link tensors can be optimally constructed using the DMRG approach or tensor-based renormalization methods \cite{CiracVerstraeteJPA09}.
   In this subsection, we discuss a simple example of tensors that represent a particular class of
   spin states called valence-bond solid (VBS) states.
   To begin, we introduce a well-known example of a valence-bond solid state - the Affleck-Kennedy-Lieb-Tasaki (AKLT) state \cite{AKLT}.

\begin{figure}[tbph]
    \includegraphics[width=8.0cm]{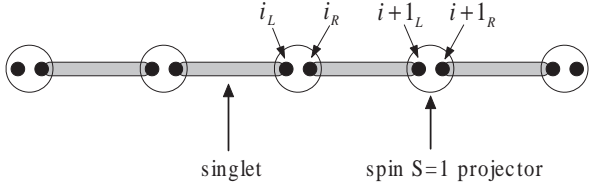}
    \caption{ A valence-bond solid construction of the AKLT state.}
    \label{fig:AKLT}
    \end{figure}

   The AKLT state is an example of a VBS state in which only one spin-singlet configuration is allowed in the wavefunction given in Eq.~\eqref{RVB1}.
   It is a one-dimensional VBS state constructed for a $S=1$ spin chain, represented pictorially in Fig. \ref{fig:AKLT},
   where each gray bond represents a spin singlet formed by two $S=1/2$ spins, i.e., Eq.~\eqref{valencebond}.
   Each lattice site is connected to two other sites by two valence bonds and is occupied by two $S=1/2$ spins.
   The AKLT wavefunction is formed by projecting the spin-$1/2 \bigotimes 1/2=1 \bigoplus 0$ quartet states into the spin $S=1$ triplet states.
   This is represented graphically in Fig. \ref{fig:AKLT} by the circles, which represent projection operators tying together two $S=1/2$ spins,
   projecting out the spin $S=0$ or singlet state and preserving only the spin $S=1$ or triplet states.

  For every adjacent pair of $S=1$ spins, two of the four constituent $S=1/2$ spins are projected into a state with a total spin of zero by the valence bond.
 Therefore, the pair of $S=1$ spins is forbidden from existing in a combined spin $S=2$ state. This condition can be realized by considering a Hamiltonian that
 is a sum of projectors $P_{i,i+1}$ that project the pairs of $S=1$ spins from the $1\bigotimes 1=2\bigoplus 1 \bigoplus 0$ space into the spin $S=2$ subspace,
 \begin{subequations}
 \begin{equation}
 \label{haklt}
  H_{\rm AKLT}=\sum_iP_{i,i+1}.
  \end{equation}
  Because the projection operators $P_{i,i+1}$ are positive semi-definite, the ground state satisfies $H_{AKLT}|\Psi_G\rangle=0$ and is simply the
  AKLT state. The projection operator $P_{i,i+1}$ can be written in terms of spin-$1$ operators as follows \cite{AKLT}:
 \begin{equation}
 \label{paklt}
  P_{i,i+1}={1\over3}+{1\over2}(\mathbf{S}_i\cdot\mathbf{S}_{i+1})+{1\over6}(\mathbf{S}_i\cdot\mathbf{S}_{i+1})^2.
 \end{equation}
 \end{subequations}
  The AKLT state is important because it is an explicit spin wavefunction that realizes the Haldane phase for integer spins (see section II). In particular,
 it is easy to see from Fig.  \ref{fig:AKLT} that an unpaired $S=1/2$ spin will be left at each end of the spin chain, which is a realization of the
 end state discussed in section II for $S=1$ Heisenberg spin chains. In the following, we show how the AKLT state can be written as an MPS state.

  The AKLT state can be constructed in two steps. First, we split each site $i$ in the spin-$1$ chain into two sites $i_L$ and $i_R$, thereby
  forming a spin-$1/2$ chain with $2N$ sites, as in Fig. \ref{fig:AKLT} (where $N$ is the number of sites in the parent spin-$1$ chain)
  and construct a dimerized chain in which the spins at sites $i_R$ and ${i+1}_L$ ($i=1,2,\cdots,N$) are joined by a valence bond
 (see Eq.~\eqref{valencebond}). The singlet bond between sites $i_R$ and ${i+1}_L$ can be written as
  \begin{equation}
  \label{mps5}
   (i,i+1)=\sum_{\sigma_{i_R},\sigma_{{i+1}_L}}R_{\sigma_{i_R},\sigma_{{i+1}_L}}|\sigma_{i_R}\rangle|\sigma_{{i+1}_{L}}\rangle,
  \end{equation}
   where $\sigma=\uparrow,\downarrow$ and the $R_{\sigma\sigma'}$ are the components of a $2\times2$ matrix:
   \begin{equation}
  \mathbf{R} =\left(
 \begin{array}{cc}
  0 & {1\over\sqrt{2}} \\
  -{1\over\sqrt{2}} & 0 \end{array}\right).  \label{R}
  \end{equation}
 In this representation, the wavefunction of the dimerized spin-$1/2$ chain can be written as
  \begin{equation}
  \label{mps6}
  |\Psi\rangle=\sum_{\sigma_{1_R},\cdots,\sigma_{N_L}}R_{\sigma_{1_R}\sigma_{2_L}}\cdots R_{\sigma_{{N-1}_R}\sigma_{N_L}}
  |\sigma_{1_R},\cdots,\sigma_{N_L}\rangle.
  \end{equation}
   Note that this state is a direct product state of $S=1/2$ RVB singlet pairs with the two end spins ($\sigma_{1_L}$ and $\sigma_{N_R}$) unspecified.

  Next, we project the two $S=1/2$ spins at sites $i_L$ and $i_R$ to the spin-$1$ states $|1,m\rangle$ ($m=0,\pm1$) with
  \begin{eqnarray}
   \label{pro1}
  |1,1\rangle & = & |\uparrow\uparrow\rangle, \\ \nonumber
  |1,0\rangle & = & {1\over\sqrt{2}}\left(|\uparrow\downarrow\rangle+|\downarrow\uparrow\rangle\right), \\ \nonumber
  |1,-1\rangle & = & |\downarrow\downarrow\rangle.
  \end{eqnarray}
  This projection can be expressed in terms of three matrices, $\mathbf{M}^{0,\pm1}$, where
  \begin{equation}
  \label{pro1m}
  |1,m\rangle=\sum_{\sigma,\sigma'}M^m_{\sigma\sigma'}|\sigma\rangle|\sigma'\rangle
  \end{equation}
  with
  \begin{subequations}
  \begin{equation}
  \mathbf{M}^1 =\left(
 \begin{array}{cc}
  1 & 0 \\
  0 & 0 \end{array}\right),  \label{m1}
  \end{equation}
  \begin{equation}
  \mathbf{M}^{-1} =\left(
 \begin{array}{cc}
  0 & 0 \\
  0 & 1 \end{array}\right),  \label{m-1}
  \end{equation}
 and
 \begin{equation}
  \mathbf{M}^0 =\left(
 \begin{array}{cc}
  0 & {1\over\sqrt{2}} \\
  {1\over\sqrt{2}} & 0 \end{array}\right).  \label{m0}
  \end{equation}
  \end{subequations}
  Thus, the AKLT state can be written as
  \begin{equation}
  \label{mpaklt}
  |\Psi_{\rm AKLT}\rangle=\sum_{s_1,s_2,\cdots,s_N}\phi_{\rm AKLT}(s_1,\cdots,s_N)|s_1,s_2,\cdots,s_N\rangle,
  \end{equation}
   where $s_i=0,\pm1$ and
  \begin{subequations}
  \begin{eqnarray}
  \phi_{\rm AKLT}(s_1,\cdots,s_N) & = & \sum_{\sigma_{1_R},\cdots,\sigma_{N_L}} [M^{s_1}_{\sigma_{1_L}\sigma_{1_R}} R_{\sigma_{1_R}\sigma_{2_L}} \nonumber \\
   & & \times M^{s_2}_{\sigma_{2_L}\sigma_{2_R}}\cdots R_{\sigma_{{N-1}_R} \sigma_{N_L}} ]  \nonumber \\
   & = & [\mathbf{A}^{s_1}\mathbf{A}^{s_2}\cdots\mathbf{A}^{s_N}]_{\sigma_{1_L}\sigma_{N_R}}.
  \label{mps7}
  \end{eqnarray}
  Here, $\mathbf{A}^{s}=\mathbf{M}^s\mathbf{R}$, and $\sigma_{1_L},\sigma_{N_R}=\uparrow,\downarrow$ correspond to four degenerate ground states on an open chain.
  Imposing the periodic boundary condition gives rise to a non-degenerate ground state with
  \begin{equation}
  \phi_{\rm AKLT}(s_1,\cdots,s_N)  =  {\rm Tr} [\mathbf{A}^{s_1}\mathbf{A}^{s_2}\cdots\mathbf{A}^{s_N}].
  \label{mps8}
  \end{equation}
  \end{subequations}

\subsubsection{PEPSs in higher dimensions and beyond}

   The AKLT construction can be extended to construct other types of VBS states and states in higher dimensions. Straightforward examples
  include $S=2$ VBS states on a square lattice and $S=3/2$ VBS states on a honeycomb lattice \cite{AKLTCMP}. These states can be written as PEPSs in their
  respective lattices.

    \begin{figure}[tbph]
    \includegraphics[width=7.8cm]{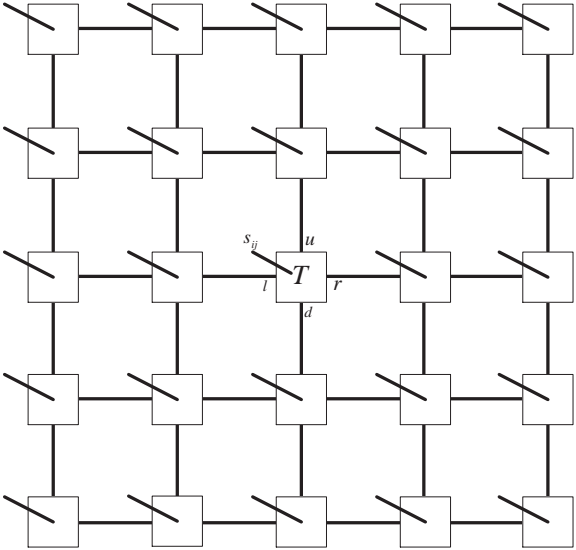}
    \caption{ Graphical representation of a PEPS in terms of contracted tensors (tensor network).
    Each box denotes a tensor $T$ with components $T^{s_{ij}}_{lrud}$ at site $ij$, where $l$, $u$, $r$, and $d$ are tensor indices related to
    left, right, up and down bonds, respectively, linking to their neighbors;
    the open lines represent the physical spin states $s_{ij}$; and the connected lines represent the contraction of the tensors.}
    \label{fig:PEPS}
    \end{figure}

  For instance, on a square lattice with a coordination number of 4, a generic PEPS wavefunction can be written in terms of
  rank 4 tensors as follows:
  \begin{subequations}
  \begin{equation}
  \label{peps1}
  |\Psi \rangle=\sum_{[s_{ij}]}\phi([s_{ij}])|[s_{ij}]\rangle,
  \end{equation}
  where $i,j=1,\cdots,N$ for an $N\times N$ system, $[s_{ij}]=(s_{11},\cdots ,s_{1N},s_{21},\cdots ,s_{2N},\cdots,s_{N1},\cdots,s_{NN})$
  denotes a spin configuration, and
 \begin{equation}
  \label{peps2}
  \phi([s_{ij}])={\rm Tr}[T^{s_{11}}\cdots T^{s_{1N}}T^{s_{21}}\cdots T^{s_{NN}}].
  \end{equation}
  where, the $T^s$s are rank 4 tensors with components
  $$T_{lrud}^{s_{ij}},$$
  where $s_{ij}$ is the physical spin index; $l$, $r$, $u$, and $d$ represent links connected to the tensors at the left, right, up and down neighboring sites
  $(i-1,j)$, $(i+1,j)$, $(i,j-1)$, and $(i,j+1)$, respectively; and ``Tr"  means tensor contraction. The above mathematical expression of tensor contraction is usually
  represented by diagrams such as that shown in Fig. \ref{fig:PEPS} for a square lattice, where connected lines represent the contraction of tensors with the same index and open lines
  represent the physical spin states $s_{ij}=-S,\cdots,S$.
  \end{subequations}

    \begin{figure}[tbph]
    \includegraphics[width=8.0cm]{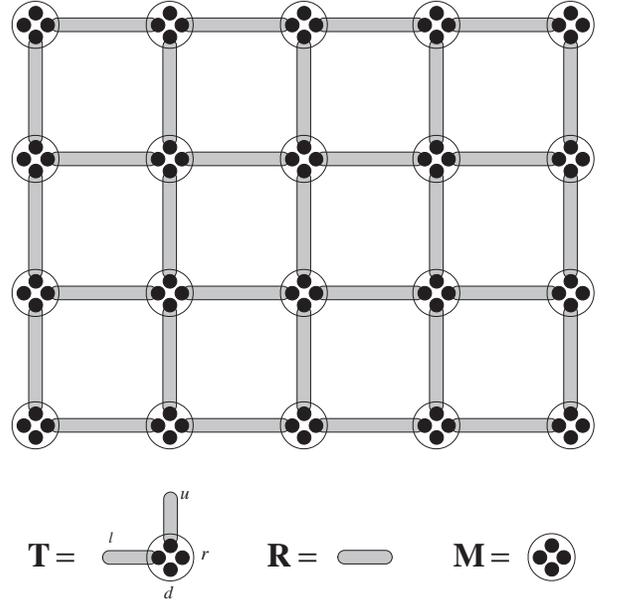}
    \caption{ The VBS construction of an $S=2$ AKLT state on a square lattice and the corresponding tensors.}
    \label{fig:AKLT2D}
    \end{figure}

  As an example, a spin $S=2$ AKLT state on a square lattice can be written in PEPS form as shown in Fig. \ref{fig:AKLT2D}.
  The tensors $T^s$ can be obtained using the VBS construction with the tensors $\mathbf{R}$ and $\mathbf{M}^{s}$, as in one dimension.
  The tensor $\mathbf{R}$ is still defined by Eq.~\eqref{R}. The tensors $\mathbf{M}^{s}$, $s=0,\pm 1, \pm 2$, project
  a state consisting of four $S=1/2$ spins in the auxiliary Hilbert space ${1\over2}\bigotimes{1\over2}\bigotimes{1\over2}\bigotimes{1\over2}=2\bigoplus 1\bigoplus 0$
  into the physical $S=2$ spin space, whose components are given by
  \begin{equation}
  M^{s}_{\sigma_l \sigma_r \sigma_u \sigma_d}=\langle s|\sigma_l \sigma_r \sigma_u \sigma_d\rangle, \label{M:AKLT2}
  \end{equation}
  where $\sigma_l,\sigma_r,\sigma_u,\sigma_d=\uparrow,\downarrow$. The tensor $\mathbf{T}$ is given by
  \begin{equation}
  T^{s}_{\sigma_l \sigma_r \sigma_u \sigma_d}=\sum_{\sigma_{l'},\sigma_{u'}} M^{s}_{\sigma_{l'} \sigma_r \sigma_{u'} \sigma_d}
  R_{\sigma_l \sigma_{l'}} R_{\sigma_u \sigma_{u'}}. \label{T:AKLT2}
  \end{equation}
  The tensor product state constructed from the above $T^{s}$s give rise to the $S=2$ AKLT state on a square lattice.

  The VBS construction can be further extended by ``fractionalizing" the spins in more exotic ways (for example, using the Majorana fermion
  representation of spins). In this way, we can write the toric code model \cite{Kitaev03} in the PEPS form as well as the Kitaev honeycomb model \cite{Kitaev06}
  (with a residual fermionic degree of freedom at each site; see section \ref{sec:Kitaev}).
  The relation between RVB states and PEPSs has also been exploited to show that some RVB states can be written as PEPSs \cite{Verstraete06a,Schuch12,Wang13,Poiblanc13}.
  However, the general relation between RVB states and PEPSs remains unclear.

  The PEPS construction provides a way to describe entanglement among local spins based on the construction of local pairs, and its application
  to geometrically frustrated lattices is limited. To overcome this limitation, researchers have extended the pair construction procedure to
  consider entanglement between more than two sites, say, a cluster or a simplex, to construct projected states.
  These projected entangled simplex states form the basis for more elaborate numerical approaches \cite{Xie14}.
  Combined with numerical techniques (tensor-based renormalization), these tensor-network methods now provide an alternative means of
  constructing variational wavefunctions.
  Readers can refer to references \cite{CiracVerstraeteJPA09,VCM08,Orus13} for details.

\subsection{Kitaev honeycomb model and related issues}\label{sec:Kitaev}

It was previously believed that spin rotational symmetry is essential for a QSL state that supports fractional spinon
 excitations. If the spin rotational symmetry is broken, the system tends to approach an ordered state. Kitaev \cite{Kitaev06} provided a counterexample to this belief through an unusual, exactly solvable model in two dimensions with strong spin-orbit coupling, which destroys
 the spin rotational symmetry, but in which deconfined spinons nevertheless exist on top of the QSL ground states. This famous model is now called the Kitaev honeycomb model. In this section, we briefly review the Kitaev honeycomb model to see how exotic ground states and low-energy excitations emerge
 from this model with broken rotational symmetry. The possibility of the realization of Kitaev-like models in realistic materials is also discussed.

\begin{figure}[tbph]
\includegraphics[width=8.0cm]{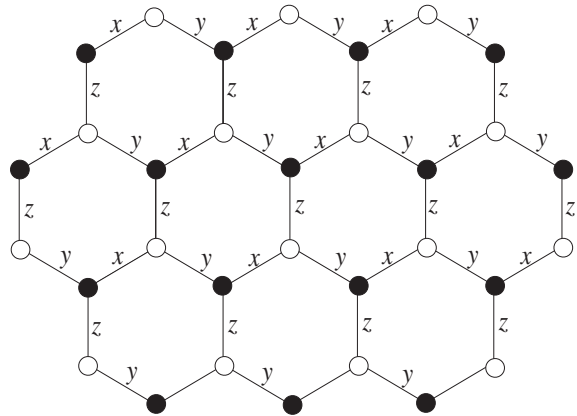}
\caption{ Kitaev honeycomb model. $x$, $y$ and $z$ denote three types of
links in the honeycomb lattice.}
\label{KitaevHoneycomb}
\end{figure}

 Kitaev considered a spin-$1/2$ model on a honeycomb lattice with spin-orbit coupling \cite{Kitaev06}. He divided all nearest neighbor bonds in the honeycomb lattice
into three types, called \textquotedblleft $x$-links\textquotedblright,
\textquotedblleft $y$-links\textquotedblright\ and \textquotedblleft $z$%
-links\textquotedblright\, as shown in Fig. \ref{KitaevHoneycomb}. The
Hamiltonian is given as follows:
\begin{equation}
H=-J_{x}\sum_{x\text{-link}}K_{ij}-J_{y}\sum_{y\text{-link}%
}K_{ij}-J_{z}\sum_{z\text{-link}}K_{ij},  \label{H-Kitaev}
\end{equation}%
where $K_{ij}$ is defined as
\begin{equation}
K_{ij}=\left\{
\begin{array}{cc}
\sigma _{i}^{x}\sigma _{j}^{x}, & \text{if }(i,j)\text{ is a }x\text{-link,}
\\
\sigma _{i}^{y}\sigma _{j}^{y}, & \text{if }(i,j)\text{ is a }y\text{-link,}
\\
\sigma _{i}^{z}\sigma _{j}^{z}, & \text{if }(i,j)\text{ is a }z\text{-link.}%
\end{array}%
\right.  \label{Kij}
\end{equation}
  Note the strong anisotropy in the spin-spin couplings $K_{ij}$.

\begin{figure}[tbph]
\includegraphics[width=4.0cm]{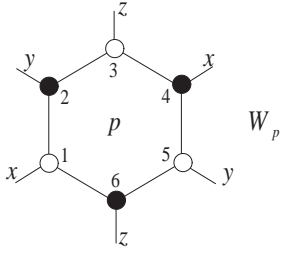}
\caption{ Loop operator $W_{p}=\protect\sigma _{1}^{x}\protect\sigma _{2}^{y}\sigma _{3}^{z}\protect\sigma _{4}^{x}\protect\sigma _{5}^{y}\sigma _{6}^{z}$ on a lattice plaquette (hexagon).}
\label{KitaevLoop}
\end{figure}

  We first consider the following loop operators $W_p$ defined for a hexagonal loop:
\begin{equation}
W_{p}\equiv \sigma _{1}^{x}\sigma _{2}^{y}\sigma _{3}^{z}\sigma
_{4}^{x}\sigma _{5}^{y}\sigma _{6}^{z}=K_{12}K_{23}K_{34}K_{45}K_{56}K_{61},
\label{W-plquette}
\end{equation}%
 where $p$ is used to label the lattice plaquettes (hexagons), as shown in Fig. \ref{KitaevLoop}. It is easy to verify that $[W_{p},K_{ij}]=0$; therefore, $%
 [H,W_{p}]=0$. Hence, the $W_{p}$s serve as good quantum numbers for the Hamiltonian given in Eq.~\eqref{H-Kitaev}, and the total Hilbert space for spins can be divided into
 a direct product of sectors that are eigenspaces of $\{W_{p}\}$. However, the eigenvalue problem cannot be completely solved by determining the eigenspaces of $\{W_p\}$.
 Each $W_{p}$ has only two eigenvalues, $w_{p}=\pm 1$. Each plaquette contains six sites, and each site is shared by three plaquettes.
 Therefore, the number of plaquettes is given by $m=N/2$, where $N$ is the number of sites. It follows that the dimension of each eigenspace of $\{W_p\}$ is
 $2^{N}/2^{m}=2^{N/2}$, i.e., splitting the Hilbert space into eigenspaces of $\{W_{p}\}$ cannot solve the eigenvalue problem completely.

 Kitaev realized that to solve the model Hamiltonian given in Eq.~\eqref{H-Kitaev}, spins can be written in terms of four Majorana fermions, because a Majorana fermion
 can be viewed as the real or imaginary part of a complex fermion. To illustrate this approach, we rewrite the complex fermions $f_{\uparrow }$ and
$f_{\downarrow }$ in Eq.~\ (\ref{slaveparticle}) in terms of four Majorana fermions $c_{1}$, $c_{2}$, $c_{3}$ and $c_{4}$:%
\begin{subequations}
\begin{equation}
\begin{array}{cc}
f_{\uparrow }=\frac{1}{2}(c_{1}+ic_{2}), & f_{\uparrow }^{\dag }=\frac{1}{2}%
(c_{1}-ic_{2}), \\
f_{\downarrow }=\frac{1}{2}(c_{3}+ic_{4}), & f_{\downarrow }^{\dag }=\frac{1%
}{2}(c_{3}-ic_{4}),%
\end{array}
\label{MFCF}
\end{equation}%
where the operators $c_{\alpha }$ ($\alpha =1,2,3,4$) are Hermitian and satisfy%
\begin{equation}
c_{\alpha }c_{\beta }+c_{\beta }c_{\alpha }=2\delta _{\alpha \beta }.
\label{MF1}
\end{equation}%
\end{subequations}
 Thus, the three spin components read $\sigma ^{x}=\frac{i}{2}(c_{1}c_{4}-c_{2}c_{3})$, $\sigma ^{y}=\frac{i}{2}(c_{3}c_{1}-c_{2}c_{4})$,
 and $\sigma ^{z}=\frac{i}{2}(c_{1}c_{2}-c_{3}c_{4})$. The single-occupancy condition $f_{\uparrow }^{\dag }f_{\uparrow }+f_{\downarrow }^{\dag
 }f_{\downarrow }=1$ (and $f_{\uparrow }^{\dag }f_{\downarrow}^{\dag}=f_{\uparrow }f_{\downarrow }=0$) becomes%
\begin{equation}
c_{1}c_{2}+c_{3}c_{4}=c_{1}c_{3}+c_{2}c_{4}=c_{1}c_{4}+c_{3}c_{2}=0,
\label{single_occupancyMF}
\end{equation}%
 which can be simplified to the single equation $c_{1}c_{2}c_{3}c_{4}=1$. Using these constraints, the spin operators can be written as
 $\sigma^{x}=ic_{1}c_{4}$, $\sigma ^{y}=-ic_{2}c_{4}$, and $\sigma^{z}=-ic_{3}c_{4}$. Rewriting $b_{x}=c_{1}$, $b_{y}=-c_{2}$,
 $b_{z}=-c_{3}$ and $c=c_{4}$, we arrive at the Kitaev representation
\begin{eqnarray}
\sigma ^{x} &=&ib^{x}c,  \notag \\
\sigma ^{y} &=&ib^{y}c,  \label{Majorana4} \\
\sigma ^{z} &=&ib^{z}c,  \notag
\end{eqnarray}%
with the constraint%
\begin{equation}
D\equiv b^{x}b^{y}b^{z}c=1.  \label{Majorana4D}
\end{equation}%
 The Majorana representation without constraints is redundant and enlarges the physical spin Hilbert space. Note that $D^{2}=1$ and that $D$ has two eigenvalues, $D=\pm 1$,
 thereby splitting the local Hilbert space into two sectors. The physical spin Hilbert space corresponds to the sector with all $D_{j}=1$. Therefore, the physical
 spin wavefunction $\left\vert \Psi _{spin}\right\rangle $ can be obtained from the Majorana
 fermion wavefunction $\left\vert \Psi _{Majorana}\right\rangle $ through the projection%
\begin{equation}
\left\vert \Psi _{spin}\right\rangle =\prod_{j}\frac{1+D_{j}}{2}\left\vert
\Psi _{Majorana}\right\rangle ,  \label{ProjectionMF}
\end{equation}%
 which retains the $D_{j}\equiv 1$ sector and removes all other sectors in the enlarged Hilbert space. Note that
 $\frac{1+D_{j}}{2}=n_{j\uparrow}+n_{j\downarrow }-2n_{j\uparrow }n_{j\downarrow }$ and that Eq.~\eqref{ProjectionMF} is nothing but the Gutzwiller projection.
 In addition, note that $D_{j}$ serves as a $Z_{2}$ gauge transformation in the enlarged Hilbert space ($D_{j}b_{j}^{\alpha }D_{j}=-b_{j}^{\alpha}$,
 $D_{j}c_{j}D_{j}=-c_{j}$) and commutes with the spin operators ($[D_j,\sigma^{\alpha}_j]=0$, $\alpha=x,y,z$) and thus with the Hamiltonian. As
 a result, the Gutzwiller projection is ``trivial" in the sense that $\prod_{j}\frac{1+D_{j}}{2}\left\vert\Psi _{Majorana}\right\rangle$ is an eigenstate of
 $H$ in the projected Hilbert space as long as $\left\vert\Psi_{Majorana}\right\rangle$ is an eigenstate of $H$ in the ``unprojected" Hilbert space and
 $\prod_{j}\frac{1+D_{j}}{2}\left\vert\Psi _{Majorana}\right\rangle\neq0$.

\begin{figure}[tbph]
\includegraphics[width=8.4cm]{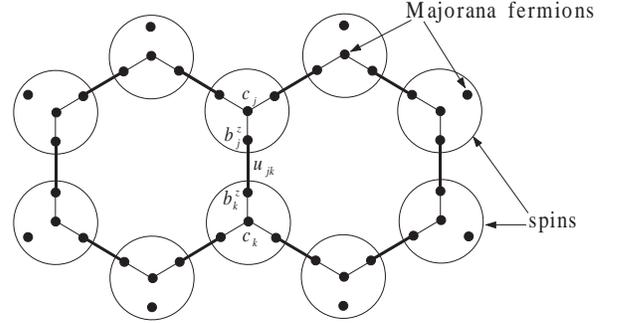}
\caption{ Graphic representation of the four-Majorana-fermion decomposition of the Hamiltonian expressed in Eq.~\eqref{H-Kitaev}.}
\label{KitaevMajoranaF}
\end{figure}

In the Majorana fermion representation, $K_{ij}$ in Eq.~\eqref{Kij} becomes%
\begin{equation}
K_{ij}=-i(ib_{i}^{\alpha }b_{j}^{\alpha })c_{i}c_{j},  \label{Kij2}
\end{equation}%
where $\alpha =x,y,z$ depends on the type of link $(ij)$. The operator $%
ib_{i}^{\alpha }b_{j}^{\alpha }$ is Hermitian, and we denote it by $\hat{u}%
_{ij}=ib_{i}^{\alpha }b_{j}^{\alpha }$. Thus, we may write%
\begin{subequations}
\begin{equation}
H=\frac{i}{4}\sum_{\langle j,k\rangle }\hat{A}_{jk}c_{j}c_{k},
\label{H-KitaevMF}
\end{equation}%
with%
\begin{equation}
\hat{A}_{jk}=2J_{\alpha (jk)}\hat{u}_{jk},\, \hat{u}_{jk}=ib_{j}^{\alpha
}b_{k}^{\alpha },  \label{Ajk}
\end{equation}%
\end{subequations}
 where $\langle j,k\rangle $ denotes nearest neighbor links on the honeycomb lattice and, by definition, $\hat{u}_{jk}=-\hat{u}_{kj}$ and $\hat{A}_{jk}=-\hat{A}_{kj}$. The Hamiltonian structure in this Majorana fermion representation is shown schematically in Fig. \ref{KitaevMajoranaF}. Note that $[H,\hat{u}_{jk}]=0$
 and $[\hat{u}_{jk},\hat{u}_{j^{\prime }k^{\prime }}]=0$. The enlarged Hilbert space of Majorana fermions can be decomposed into common eigenspaces of
 $\{\hat{u}_{jk}\}$ indexed by the corresponding eigenvalues $u_{jk}=\pm 1$. Thus, the Hamiltonian in the invariant subspace indexed by $u=\{u_{jk}\}$ becomes%
\begin{equation}
H_{u}=\frac{i}{4}\sum_{\langle j,k\rangle
}A_{jk}c_{j}c_{k},A_{jk}=2J_{\alpha (jk)}u_{jk},  \label{H-KitaevMFu}
\end{equation}%
 where we have replaced $\hat{A}_{jk}$ and $\hat{u}_{jk}$ with their eigenvalues. Note that $u_{jk}\rightarrow-u_{jk}$ upon the $Z_2$ gauge transformation
 $u_{jk}\rightarrow D_ju_{jk}D_j$, and it is more convenient to classify the eigenstates of $H$ in terms of the gauge-invariant loop operator $W(j_{0},\cdots
,j_{n})=K_{j_{n}j_{n-1}}\cdots K_{j_{1}j_{0}}$, which can be written as%
\begin{equation}
W(j_{0},\cdots ,j_{n})=\left( \prod_{s=1}^{n}-i\hat{u}_{j_{s}j_{s-1}}\right)
c_{n}c_{0}.  \label{W-Loop-Majorana}
\end{equation}%
The closed-loop operator $W_{p}$ (see Eq.~\eqref{W-plquette}) is gauge
invariant under the $Z_2$ transformation because $c_{n}=c_{0}$, and the gauge-invariant
quantities $w=\{w_{p}\}$ can be used instead of $%
u=\{u_{jk}\}$ to parameterize the eigenstates, i.e.,%
\begin{equation}
H_{w}=\frac{i}{4}\sum_{\langle j,k\rangle }A_{jk}c_{j}c_{k}.
\label{H-KitaevMFw}
\end{equation}%
For a given set of $A_{ij}$ fixed by $\{w_p\}$, the quadratic Hamiltonian as expressed in Eq.~\eqref{H-KitaevMFu}
and Eq.~\eqref{H-KitaevMFw} can be diagonalized into the following canonical form:%
\begin{equation}
H_{canonical}=\frac{i}{2}\sum_{m}\epsilon _{m}c_{m}^{\prime }c_{m}^{\prime
\prime }=\sum_{m}\epsilon _{m}\left( f_{m}^{\dagger }f_{m}-\frac{1}{2}%
\right) ,  \label{H-canonical-Majorana}
\end{equation}%
where $\epsilon _{m}\geq 0$, $c_{m}^{\prime }$ and $c_{m}^{\prime \prime }$
are normal Majorana modes, and $f_{m}^{\dagger }=\frac{1}{2}(c_{m}^{\prime
}-ic_{m}^{\prime \prime })$ and $f_{m}=\frac{1}{2}(c_{m}^{\prime
}+ic_{m}^{\prime \prime })$ are the corresponding complex fermion operators. The
ground state of the Majorana system has an energy of%
\begin{equation}
E=-\frac{1}{2}\sum_{m}\epsilon _{m}.  \label{Eg-MF}
\end{equation}

 We now discuss the system of Majorana fermions on the honeycomb lattice.
 First, we note that the global ground-state energy does not depend on the signs of the
exchange constants $J_{x}$, $J_{y}$, and $J_{z}$. For instance, if $J_{z}$ is replaced
with $-J_{z}$, we can compensate for this sign change by changing the signs of the
variables $u_{jk}$ for all $z$-links using the gauge operator $D_{j}$,
leaving the values of $A_{jk}$ and $w_{p}$ unchanged. Therefore, as far as solving for
the ground-state energy and the excitation spectrum is concerned, the signs of the exchange
constants $J$ do not matter. However, such a sign change does affect other measurable physical
quantities.

Second, it was proven by Lieb \cite{Lieb94} and numerically investigated by Kitaev himself
that the ground state of the Majorana system is achieved when the system is in the vortex-free
configuration, namely, $w_{p}=1$ for all plaquettes $p$. In this vortex-free
configuration, one can solve for the (fermionic) energy spectrum of the Hamiltonian by
 directly Fourier transforming Eq.~\eqref{H-KitaevMFw} to obtain%
\begin{equation}
\epsilon _{\mathbf{q}}=\pm |J_{x}e^{i\mathbf{q}\cdot \mathbf{a}}+J_{y}e^{i%
\mathbf{q}\cdot \mathbf{b}}+J_{z}|,  \label{Eq-MajoranaF}
\end{equation}%
 where $\mathbf{a=(}\frac{1}{2},\frac{\sqrt{3}}{2}\mathbf{)}$ and $\mathbf{b=(-}\frac{1}{2},\frac{\sqrt{3}}{2}\mathbf{)}$ are two basis vectors in the $%
 xy$ coordinates. The fermionic spectrum may or may not be gapped, depending on whether a solution to the equation $\epsilon _{\mathbf{q}}=0$ exists.
 $\epsilon _{\mathbf{q}}=0$ has a solution if and
only if $|J_{x}|$, $|J_{y}|$, and $|J_{z}|$ satisfy the triangle inequalities:%
\begin{equation}
|J_{x}|\leq |J_{y}|+|J_{z}|,|J_{y}|\leq |J_{z}|+|J_{x}|,|J_{z}|\leq
|J_{x}|+|J_{y}|.  \label{Jxyz-phase}
\end{equation}

\begin{figure}[tbph]
\includegraphics[width=6.4cm]{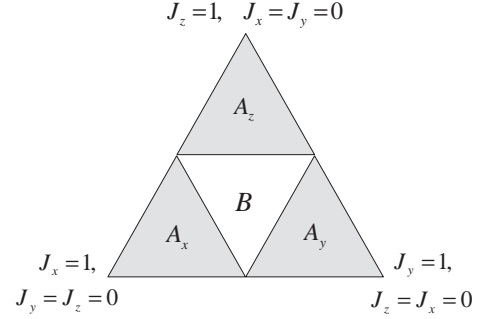}
\caption{ Phase diagram of the Kitaev honeycomb model. The triangle is the
section of the positive octant ($J_x,J_y,J_z\geq 0$) that lies in the plane $J_x+J_y+J_z=1$. The $A$ phase contains three gapped subphases. The $B$ phase is
gapless.}
\label{KitaevPhase}
\end{figure}

As a result, two phases exist in the system of Majorana fermions on the honeycomb lattice, with the phase diagram shown in Fig. \ref{KitaevPhase}. The
 first phase, called the $A$ phase, is gapped and contains three subphases ($%
A_{x}$, $A_{y}$, and $A_{z}$) in the phase diagram. The second, called the $%
B$ phase, is gapless. In the $A$ phase, for example, in the $A_{z}$ subphase, the Hamiltonian expressed in Eq.~\eqref{H-Kitaev}) can be mapped to the Kitaev toric code model
in the limit $|J_{x}|,|J_{y}|\ll |J_{z}|$, and the phase hosts Abelian anyonic excitations. The $B$
 phase acquires an energy gap in the presence of a magnetic field. Very interestingly, it hosts stable non-Abelian anyons when the energy gap is opened
 up by a magnetic field. The $B$ phase is a very attractive state in the context of topological
quantum computation. Readers can refer to the recent review article \cite{NayakRMP08} for details.

 In addition to the elegant Majorana decomposition method pioneered by Kitaev, other insightful approaches
to the Kitaev honeycomb model also exist. For instance, Feng, Zhang and Xiang \cite{Feng07} and Chen and Nussinov \cite{ChenNussinov08}
found that the original Kitaev honeycomb model can be exactly solved with the help of the Jordan-Wigner transformation.
This approach provides a topological characterization of the quantum phase transition from the $A$ phase to the $B$ phase.
A nonlocal string order parameter can be defined in one of these two phases \cite{Feng07,ChenNussinov08}. In the appropriate dual
representations, these string order parameters become local order parameters {\em after some singular transformation},
and a description of the phase transition in terms of Landau's theory of continuous phase transitions becomes applicable \cite{Feng07}.
The Jordan-Wigner transformation also enables a fermionization of the Kitaev honeycomb model, allowing it to be mapped to
a $p$-wave-type BCS pairing problem. The spin wavefunction can be obtained from the fermion model, and
the anyonic character of the vortex excitations in the gapped phase also has an explicit fermionic construction \cite{ChenNussinov08}.

The Kitaev honeycomb model can also be understood within the framework of fermionic RVB theory.
Both confinement-deconfinement transitions from spin liquids to AFM or stripy AF/FM phases and topological quantum phase transitions between gapped
and gapless spin liquid phases can be described within the framework of $Z_2$ gauge theory \cite{Baskaran07,Mandal11,Mandal12}.

Exact diagonalization has been applied to study the Kitaev honeycomb model on small lattices \cite{ChenWangDasSarma10}.
Perturbative expansion methods have been developed to study the gapped phases of the Kitaev honeycomb model and its generalization \cite{Schmidt08,Vidal08,Dusuel08}.
Several papers \cite{Lee07,Yu08a,Yu08b,Kells09} have noted the existence of an analogy between the $Z_2$ vortices in the Kitaev honeycomb model
and the vortices in $p+ip$ superconductors.

 Enormous efforts have been devoted to searching for exactly solvable generalizations of the Kitaev honeycomb model.
 It has been proposed that the exact solvability will not be spoiled when the fermion gap is opened for the non-Abelian phase \cite{Lee07,Yu08b}.
 Generalizations to other lattice models and even to three dimensions have also been developed \cite{Yao07,Yang07,Yao09,Nussinov09,Wu09,Ryu09,Baskaran09,Tikhonov10,Yao11,Lai11}.
 Non-trivial emergent particles, such as chiral fermions \cite{Yao07}, have been constructed in these exactly solvable lattice models.
 These developments have significantly advanced our understanding of emergent phenomena based on solvable models in dimensions greater than one.

 The exotic properties of the Kitaev honeycomb model have motivated researchers to search for realizations of this model in realistic materials.
 It has been demonstrated by Jackeli and Khaliullin \cite{Jackeli09} and by Chaloupka, Jackeli and Khaliullin \cite{Chaloupka10} that a generalization of the Kitaev honeycomb model may
indeed arise in layered honeycomb lattice materials in the presence of strong spin-orbit coupling.
 These authors showed that in certain iridate magnetic insulators (A$_2$IrO$_3$, A=Li, Na), the effective low-energy Hamiltonian for
the effective $J_{eff}=1/2$ iridium moments is given by a linear combination of the AFM Heisenberg model ($H_H$) and the Kitaev honeycomb model ($H_K$),
\begin{equation}
H = (1 - \alpha)H_H + 2\alpha H_K, \label{HHK}
\end{equation}
where $\alpha$, expressed in terms of the microscopic parameters, determines the relative strength of the Heisenberg and Kitaev interactions.
Interestingly, the Kitaev honeycomb model can also be realized as the exact low-energy effective Hamiltonian of a spin-$1/2$ model with
spin rotational and time-reversal symmetries \cite{Wang10}.
The Heisenberg-Kitaev model (\ref{HHK}) exhibits a rich phase diagram. Readers who are interested in these developments may refer to, for example,
references \cite{Chaloupka10,Jiang11,Reuther11,Kimchi11,Singh12,Price12,Schaffer12,Yu13,Lee14,Kimchi14} for details.
 A comprehensive review on this topic has also been published by Nussinov and van den Brink \cite{Nussinov13}.

\section{QSL states in real materials}

   Experimental studies of interacting spins in geometrically frustrated lattices aim at identifying non-trivial and exotic ground states.  Among these ground states, spin liquid states have been sought ever since the proposal of the RVB state \cite{Anderson73}.
   This issue has been intensively debated in the context of the spin states behind the high-Tc superconductivity of cuprates.
  However, before this century, there was no direct observation of spin liquid states. The situation changed in
   2003, when an organic Mott insulator with a quasi-triangular lattice was found to exhibit no magnetic ordering even at tens of mK, four orders of magnitude lower than the energy scale of the exchange interactions \cite{Kanoda03}.
   The low-temperature state is most likely a form of the sought-after spin liquids. Since then, what can be called spin liquids have been successively reported for quasi-triangular, kagome and hyperkagome lattices. In this section, we review the experimental studies mainly with respect to the magnetic and thermodynamic properties of the materials for which sound experimental data have been accumulated in discussing the presence of spin liquids.

 \subsection{Anisotropic triangular lattice systems:  $\kappa$-(ET)$_{2}$Cu$_{2}$(CN)$_{3}$ and EtMe$_{3}$Sb[(Pd(dmit)$_{2}$]$_{2}$}

\begin{figure}[tbph]
\includegraphics[width=8cm]{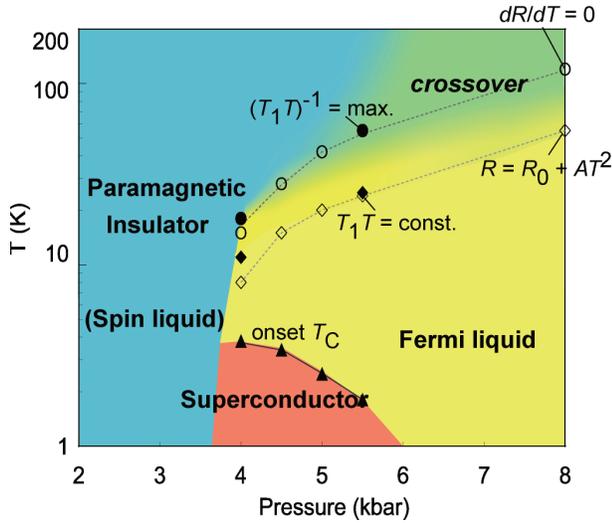}
    \caption{\cite{Kurosaki05} Temperature-pressure phase diagram of the spin-liquid compound with a quasi-triangular lattice, $\kappa$-(ET)$_{2}$Cu$_{2}$(CN)$_{3}$, which undergoes a Mott transition at moderate pressure.} \label{Kuroaki_k_Cu2CN3_Phase_Diagram}
\end{figure}

  Both are half-filled band systems with anisotropic triangular lattices, which are isosceles for $\kappa$-(ET)$_{2}$Cu$_{2}$(CN)$_{3}$ and three different laterals for EtMe$_{3}$Sb[(Pd(dmit)$_{2}$]$_{2}$ \cite{Kanoda06b, Kanoda11, Kato14}.
  At ambient pressure, they are Mott insulators; however, the spins are not ordered at low temperatures on the order of tens of mK.
  A noticeable feature of both systems is that they undergo Mott transitions at moderate pressures 0.4 GPa for $\kappa$-(ET)$_{2}$Cu$_{2}$(CN)$_{3}$ {\cite{Komatsu96, Kurosaki05, Furukawa15a}} and 0.5 GPa for EtMe$_{3}$Sb[(Pd(dmit)$_{2}$]$_{2}$ \cite{Kato06}.
  (Note that these pressure values indicate pressures applied at room temperature and are reduced by approximately 0.2 GPa at low temperatures.)
  The temperature-pressure phase diagram of $\kappa$-(ET)$_{2}$Cu$_{2}$(CN)$_{3}$ is depicted in Fig. \ref{Kuroaki_k_Cu2CN3_Phase_Diagram}.
  A spin liquid phase resides in proximity to the Mott transition; this feature appears to be a key
to the stability of spin liquids and can be closely linked to the metal-insulator transition \cite{Senthil08, ZhouNg13}.
  According to the numerical studies of the anisotropic triangular-lattice Hubbard model, the ground states near to the Mott transition are controversial {\cite{Morita02, Kyung06, Watanabe08, Tocchio13, Laubach15}}, implying that spin-liquid and magnetic phases are competing very closely and can be easily imbalanced by a tiny perturbation.

i)  $\kappa$-(ET)$_{2}$Cu$_{2}$(CN)$_{3}$

\begin{figure}[tbph]
    \includegraphics[width=8cm]{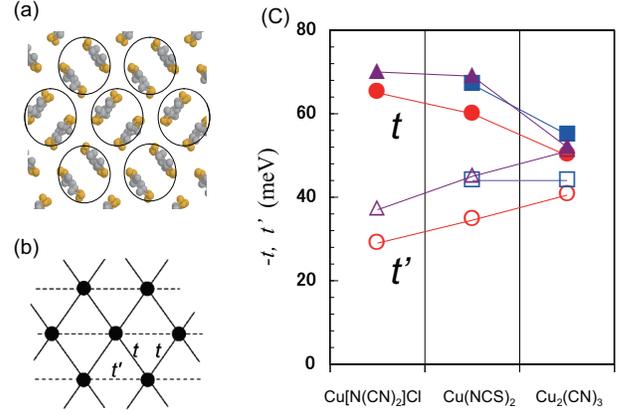}
    \caption{(a) In-plane structure of the ET layer in $\kappa$-(ET)$_{2}$X. It is modelled to (b) an anisotropic triangular lattice. (c) First principles calculations of transfer integrals in $\kappa$-(ET)$_{2}$X for X=Cu[N(CN)$_{2}$]Cl, Cu(NCS)$_{2}$ and Cu$_{2}$(CN)$_{3}$ ; squares \cite{Nakamura09}, circles \cite{Kandpal09}, and triangles \cite{Koretsune14}.}
    \label{Kandpal_Nakamura_transfer}
    \end{figure}

    $\kappa$-(ET)$_{2}$Cu$_{2}$(CN)$_{3}$ is a layered compound, where $\kappa$-(ET)$_{2}$X has a variety of anions X and ET is bis(ethylenedithio)tetrathiafulvalene \cite{Komatsu96}.
    $\kappa$-(ET)$_{2}$X is composed of the ET layers with 1/2 hole per ET and the layers of monovalent anions X$^{-}$,
    which have no contribution to the electronic conduction or magnetism.
    In the ET layer, strong ET dimers are formed (ET)$_{2}$, each of which accommodates a hole
    in an anti-bonding orbital of the highest occupied molecular orbital (HOMO) of the ET.
    As the anti-bonding band is half-filled and the Coulomb repulsive energy is comparable to the band width,
    the family of $\kappa$-(ET)$_{2}$X is good model system to study Mott physics \cite{Kino95,Kanoda97a,Kanoda97b,Kanoda06, Powell11}.
    The estimates of the transfer integrals between the adjacent anti-bonding orbitals on the isosceles triangular lattices, $t$ and $t^{\prime}$,
    are in a range of 50 meV, depending on the method of calculation, e.g., either the molecular orbital (MO)-based tight-binding calculation \cite{Mori84, Mori99, Komatsu96} or the first principles calculation \cite{Nakamura09, Kandpal09, Koretsune14}.
    Nevertheless, one can see that the values have clear systematic variation in terms of anion X, as shown in Fig. \ref{Kandpal_Nakamura_transfer},
    where the values of $t$ and $t^{\prime}$ are calculated via the latter method: the $t^{\prime}/t$ value of $\kappa$-(ET)$_{2}$Cu[N(CN)$_{2}$]Cl is 0.75 (the MO-based calculations) and 0.44-0.52 (first principles calculations),
    while that of $\kappa$-(ET)$_{2}$Cu$_{2}$(CN)$_{3}$ is 1.06 and 0.80-0.99, respectively, suggestive of high geometrical frustration.

\begin{figure}[tbph]
    \includegraphics[width=8cm]{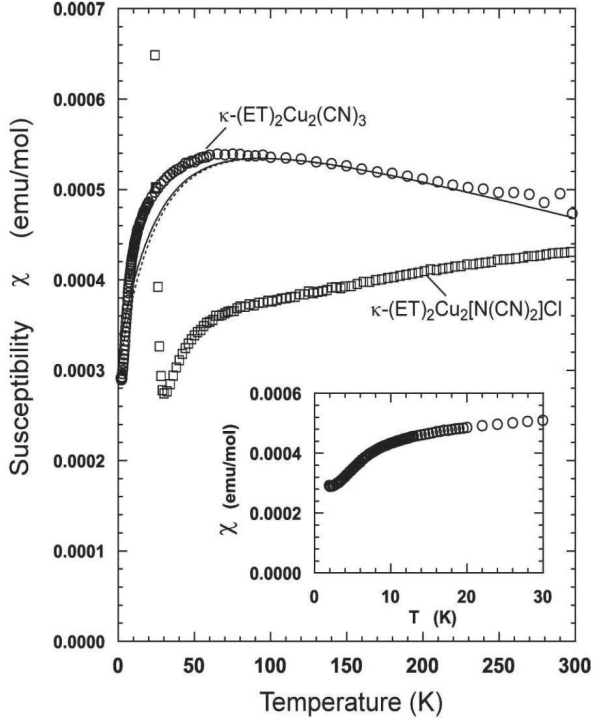}
    \caption{\cite{Kanoda03} Magnetic susceptibility of poly crystalline $\kappa$-(ET)$_{2}$Cu$_{2}$(CN)$_{3}$ and $\kappa$-(ET)$_{2}$Cu[N(CN)$_{2}$]Cl. The core diamagnetic susceptibility has already been subtracted. The solid and dotted lines represent the result of the series expansion of the triangular-lattice Heisenberg model using [6/6] and [7/7] Pad\'{e} approximations with $J$ = 250 K. The susceptibility of $\kappa$-(ET)$_{2}$Cu$_{2}$(CN)$_{3}$ below 30 K is expanded in the inset.}
    \label{Shimizu_sus}
    \end{figure}

   The temperature dependence of the spin susceptibility, $\chi$, of $\kappa$-(ET)$_{2}$Cu$_{2}$(CN)$_{3}$ differs from that of the less frustrated compound
   $\kappa$-(ET)$_{2}$Cu[N(CN)$_{2}$]Cl, as seen in Fig. \ref{Shimizu_sus} \cite{Kanoda03}.
   An abrupt upturn at 27 K in the latter is a manifestation of the antiferromagnetic transition, with a slight spin canting of approximately 0.3 degree \cite{Miyagawa95}.
   However, $\kappa$-(ET)$_{2}$Cu$_{2}$(CN)$_{3}$ has no anomaly in $\chi(T)$. Its overall behavior features a broad peak,
   which is reconciled by the triangular-lattice Heisenberg model with an exchange interaction of $J \sim$ 250 K.
   In contrast to  $\kappa$-(ET)$_{2}$Cu[N(CN)$_{2}$]Cl, the magnetic susceptibility of  $\kappa$-(ET)$_{2}$Cu$_{2}$(CN)$_{3}$ may be described by the Heisenberg model
   because it is situated further from the Mott boundary, while $\kappa$-(ET)$_{2}$Cu[N(CN)$_{2}$]Cl undergoes a Mott transition
   at a low pressure (25 MPa) as it is about to enter a metallic state \cite{Lefebvre00,Kagawa05}.
   There is no indication of magnetic ordering in the susceptibility of $\kappa$-(ET)$_{2}$Cu$_{2}$(CN)$_{3}$, at least down to 2 K, the lowest temperature measured.
   Furthermore, no Curie-like upturn can be identified; the concentration of Cu$^{2+}$ impurity spins detected by ESR is estimated
   to be less than 0.01 \% for $\kappa$-(ET)$_{2}$Cu$_{2}$(CN)$_{3}$ \cite{Kanoda06}.

    \begin{figure}[tbph]
    \includegraphics[width=8cm]{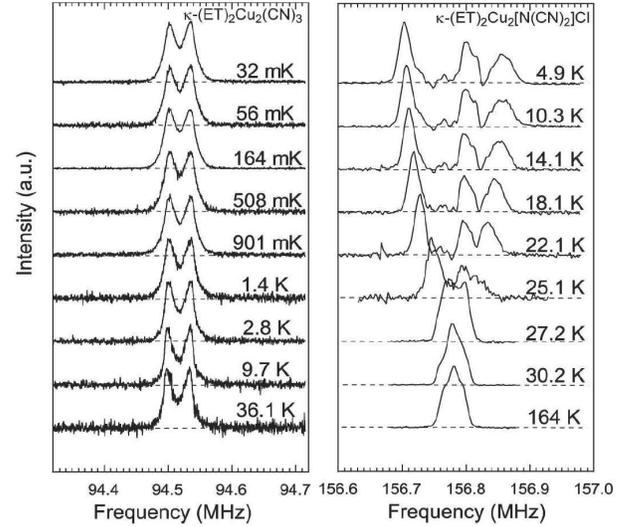}
    \caption{\cite{Kanoda03} $^{1}$H NMR spectra for single crystals of $\kappa$-(ET)$_{2}$Cu$_{2}$(CN)$_{3}$ and $\kappa$-(ET)$_{2}$Cu[N(CN)$_{2}$]Cl.}
    \label{Shimizu_1H_spc}
    \end{figure}

\begin{figure}[tbph]
  \includegraphics[width=8cm]{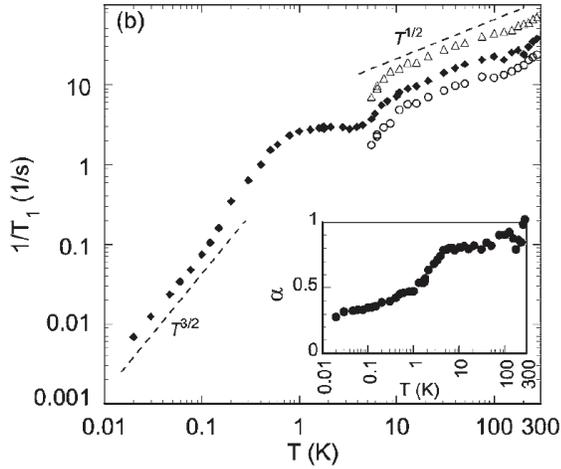}
  \caption{\cite{Kanoda06} $^{13}$C nuclear spin-lattice relaxation rate for a single crystal of $\kappa$-(ET)$_{2}$Cu$_{2}$(CN)$_{3}$. The open triangles and circles represent the relaxation rates of two separated lines coming from two non-equivalent carbon sites in an ET. At low temperatures below 5 K, the two lines merge and are not distinguished. The inset shows the exponent in the stretched exponential fitting to the relaxation curves of the whole spectra, whose relaxation rates are plotted using closed diamonds.}
  \label{Shimizu_13C_T1}
  \end{figure}

  The detailed spin states can be examined by performing NMR measurements, which probe the static and dynamical hyperfine fields at the nuclear sites.
  Fig. \ref{Shimizu_1H_spc} shows the single-crystal $^{1}$H NMR spectra for the two compounds \cite{Kanoda03}.
  A clear line splitting in $\kappa$-(ET)$_{2}$Cu[N(CN)$_{2}$]Cl at 27 K is evidence for commensurate antiferromagnetic ordering,
  with the moment estimated to be 0.45 $\mu_{\textrm{B}}$ per ET dimer in separate $^{13}$C NMR studies \cite{Miyagawa04}.
  However, the spectra for $\kappa$-(ET)$_{2}$Cu$_{2}$(CN)$_{3}$ shows neither a distinct broadening nor splitting down to 32 mK,
  which is four orders of magnitude lower than the $J$ value of 250 K.
  This indicates the absence of long-range magnetic ordering.
  The absence of ordering is also corroborated by zero field $\mu$SR experiments \cite{Pratt11}.
  The nuclear spin lattice relaxation rate, $1/T_{1}$, which probes the spin dynamics, behaves similarly at the $^{1}$H and $^{13}$C sites.
  Fig. \ref{Shimizu_13C_T1} shows $1/T_{1}$ at the $^{13}$C sites, which decreases monotonically with a square-root temperature dependence
  down to 10 K and exhibits a dip-like anomaly at approximately 6 K \cite{Kanoda06}.
  Below 6 K, $1/T_{1}$ levels off down to 1 K or lower, followed by a steep decrease approximated by $T^{3/2}$ at even lower temperatures.
  The two anomalies at 6 K and 1.0 K are obvious. However, they are not so sharp as to be considered as phase transitions.
  Due to the large hyperfine coupling of the $^{13}$C sites located in the central part of ET,
  an electronic inhomogeneity gradually developing on cooling is captured by spectral broadening,
  which is enhanced at approximately 6 K and saturates below 1 K \cite{Kanoda06,Kawamoto06}.
  The detailed NMR \cite{Kanoda06} and $\mu$SR \cite{Pratt11} measurements point to the field-induced emergence of staggered-like moments,
  which is distinct from the conventional magnetic order.
  A separate $\mu$SR study \cite{Goto12} suggests a phase separation.
  The degree of inhomogeneity in the $^{13}$C relaxation curve, which is characterized by the deviation of the exponent
  in the stretched exponential fitting of the relaxation curve (see Inset of Fig. \ref{Shimizu_13C_T1}), increases below 5-6 K \cite{Kanoda06}.
  The $^{1}$H relaxation curve also starts to bend at the much lower temperatures, e.g., below 0.4 K, and fits to a roughly equally weighed sum of two exponential functions,
the 1/$T_{1}$'s of which are proportional to $T$ and $T^{2}$.
  No appreciable field dependence of the $^{13}$C relaxation rate is observed between 2 T and 8 T.
  There is no experimental indication of a finite excitation gap in any of the magnetic measurements.

\begin{figure}[tbph]
  \includegraphics[width=8cm]{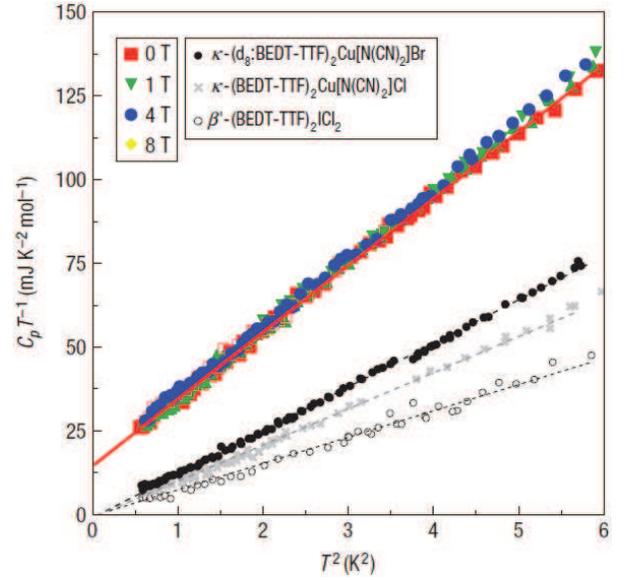}
  \caption{\cite{SYamashita} Low-temperature specific heat $C_p$ of $\kappa$-(ET)$_{2}$Cu$_{2}$(CN)$_{3}$ for several magnetic fields up to 8 T in $C_p/T$ versus $T^{2}$ plots. Those of antiferromagnetic insulators $\kappa$-(ET)$_{2}$Cu[N(CN)$_{2}$]Cl, deuterated $\kappa$-(ET)$_{2}$Cu[N(CN)$_{2}$]Br and $\beta$-(ET)$_{2}$ICl$_{2}$ are also plotted for comparison.}
  \label{SYamashita_specific_heat}
  \end{figure}

   Thermodynamic investigations were conducted by means of the specific heat and thermal conductivity measurements.
   Fig. \ref{SYamashita_specific_heat} shows the specific heat for $\kappa$-(ET)$_{2}$Cu$_{2}$(CN)$_{3}$ and
   several Mott insulators with antiferromagnetic spin ordering \cite{SYamashita}.
   For all of the antiferromagnetic materials, the electronic specific heat coefficient, $\gamma$, is vanishing, as expected for insulators.
   For the $\kappa$-(ET)$_{2}$Cu$_{2}$(CN)$_{3}$ spin liquid system, however, the extrapolation of the $C/T$ vs. $T^{2}$ line
   to absolute zero yields $\gamma$=12$\sim$15 mJ/K$^{2}$mol.
   The linearity holds down to 0.3 K, below which a nuclear Schottky contribution overwhelms the electronic contribution to $C$.
   The finite value despite the Mott insulating state is a marked feature of spin liquids and suggests fermionic excitations in the spin degrees of freedom.
   Interestingly, the low-temperature susceptibility and the $\gamma$ value give the Wilson ratio on the order of unity.
   A spinon Fermi sea is an intriguing model for this phenomenon \cite{Motrunich05}. However, neither the $U(1)$ spin liquid, where $C$ follows $T^{2/3}$ scaling,
   nor the $Z_{2}$ spin liquid, where $C$ is gapped, reconciles the observed features in their original forms.
   Randomness may be an optional parameter to modify the temperature dependence. Another interesting feature is the field-insensitivity up to 8 T,
   which appears incompatible with the $U(1)$ spin liquid states with Dirac cones.

\begin{figure}[tbph]
  \includegraphics[width=8cm]{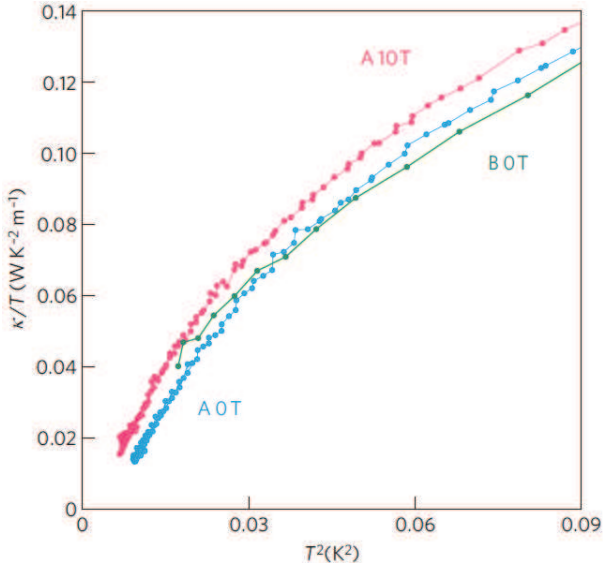}
  \caption{\cite{MYamashita09} Low-temperature thermal conductivity $\kappa$ of $\kappa$-(ET)$_{2}$Cu$_{2}$(CN)$_{3}$ (samples A and B) in $\kappa /T$ versus $T^{2}$ plots. Sample A was investigated at 10 T applied perpendicular to the basal plane, as well as at 0 T.}
  \label{MYamashita_TC}
  \end{figure}

   Thermal transport measurements result in somewhat controversial consequences \cite{MYamashita09}.
   The thermal conductivity divided by the temperature tends to vanish with decreasing temperature, as shown in Fig. \ref{MYamashita_TC}.
   The gap, if one is present, is estimated to be 0.43 K, which is quite small compared with the exchange energy of 250 K.
   The extremely small gap may indicate a gapped $Z_{2}$ spin liquid located near a quantum critical point.
   The discrepancy between the thermal transport and NMR and specific heat data remains an open issue.
   It may be attributed to the Anderson localization of spinons.

   The 6-K anomaly in the NMR spectrum and relaxation rate also manifests itself in the specific heat \cite{SYamashita}
   and thermal conductivity \cite{MYamashita09} as a hump and a shoulder, respectively, indicating that the anomaly is thermodynamic, as well as magnetic.
   However, the thermal expansion coefficient shows a cusp \cite{Manna2010} and the ultrasonic velocity shows a dip-like minimum,
   signifying lattice softening at approximately 6 K \cite{Poirier14}.
   In view of these results, this anomaly is likely associated with spin-lattice coupling.
   Instabilities of the spinon Fermi surfaces (e.g., \cite{Lee05,Galitski07,Grover10,YZhou2011}) are among the possible origins of the anomaly.

   Although the spin liquid is insulating, anomalous charge dynamics are suggested for the low-energy optical and dielectric responses.
   The optical gap for $\kappa$-(ET)$_{2}$Cu$_{2}$(CN)$_{3}$ is much smaller than that for $\kappa$-(ET)$_{2}$Cu[N(CN)$_{2}$]Cl,
   although the former system is situated further from the Mott transition than the latter \cite{Ketzmarki06}.
   It is proposed that gapless spinons are responsible for low-energy optical absorption inside the Mott gap \cite{NgLee07}.
   The dielectric \cite{Abdel-Jawad10}, microwave \cite{Poirier12} and terahertz \cite{Itoh13} responses are enhanced at low temperatures.
   The possible charge-imbalance excitations within the dimer are theoretically proposed \cite{Hotta10, Naka10, Dayal11}.
   Relaxor-like dielectric, transport and optical properties are discussed in terms of coupling with disordered anion layers \cite{Pinteric14,Dressel16}.

ii) EtMe$_{3}$Sb[(Pd(dmit)$_{2}$]$_{2}$

\begin{figure}[tbph]
  \includegraphics[width=8cm]{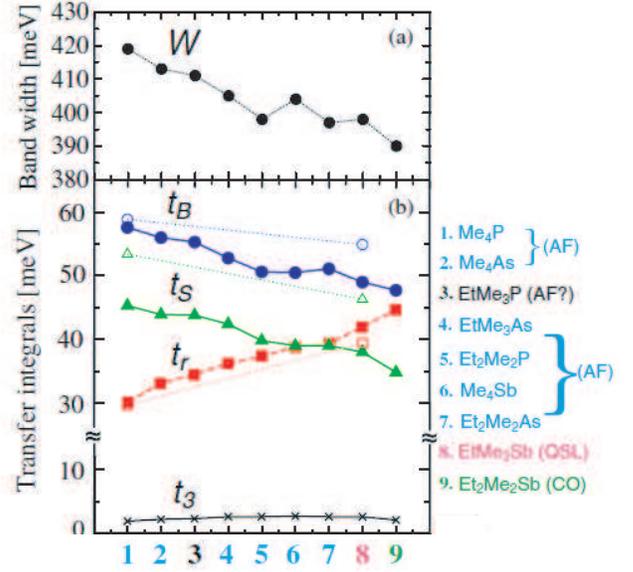}
  \caption{\cite{Tsumuraya13} First principles calculations of band width $W$ (a) and transfer integrals (b) in A[(Pd(dmit)$_{2}$]$_{2}$ for various cations, A. The Pd(dmit)$_{2}$ layers are modeled to triangular lattices characterized by transfer integrals, $t_{\textrm{B}}$, $t_{\textrm{S}}$ and $t_{r}$. $t_{3}$ is the interlayer transfer integral. AF, QSL and CO stand for antiferromagnet, quantum spin liquid and charge-ordered insulator.}
  \label{Tsumuraya_band_cals}
  \end{figure}

   This compound is a member of the A[(Pd(dmit)$_{2}$]$_{2}$ family of materials, which contain a variety of monovalent cations such as A$^{+}$=Et$_{x}$Me$_{4-x}$Z$^{+}$ (Et = C$_{2}$H$_{5}$, Me= CH$_{3}$, Z =N, P, As, Sb, and $x$ = 0, 1, 2), where dmit is 1,3-dithiole-2-thione-4,5-dithiolate \cite{Kato14}
   A[(Pd(dmit)$_{2}$]$_{2}$ is a layered system composed of conducting Pd(dmit)$_{2}$ layers and insulating A layers.
   In the conducting layers, Pd(dmit)$_{2}$ is strongly dimerized as in $\kappa$-(ET)$_{2}$X, whereas the [Pd(dmit)$_{2}$]$_{2}$ dimer accepts an electron from cation A$^{+}$ instead of the hole in ET$_{2}^{+}$.
   A prominent feature of the A[(Pd(dmit)$_{2}$]$_{2}$ family is that the transfer integrals of the three laterals in the triangular lattice can be finely tuned via chemical substitution of A$^{+}$=Et$_{x}$Me$_{4-x}$Z$^{+}$ (Kato, 2014).
   Their first principles calculations are shown in Fig. \ref{Tsumuraya_band_cals} \cite{Tsumuraya13}
   The spin liquid material EtMe$_{3}$Sb[(Pd(dmit)$_{2}$]$_{2}$ is in a region where the three transfer integrals are equalized.
   As expected, the materials situated outside of this region have antiferromagnetic ground states.
   The alloying of the boundary materials offers the chance to study possible critical regions between spin liquids and ordered states \cite{Kato14}.
   There is a charge-ordered material near the spin liquid, suggesting that the charge cannot always be assumed to be separate degrees of freedom from the spin physics.

   Below, we review the properties of EtMe$_{3}$Sb[(Pd(dmit)$_{2}$]$_{2}$ and other related materials.

\begin{figure}[tbph]
  \includegraphics[width=8.4cm]{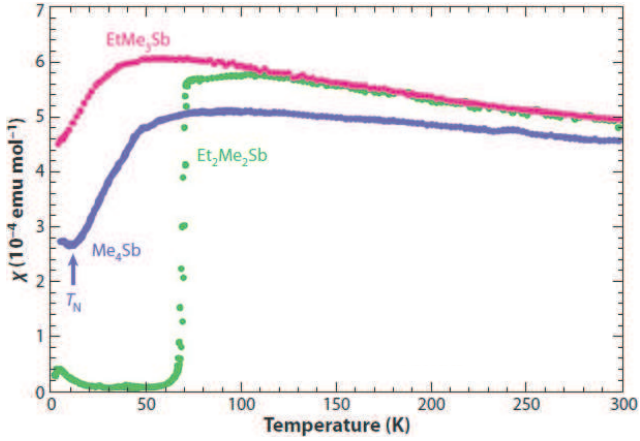}
  \caption{\cite{Kato14} Magnetic susceptibility of an antiferromagnet Me$_{4}$Sb[(Pd(dmit)$_{2}$]$_{2}$, a spin liquid EtMe$_{3}$Sb[(Pd(dmit)$_{2}$]$_{2}$ and a charge-ordered insulator Et$_{2}$Me$_{2}$Sb[(Pd(dmit)$_{2}$]$_{2}$. The core diamagnetic susceptibility has already been subtracted.}
  \label{Kato_sus}
  \end{figure}

   The magnetic susceptibility of EtMe$_{3}$Sb[(Pd(dmit)$_{2}$]$_{2}$ shows a broad peak at approximately 50 K and points to a finite value in the low-temperature limit without any anomaly down to 2K, as shown in Fig. \ref{Kato_sus} \cite{Kato14,Kanoda11}, which is reminiscent of $\kappa$-(ET)$_{2}$Cu$_{2}$(CN)$_{3}$.
   The fitting of the triangular lattice Heisenberg model to the data yields an exchange interaction of 220 K to 280 K, which is nearly the same as for $\kappa$-(ET)$_{2}$Cu$_{2}$(CN)$_{3}$.
   Also shown are the susceptibilities of antiferromagnetic and charge-ordered insulators, which exhibit small kink signaling of magnetic ordering and a sudden decrease indicative of a spin gapful state, respectively, despite their similar behaviors at high temperatures \cite{Tamura02}.
   This indicates that the diversity in the ground states is an outcome of low-energy physics, while the same diversity is not distinguished at high energy scales.

  \begin{figure}[tbph]
  \includegraphics[width=8.4cm]{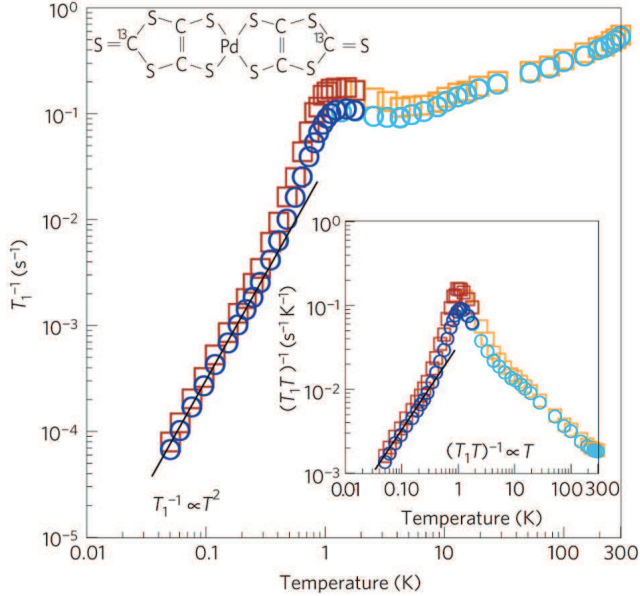}
  \caption{\cite{Itou10} $^{13}$C nuclear spin-lattice relaxation rate $1/T_{1}$ of EtMe$_{3}$Sb[Pd(dmit)$_{2}$]$_{2}$. Inset shows the $1/T_{1}T$ versus $T$ plots. The circles indicate the values determined from the stretched-exponential fitting to the relaxation curves and the squares denote the values determined from the initial decay slopes of the relaxation curves.}
  \label{Itou_13C_T1}
  \end{figure}

   The $^{13}$C NMR captures no signature of magnetic ordering down to 20 mK, although a slight broadening equivalent to the broadening for $\kappa$-(ET)$_{2}$Cu$_{2}$(CN)$_{3}$ is observed at low temperatures \cite{Itou10}.
   The temperature dependence of the $^{13}$C nuclear spin-lattice relaxation rate is shown in Fig. \ref{Itou_13C_T1} \cite{Itou10}.
   It exhibits a non-monotonic temperature dependence.
   At low temperatures below 1 K, it follows a $T^{2}$ dependence, suggesting no finite gap. However, the power of 2 implies a complicated nodal gap, which is not obviously consistent with the finite susceptibility value and the thermodynamic measurements described below.
   Furthermore, $1/T_{1}$ forms a shoulder or a kink at approximately 1 K and becomes moderate in temperature dependence above 1 K.
   The kink temperature increases for higher magnetic fields or frequencies.
   The relaxation curve becomes a non-single exponential curve below 10 K but reverses below 1 K, indicating that the inhomogeneity increases below 10 K \cite{Itou10,Itou11}.
   The reversal at 1 K can be an indication of either a recovery in the homogeneity below 1 K or the microscopic nature of the inhomogeneity, which is subject to spin-diffusion averaging of the heterogeneous relaxation time that is longer at lower temperatures.
   The 1-K relaxation-rate anomaly in Et$_{2}$Me$_{2}$Sb[(Pd(dmit)$_{2}$]$_{2}$ may be compared to the broad anomaly
   around nearly the same temperature for $\kappa$-(ET)$_{2}$Cu$_{2}$(CN)$_{3}$. However, they appear different with respect to field (or frequency) dependence and spatial scale of  inhomogeneity.

\begin{figure}[tbph]
  \includegraphics[width=8cm]{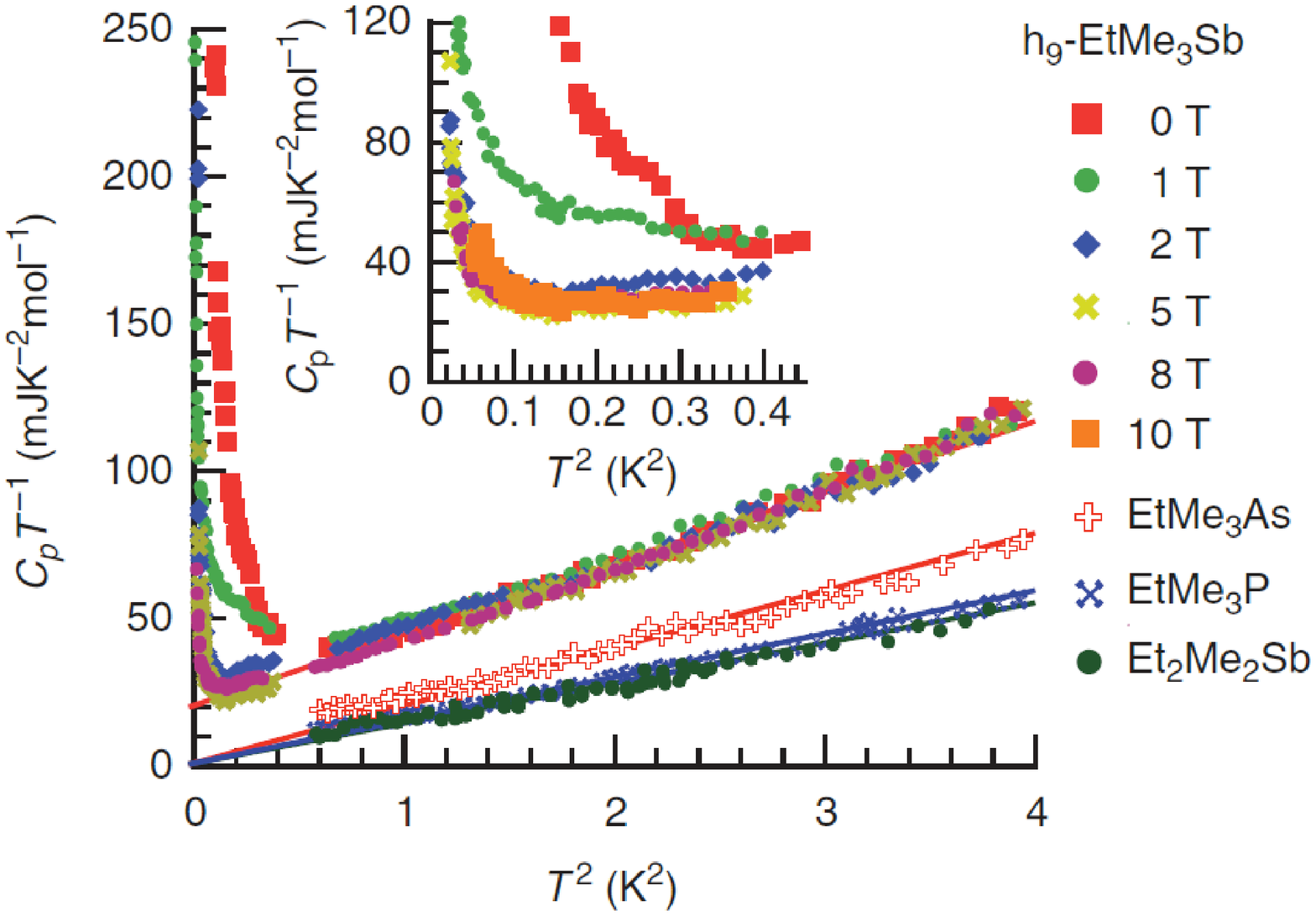}
  \caption{\cite{SYamashita11} Low-temperature specific heat $C_p$ of EtMe$_{3}$Sb[Pd(dmit)$_{2}$]$_{2}$ for several magnetic fields up to 10 T in $C_p/T$ versus $T^{2}$ plots. The data of other insulating systems, i.e., Et$_{2}$Me$_{2}$As[(Pd(dmit)$_{2}$]$_{2}$, EtMe$_{3}$As[(Pd(dmit)$_{2}$]$_{2}$ and EtMe$_{3}$P[(Pd(dmit)$_{2}$]$_{2}$, are also plotted for comparison. A large upturn below 1 K is probably attributable to the rotational tunneling of Me groups. The low-temperature data are expanded in the inset.}
  \label{SYamashita_specific_heat2}
  \end{figure}

\begin{figure}[tbph]
  \includegraphics[width=8cm]{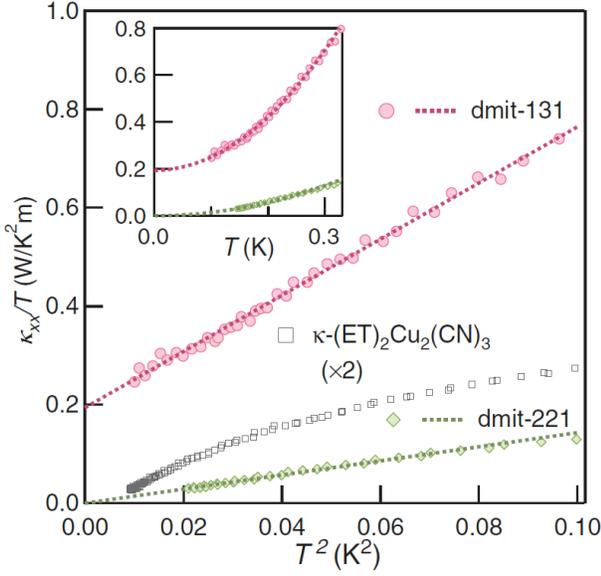}
  \caption{\cite{MYamashita10} Low-temperature thermal conductivity $\kappa$ of EtMe$_{3}$Sb[Pd(dmit)$_{2}$]$_{2}$ (dmit-131) in $\kappa /T$ versus $T^{2}$ and  $\kappa /T$ versus $T$ (inset) plots. The data of other insulators, i.e., Et$_{2}$Me$_{2}$Sb[Pd(dmit)$_{2}$]$_{2}$ (dmit-221, non-magnetic) and $\kappa$-(ET)$_{2}$Cu$_{2}$(CN)$_{3}$, are also plotted for comparison.}
  \label{MYamashita_TC2}
  \end{figure}

   The thermodynamic measurements are indicative of fermionic low-energy excitations.
   Fig. \ref{SYamashita_specific_heat2} shows the temperature dependence of the specific heat \cite{SYamashita11}.
   The linearity of $C/T$ versus $T^{2}$ in EtMe$_{3}$Sb[(Pd(dmit)$_{2}$]$_{2}$ is extrapolated to a zero Kelvin
   to give a finite value of $\gamma$, whereas other Mott insulators appear to have vanishing  $\gamma$, as expected for conventional insulators.
   There is no field dependence in $C/T$ in EtMe$_{3}$Sb[(Pd(dmit)$_{2}$]$_{2}$ up to 8 T, as in $\kappa$-(ET)$_{2}$Cu$_{2}$(CN)$_{3}$.
   The thermal conductivity results are consistent with the specific heat results, as seen in Fig. \ref{MYamashita_TC2},
   where the low-temperature $\kappa /T$ value for EtMe$_{3}$Sb[(Pd(dmit)$_{2}$]$_{2}$ is as high as 0.2 WK$^{-2}$m in the zero-Kelvin limit,
   implying the presence of gapless thermal transporters with fermionic statistics \cite{MYamashita10}.
   The mean free path for thermal transport is estimated to be of the order of 1 $\mu$m.
   $\kappa$ is enhanced by the application of a magnetic field above a threshold value, suggesting that the gapped excitations coexist with the gapless excitations \cite{MYamashita10}.

 \subsection{Kagome-lattice system: ZnCu$_{3}$(OH)$_{6}$Cl$_{2}$}

   The kagome lattice is constructed by using corner-sharing triangles in contrast to the edge-sharing in the triangular lattices, as shown in Fig. \ref{Kagome-classical-GS}.
   Thus, the spin states in the kagome lattice have larger degeneracy than those in the triangular lattices, leading to high potential for hosting a spin liquid.
   Actually, the theoretical perspective of seeking a spin liquid is more promising for the kagome-lattice Heisenberg model than for the triangular lattice \cite{Sachdev92,Lecheminant97,Mila98,Misguich04}.
   Among several candidates for the kagome spin systems, we select a spin-1/2 system, ZnCu$_{3}$(OH)$_{6}$Cl$_{2}$, which is known as herbertsmithite, whose magnetism has been extensively investigated.
   This is a member of a family of materials with variable compositions, i.e., Zn$_{x}$Cu$_{4-x}$(OH)$_{6}$Cl$_{2}$ ($0<x<1$ ).
   As an end material, Cu$_{4}$(OH)$_{6}$Cl$_{2}$ has a distorted pyrochlore lattice of $S=1/2$ Cu$^{2+}$ spins, whereas the other end material,
   ZnCu$_{3}$(OH)$_{6}$Cl$_{2}$, has a two-dimensional ($a-b$ plane) perfect kagome-lattice of Cu$^{2+}$ spins separated
   by different crystallographic sites occupied by Zn$^{2+}$ \cite{Shores05}.
   The structural symmetry changes across $x=0.33$, above which Cu$^{2+}$ partially occupies the Zn sites in addition to the kagome lattice.
   There is an argument for the mixture of Zn in the kagome sites in ZnCu$_{3}$(OH)$_{6}$Cl$_{2}$.
   Magnetic susceptibility \cite{Bert07} and specific heat \cite{Vries08} suggest that approximately $6\%$ of the kagome sites are replaced by non-magnetic Zn.  The same amount of Cu is assumed to invade the nominal Zn sites.
   Thus, significant efforts have been made to extract the intrinsic properties of the kagome lattice from the experimental data.

  \begin{figure}[tbph]
  \includegraphics[width=8cm]{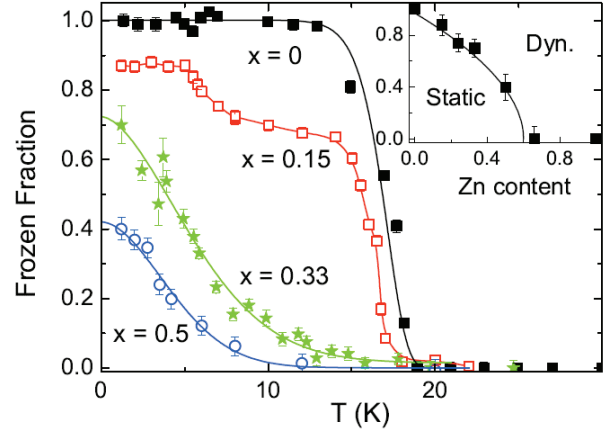}
  \caption{\cite{Mendels07} Temperature variation of the spin-frozen fraction determined by muon spin rotation experiments for Zn$_{x}$Cu$_{4-x}$(OH)$_{6}$Cl$_{2}$. Inset shows the $x$- dependence of the spin-frozen fraction at a low temperature.}
  \label{Mendels_SFf}
  \end{figure}

\begin{figure}[tbph]
  \includegraphics[width=8cm]{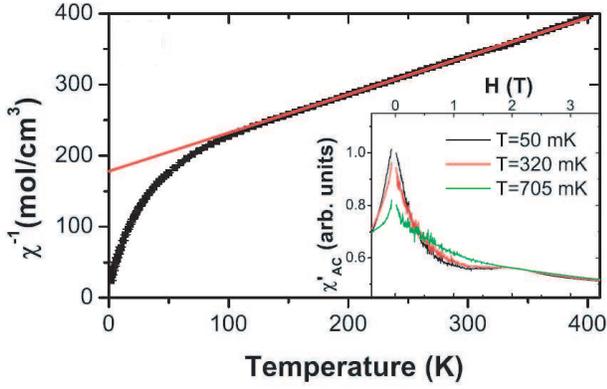}
  \caption{\cite{Kagome07} Temperature dependence of the inverse magnetic susceptibility $\chi ^{-1}$ of ZnCu$_{3}$(OH)$_{6}$Cl$_{2}$. The line denotes a Curie-Weiss fit. Inset: ac susceptibility (at 654 Hz) at low temperatures.}
  \label{Helton_sus}
  \end{figure}

   Experimental evidence for the absence of magnetic ordering in ZnCu$_{3}$(OH)$_{6}$Cl$_{2}$ can be obtained from $\mu$SR experiments \cite{Mendels07}.
   The relaxation profile shows no internal field down to 50 mK.
   The experiments for a wide range of $x$ found that the absence of an internal field was persistent in a certain range below $x=1$ (see Fig. \ref{Mendels_SFf}) \cite{Mendels07}.
   The magnetic susceptibility exhibits a Curie-Weiss behavior at high temperatures above 100 K, as shown in Fig. \ref{Helton_sus} \cite{Kagome07}.
   The Weiss temperature is $\sim$300 K, which implies an antiferromagnetic exchange interaction of $J=17$ meV.
   The dc and ac magnetic susceptibility indicates no magnetic ordering down to 0.1 K and 0.05 K, respectively, which is four orders of magnitude lower than $J$ \cite{Kagome07}.
   The susceptibility increases progressively at lower temperatures. Two mechanisms are possible.
   First, impurities from Cu/Zn inter-site mixing can give a Curie-like upturn.
   Second, Dzyaloshinsky-Moriya interactions may be present between the adjacent sites with broken inversion symmetry, as in the kagome lattice \cite{Rigol07}.
   The high-field magnetization measurements suggest that the former is mainly responsible for the increasing susceptibility \cite{Bert07}.

  \begin{figure}[tbph]
  \includegraphics[width=8cm]{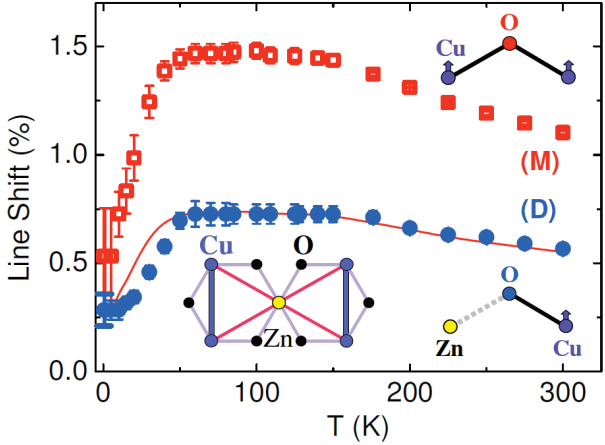}
  \caption{\cite{Olariu08} $^{17}$O NMR shift of two lines (M and D) decomposed from the observed spectra for a powder of ZnCu$_{3}$(OH)$_{6}$Cl$_{2}$. The M and D lines are considered to come from the oxygen sites depicted in the inset. The red curve represents the trace of a half of the value of the M line. The sketch in the lower left corner illustrates the environment of a Zn substituted on the Cu kagome plane, and thick lines represent Cu-Cu dimers.}
  \label{Olariu_17O_NMR_shift}
  \end{figure}

\begin{figure}[tbph]
  \includegraphics[width=8cm]{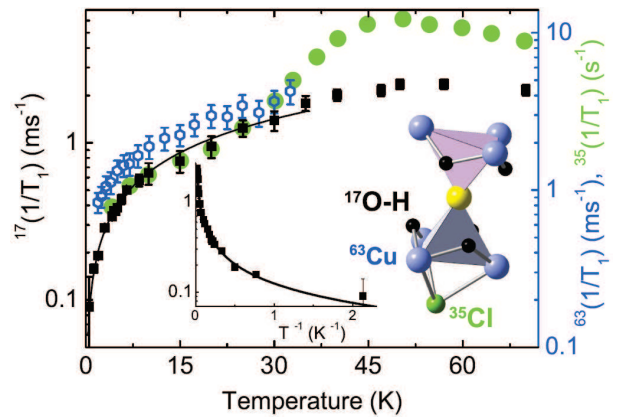}
  \caption{\cite{Olariu08} $^{17}$O, $^{63}$Cu and $^{35}$Cl nuclear spin-lattice relaxation rates $1/T_{1}$ for a powder of ZnCu$_{3}$(OH)$_{6}$Cl$_{2}$. Inset shows $^{17}$O $1/T_{1}$ versus $1/T$ plots.}
  \label{Imai_Olariu_NMR_T1}
  \end{figure}

   NMR, which probes magnetism in a site-selective manner, was informative particularly for this material because the analysis of spectra
   allows one to distinguish the intrinsic magnetism from the extrinsic one.
   The NMR spectra at $^{35}$Cl and $^{17}$O sites are broad \cite{Imai08,Olariu08}, reflecting the inhomogeneous local fields,
   supposedly due to the Zu/Cu mixture. However, the smallest shift value in the broad $^{35}$Cl spectrum follows a Curie-Weiss law down to 25 K,
   followed by a decrease at lower temperatures \cite{Imai08}.
   This is considered to indicate intrinsic magnetism for the kagome lattice \cite{Imai08}.
   The $^{17}$O probes the kagome sites more preferentially than the nominal Zn sites due to larger hyperfine coupling with the kagome sites \cite{Olariu08}.
   The $^{17}$O NMR spectra were decomposed into two components. One is from the $^{17}$O sites coordinated by two Cu$^{2+}$ ions,
   while the other is from the $^{17}$O sites coordinated by a Cu$^{2+}$ and a Zu$^{2+}$ in the kagome plane.
   The relative fraction of the two components was consistent with a 6 \% Zn admixture.
   The NMR shifts of the respective components, as shown in Fig. \ref{Olariu_17O_NMR_shift}, are considered to
    be local susceptibilities at Cu sites with and without Zn$^{2+}$ at the neighboring sites \cite{Olariu08}.
   Both decrease below 50 K and saturate to finites values, indicating the gapless nature of the spin excitations.
   The low-temperature decrease in the shift at the Cu site with a Zn neighbor is in contrast to the enhancement commonly observed
   in the neighborhood of non-magnetic impurities \cite{Olariu08}.
   This behavior also suggests that the Curie-like upturn in the bulk susceptibility at low temperatures is not from the kagome plane.
   For the NMR relaxation rate, all of the O, Cl and Cu nuclear spins exhibit power-laws against temperature down to 0.47 K for O and 2 K
   or lower for Cl and Cu, indicating a gapless spin liquid (see Fig. \ref{Imai_Olariu_NMR_T1}) \cite{Imai08,Olariu08}.
   Although the power somewhat depends on the nuclear site, the relaxation profile is overall nuclear site-insensitive,
   which is filtered by the nuclear site-specific form-factor determined by its location relative to the kagome lattice,
   suggesting non-dispersive spin dynamics. Otherwise, the temperature profile of the relaxation rate would be site-dependent \cite{Olariu08}.
   This feature is potentially relevant to the spinon excitation with the continuum.
   More recently, NMR experiments performed at low temperatures have revealed an anomaly in the relaxation rate
   at a temperature depending on the applied field, which may signify field-induced spin freezing \cite{Jeong11}.
   Very recently, a $^{17}$O NMR experiment performed with a single crystal has found different features from those observed so far in the powder samples \cite{Fu15}. According to the analysis of NMR spectra, there is no significant contamination of Zn in the Cu sites within the kagome plane, and the Knight shift shows appreciable temperature- and field-dependences, suggesting a spin gap of the order of 10 K, as shown in FIG \ref{Fu_Imai_NMR_2015}, in contradiction with the consequences of the earlier NMR and neutron (see below) experiments.

  \begin{figure}[tbph]
  \includegraphics[width=8cm]{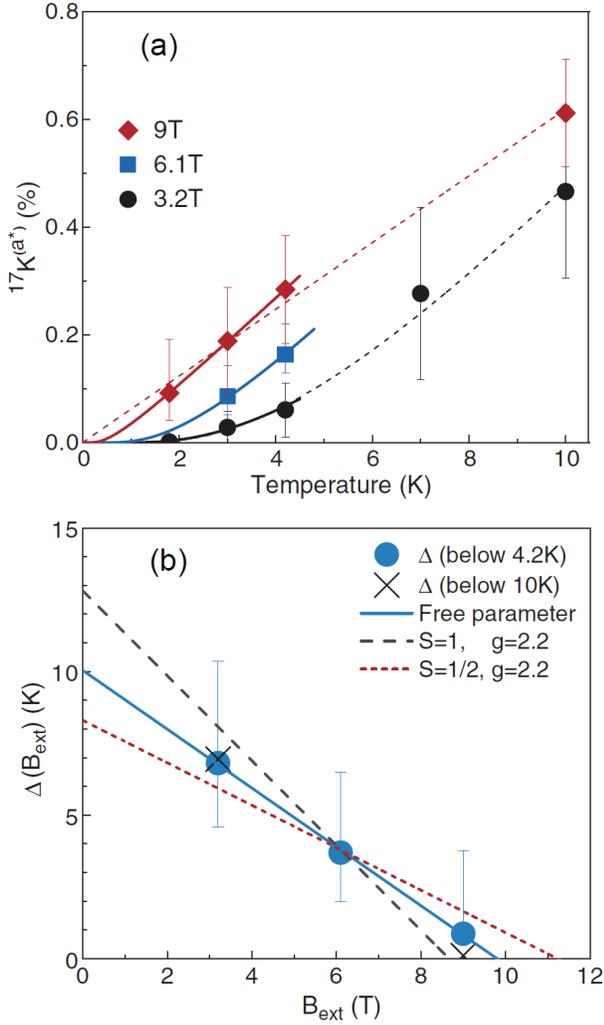}
  \caption{\cite{Fu15} (a) Temperature dependence of $^{17}$O Knight shift and (b) the field dependence of the spin gap deduced from the Knight shift for a single-crystal ZnCu$_{3}$(OH)$_{6}$Cl$_{2}$.}
  \label{Fu_Imai_NMR_2015}
  \end{figure}

  \begin{figure}[tbph]
  \includegraphics[width=8cm]{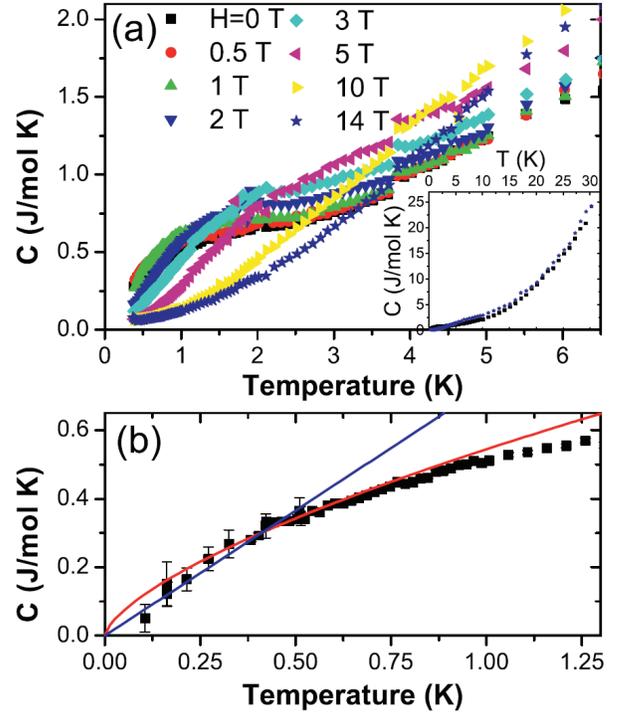}
  \caption{\cite{Kagome07}  (a) Specific heat $C$ of ZnCu$_{3}$(OH)$_{6}$Cl$_{2}$ in various applied fields. Inset shows $C$ over a wider temperature range in applied fields of 0 T (square) and 14 T (star). (b) $C$ in a zero field at low temperatures. The lines represent power law fits.}
  \label{Helton_specific_heat}
  \end{figure}

   The low-temperature specific heat was investigated under external magnetic fields \cite{Kagome07,Vries08}.
   As shown in Fig. \ref{Helton_specific_heat}(a) \cite{Kagome07}, there is an enormous field dependence.
   The temperature dependence at a zero field is approximated by a power law $T^{\alpha}$ with $\alpha$ unity or smaller (see Fig. \ref{Helton_specific_heat}(b)).
   The broad peak present even at a zero field is shifted to higher temperatures under higher fields.
   Assuming that the field-dependent peak is a Schottky contribution associated with a field-induced gap,
   the data for different fields were analyzed in detail to reveal the intrinsic specific heat of the kagome lattice \cite{Vries08}.
   The deduced Schottky component is consistent with Zeeman splitting of the $6\%$ Cu impurities in the Zn sites
   at higher fields, and the intrinsic $C/T$ follows a power law $T^{\alpha}$ with $\alpha=1.3$ as the best estimate, suggesting gapless excitations \cite{Kagome07,Vries08,Shaginyan11}.

\begin{figure}[tbph]
  \includegraphics[width=8cm]{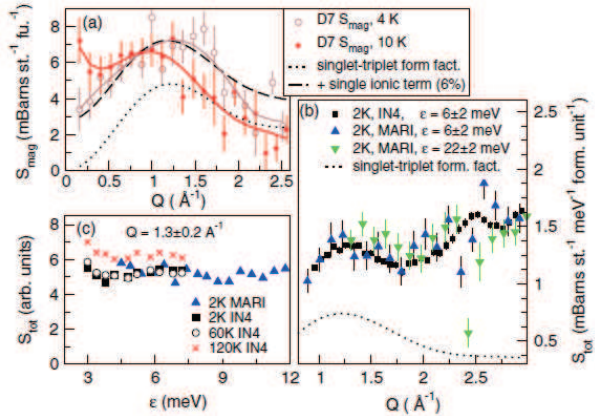}
  \caption{\cite{Vries09} (a) Instantaneous magnetic correlations at 4 K and 10 K for a time scale corresponding to approximately 6.5 meV. The solid lines are a guide to the eye. (b) The $Q$ dependence in the dynamic correlations with the energy integration interval indicated in the legend. The dotted line in panel (a) and (b) is the structure factor for dimer-like AF correlations. The dashed line, a single-ion contribution corresponding to the 6$\%$ antisite spins in this system, is added. (c) The energy and temperature dependence at $Q$=1.3 \AA$^{-1}$. D7, IN4 and MARI in the legends stand for the types of spectrometers used.}
  \label{deVries_Neutron}
  \end{figure}

\begin{figure}[tbph]
  \includegraphics[width=8cm]{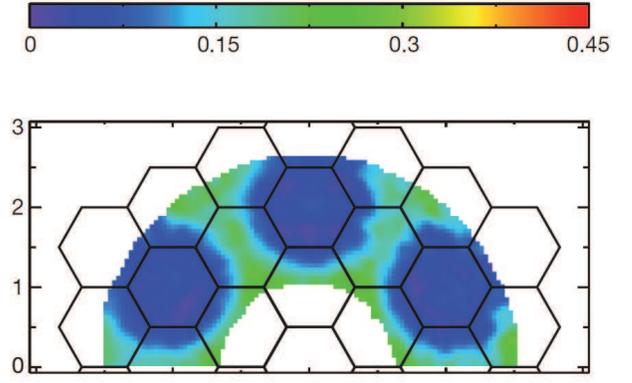}
  \caption{\cite{Han12} Contour plot of dynamical structure factor, $S_{mag} (\textbf{Q}, \omega )$, integrated over 1$\leq \hbar \omega \leq$ 9 meV for a single-crystal ZnCu$_{3}$(OH)$_{6}$Cl$_{2}$ at 1.6 K. The intense scattering is extended in a green-colored region, without peaking at any specific points.}
  \label{Han_Neutron}
  \end{figure}

   Neutron-scattering experiments, which are capable of profiling spin excitations with respect to momentum and energy transfers, are available for herbertsmithite. One of the key issues of elementary excitations in spin liquids is the possible fractionalization of $S=1$ spin excitations into $S=1/2$ spinons, which could manifest themselves as a continuum in the spin excitation spectrum, \textit{i.e.}, dynamic structure factor $S(\textbf{Q}, \omega )$, where \textbf{Q} and  $\omega$ are momentum transfer and energy transfer divided by $\hbar$, respectively.
   Such a continuum is observed in a highly anisotropic triangular-lattice system, Cs$_{2}$CuCl$_{4}$, ($J^{\prime}/J \sim$ 3 and $J^{\prime}\sim$ 0.34 meV in Fig. 17), although it undergoes a magnetic transition into a spin-spiral order at 0.62 K \cite{Coldea01, Coldea03}.
   Several features signifying the continuum are found via neutron experiments of herbertsmithite, which were first performed for polycrystalline or powder samples.
   The inelastic scattering experiments exhibit no excitation gap at least down to 0.1 meV, which corresponds to $\sim J/170$, and insensitivity of the scattering strength to \textbf{Q}, indicating gapless and local natures of spin fluctuations \cite{Kagome07}.
   Furthermore, the scattering intensity is only weakly dependent on  $\omega$ up to 25 meV and temperature up to 120 K and shifts toward lower \textbf{Q} as temperature is increased \cite{Vries09}.
   Some of the results are displayed in Fig. \ref{deVries_Neutron}.
   All these features are suggestive of a continuum in spin excitations and the persistence of the short-range nature of spin correlations even at low temperatures.
   Recent experiments on a large single crystal have succeeded in capturing the continuum nature, as seen in the green area in Fig. \ref{Han_Neutron}.
   The momentum profile of the excitation intensity (dynamic structure factor), $S(\textbf{Q}, \omega )$, is approximately reproduced by the simulated structure factor of uncorrelated dimer-singlets, which indicates to the short-ranged spin correlations at least down to 1.6 K \cite{Han12}.
   The short-range nature that persists even at low temperatures, as suggested by the powder experiments as well, is generally in favor of a gapped state, whereas there is no indication of a spin gap down to 0.25 meV at any \textbf{Q} values in the excitation spectra \cite{Han12}.
   It is puzzling that spin dynamic correlation exhibits short-range RVB nature while the spectrum is gapless. One possibility is that the Herbertsmithite is in a $Z_2$ spin liquid in close proximity to a critical point to the $U(1)$ Dirac liquid, as indicated by some recent numerical works \cite{LiTao16}, although the true ground state of the isotropic Heisenberg model on a kagome lattice is still under debate \cite{Iqbal16comment}.

 \subsection{Hyperkagome-lattice system: Na$_{4}$Ir$_{3}$O$_{8}$}

   The hyperkagome lattice is a three-dimensional network of corner-sharing triangular lattices.
   In Na$_{4}$Ir$_{3}$O$_{8}$, the Ir$^{4+}$ ion with $5d^{5}$ electrons likely takes on a low-spin state.
   These ions locate on the corners, forming a S=1/2 hyperkagome lattice \cite{Na4Ir3O8}.
   The resistivity of the ceramic sample is 10 Ohmcm at room temperature.
   The samples are semiconducting, with a charge transport gap of 500 K, implying the proximity of this system to the Mott transition,
   which is different from the kagome materials reviewed above \cite{Na4Ir3O8}.
   A  connection between the spin liquid and the metal-insulator transition, similar to the case of $\kappa$-(ET)$_{2}$Cu$_{2}$(CN)$_{3}$, is shown \cite{Podolsky09}.
   A distinct feature of Na$_{4}$Ir$_{3}$O$_{8}$ among spin liquid candidates is its large spin-orbit coupling,
   which introduces additional interest to the physics of spin liquids \cite{ChenBalents08,Zhou08}.
   Several theoretical studies propose that Na$_{4}$Ir$_{3}$O$_{8}$ is a 3D QSL with fermionic spinons \cite{Zhou08,Lawler08}.

\begin{figure}[hptb]
  \includegraphics[width=8cm]{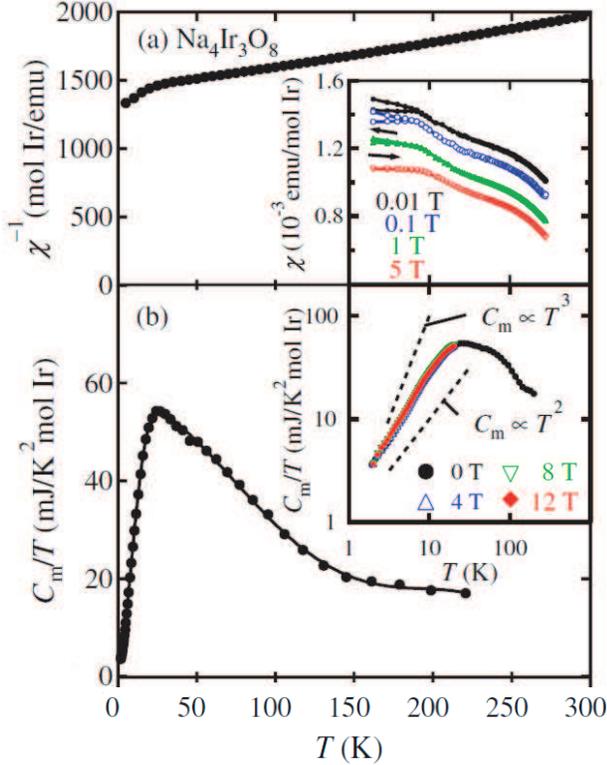}
  \caption{\cite{Na4Ir3O8} (a) Temperature dependence of the inverse magnetic susceptibility $\chi^{-1}$ of polycrystalline Na$_{4}$Ir$_{3}$O$_{8}$ under 1 T. Inset shows magnetic susceptibility $\chi$ in various fields up to 5 T; for clarity, the curves are shifted by 3, 2, and 1 $\times$ 10$^{-4}$ emu/mol Ir for 0.01, 0.1, and 1 T data, respectively. (b) Magnetic specific heat $C_{m}$ divided by temperature $T$ of polycrystalline Na$_{4}$Ir$_{3}$O$_{8}$. To estimate Cm, data for Na$_{4}$Sn$_{3}$O$_{8}$ is used as a reference of the lattice contribution. Inset shows $C_{m}/T$ versus $T$ in various fields up to 12 T. (c) Magnetic entropy. }
  \label{Okamoto_sus_specific_heat}
  \end{figure}

Fig. \ref{Okamoto_sus_specific_heat}(a) shows the magnetic susceptibility of Na$_{4}$Ir$_{3}$O$_{8}$, which weakly increases with decreasing temperature, as characterized by the Curie-Weiss temperature of -650 K \cite{Na4Ir3O8}. This implies an antiferromagnetic interaction of hundreds of Kelvin.  There is no clear indication of magnetic ordering at least down to 2 K, whereas a small anomaly reminiscent of spin glass observed in the magnetization history against the field/temperature variation is attributed to a tiny fraction of the total spins \cite{Na4Ir3O8}.

  \begin{figure}[hptb]
  \includegraphics[width=8cm]{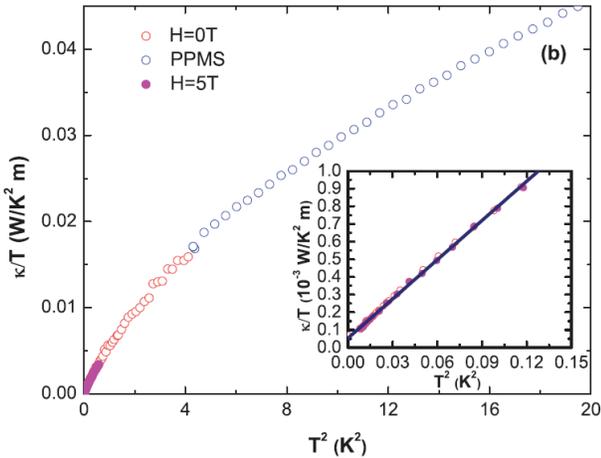}
  \caption{\cite{Singh13} Thermal conductivity $\kappa$ of Na$_{4}$Ir$_{3}$O$_{8}$ in $\kappa /T$ versus $T^{2}$ plots for magnetic fields of 0 T and 5 T. Inset shows the low-temperature part of the data.}
  \label{Singh_thermal_conductivity}
  \end{figure}

   The electronic (magnetic) contribution to the specific heat of Na$_{4}$Ir$_{3}$O$_{8}$, as shown in Fig. \ref{Okamoto_sus_specific_heat}(b),
   has a broad peak at 20 K. However, no anomaly signifying magnetic ordering is apparent \cite{Na4Ir3O8}.
   The magnetic entropy estimated by integrating the $C/T$ in Fig. \ref{Okamoto_sus_specific_heat}(b) reaches $70-80\%$ of $R\ln 2$ ($=5.7$ J/molK) at 100 K,
   a much lower temperature than the Weiss temperature of $\sim$600 K, which features frustrated magnetism.
   The $C/T$ is characterized by a curious $T^{2}$ dependence at the lowest temperatures.
   The $\gamma$ term, when present, appears on the order of 1 mJ/K$^{2}$mol Ir.
   Recent experiments extended down to 500 m K have found that $C_{\textrm{m}}/T$ is well approximated by a form of $\gamma + \beta T^{2.4}$ with $\gamma =$2.5 mJ/K$^{2}$molIr \cite{Singh13}.
   As seen in the inset of Fig. \ref{Okamoto_sus_specific_heat}(b), the applied field has no influence on the specific heat, at least up to 12 T.

   The temperature dependence of thermal conductivity is shown in Fig. \ref{Singh_thermal_conductivity} \cite{Singh13}.
   At low temperatures down to 75 mK, $\kappa/T$ is linear in $T^{2}$.
   The $\kappa /T$ value extrapolated to $T=0$ is $6.3 \times 10^{-2}$ mW/K$^{2}$m, which is a vanishingly small value,
   compared with the value of EtMe$_{3}$Sb[(Pd(dmit)$_{2}$]$_{2}$, 0.2 W/K$^2$m in Fig. \ref{MYamashita_TC2}.
   The suppression of the $\kappa /T$ value by the extrinsic grain-boundary effect is not ruled out \cite{Singh13}.
   The feature that $\gamma$ is diminished and $\kappa /T$ is vanishing at low temperatures, while both are sizable
   at high temperatures of the order of Kelvin, appears to be in accordance with a theoretical picture of spinon Fermi surfaces
   that undergo a pairing instability at low temperatures \cite{Zhou08}.
   In this context, the magnetic susceptibility, remaining large even at low temperatures, can be due to the large spin-orbit interactions of Ir \cite{Zhou08}.

   The substitution of non-magnetic Ti$^{4+}$ ions at Ir sites will give rise to a Curie-like tail in the spin susceptibility curve \cite{Na4Ir3O8},
   similar to Zn substitution for Cu in high-Tc cuprates, indicating an RVB spin background.
   The scaling analysis of magnetic Gruneisen parameters is suggestive of the proximity of Na$_{4}$Ir$_{3}$O$_{8}$ to a zero-field quantum critical point \cite{Singh13}.

\begin{figure}[tbph]
  \includegraphics[width=8cm]{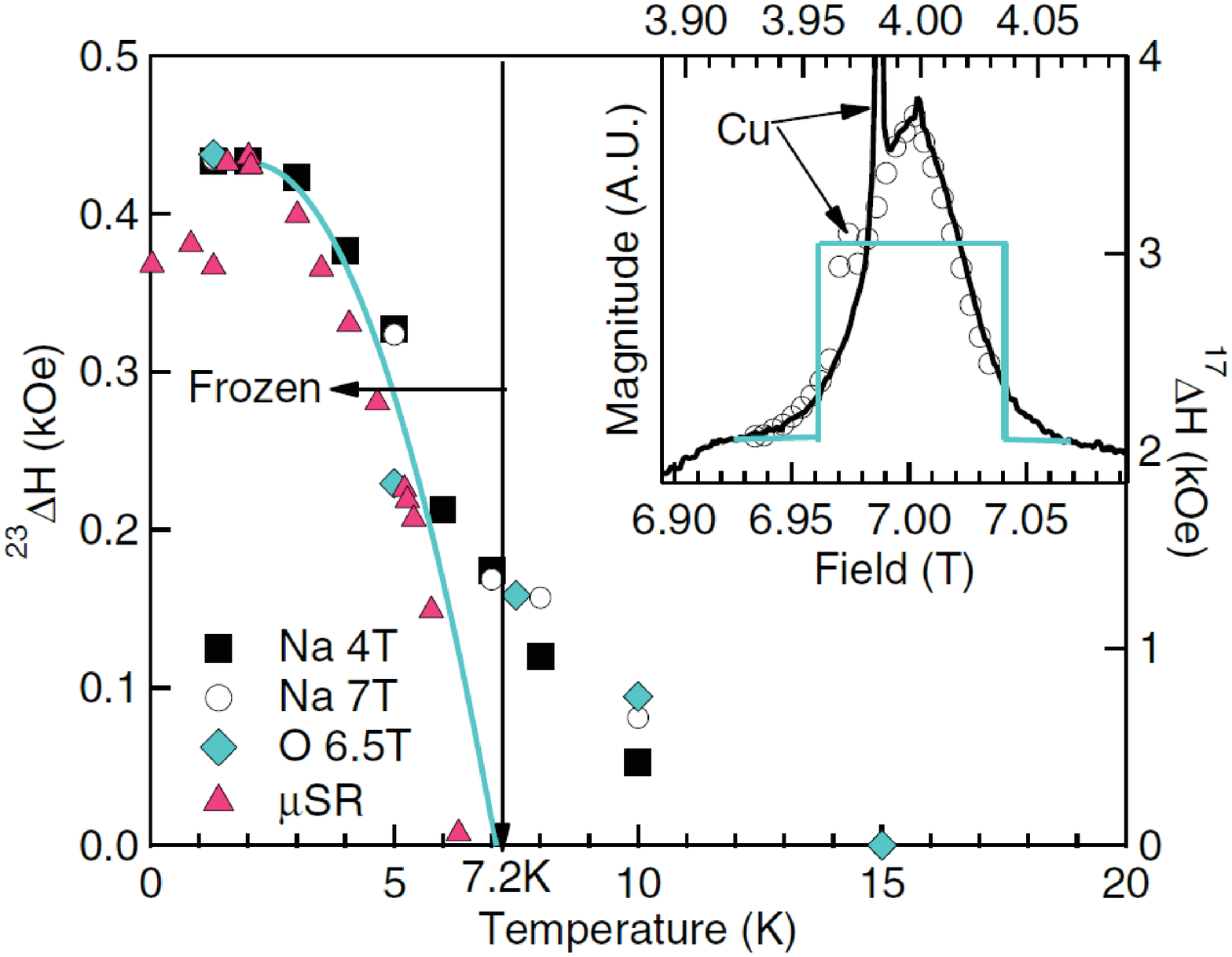}
  \caption{\cite{Shockley15}The line width (FWHM) of Gaussian-broadened $^{17}$O and $^{23}$Na NMR spectra and the mean value of the distributed local fields detected based on $\mu$SR \cite{Dally14}.
  For the NMR line width, its deviation from the value at 15 K is plotted.
  Inset: $^{23}$Na spectra at 78.937 MHz for 7 T (empty circles) and 45.046 MHz for 4 T (solid line) with the horizontal axis shifted by 3.005 T at 1.3 K. The blue line shows the expected powder pattern of the spectrum, with every Ir-site carrying the same moment. }
  \label{Shockley_NMR}
  \end{figure}

   Very recent $\mu$SR \cite{Dally14} and NMR experiments \cite{Shockley15} have found some indications that are not in accordance with the above claims.
   Both probes detected the emergence of local fields signifying the freezing of moments at low temperatures, as shown in Fig. \ref{Shockley_NMR}.
   The muons are revealed to sense an inhomogeneous local field of electronic origin that appears at 6 K, where the irreversibility in magnetization occurs, and levels off to 70 G on average, which may correspond to 0.5 $\mu_{\textrm{B}}$ on Ir.
   It is suggested, however, that the spin correlation is short-ranged (of the order of one unit-cell) and quasi-static in that the slow dynamics captured by the relaxation rate persist down to 20 mK.
    The quasi-static nature is also seen in the S=1 triangular-lattice system, NiGa$_{2}$S$_{4}$ \cite{Nakatsuji05, MacLaughlin08}.
    $^{17}$O and $^{23}$Na NMR lines show broadening, which is roughly scaled to the $\mu$SR results at low temperatures, as seen in Fig. \ref{Shockley_NMR}; the moment is estimated at 0.27 $\mu_{\textrm{B}}$ on Ir.
    The NMR line profile also suggests inhomogeneous spin freezing and slow dynamics persisting down to low temperatures although the temperature dependence of the relaxation rates on the muon and $^{23}$Na differ.
    Noticeably, the $^{23}$Na relaxation rate exhibits a peak indicative of the critical slowing down at approximately 7.5 K despite no anomaly in specific heat.
    The nature and origin of these anomalous properties are not clear at present; however, it is likely that disorder plays a vital role in this system, which can host configurationally degenerate phases with fluctuating order \cite{Dally14}.
    Considering that muon, $^{17}$O and $^{23}$Na captured the behavior of the majority of spins in the sample, the disorder effect, if any, is such that it is not restricted to finite areas but extended over the system, being reminiscent of the quantum Griffiths effect given the inhomogeneity and slow dynamics.

\begin{widetext}

\begin{table}[htbp]
\caption{Spin liquid materials summary}
\begin{tabular}{|p{7em}|p{11em}|p{11em}|p{11em}|p{11em}|} \hline
	\textbf{Material} & Triangular, \hspace{8em}$\kappa$\textbf{-(ET)$_{2}$Cu$_{2}$(CN)$_{3}$} & Triangular \textbf{M[Pd(dmit)$_{2}$]$_{2}$} & Kagome \textbf{ZnCu$_{3}$(OH)$_{6}$Cl$_{2}$} & Hyper-Kagome, \textbf{Na$_{4}$Ir$_{3}$O$_{8}$} \\ \hline \hline
	\textbf{Susceptibility} & A broad peak at 60 K, & A broad peak at 50 K, & Curie-Weiss at high-T  & Curie-Weiss  \\
	 & Finite at 2 K, $J=$250 K & Finite at 2 K, $J=220 \sim$ & $\Theta_{\textrm{W}}=$ -300 K, $J=$230 K, &  $\Theta_{\textrm{W}}=$ -650 K \\
	 & (*1) & 280 K (*7) & Upturn at low-T due to impurity sites & (*19, *20) \\
	 & & & (*11, *12) & \\ \hline
	\textbf{Specific heat} & Gapless,  & Gapless,   & Gapless, & Gapless, \\
	& $\gamma =$15 mJ/K$^{2}$ mol, & $\gamma$=20 mJ/K$^{2}$mol, & $C \sim T^{\alpha}$ & $C \sim T^{2}$ (*19), \\
	& Field-independent & Field-independent & $\alpha = 1.3$ at high fields & $C \sim \gamma T + \beta T^{2.4}$,  \\
	& (*2) & (*8) & (*13) &  $\gamma =2$ mJ/K$^{2}$mol (*20),\\
	& & & &  Field-independent \\
	& & & & (*21, *22)  \\ \hline
	\textbf{Thermal conductivity} & Gapped; $\Delta =$ 0.46 K (*3) & Gapless; finite $\kappa/T$ (*9)& & Vanishingly small $\kappa /T$ (*22) \\ \hline
	\textbf{NMR shift} & Not precisely resolved & Not precisely resolved & High-T    & $^{17}$O shift --- scales to  \\
	& (*4) & (*10) & Broad peak at at 50-60 K  & $\chi_{\textrm{bulk}}$ in 100 K - 300 K\\
	& & &for $^{17}$O  (*14,*15), &  but levels off below \\
	&  & & at 25-50 K for $^{35}$Cl (*16) &  80 K (*23) \\
	& & & Low-T & $^{17}$O,$^{23}$Na-  \\
	& & & gapless :finite value & inhomogeneous line \\
	& & & (*14) & broadening at low-$T$ \\
	& & & gapped : $\Delta \sim$ 10 K (*15) & (*23) \\ \hline
	\textbf{NMR $1/T_{1}$} & Inhomogeneous $1/T_{1}$, & Inhomogeneous $1/T_{1}$, & $1/T_{1} \sim T^{\alpha}$ & $^{23}$Na $1/T_{1}$-- a peak \\
	& Power law,  & Power law,  & $\alpha \sim 0.73$ for $^{17}$O (*14) & formation typical of \\
	& $^{1}$H $1/T_{1};\sim T$ / $\sim T^{2}$ at & $^{13}$C $1/T^{2}$ at $<$ 0.5 K & $\alpha \sim$ 0.5 for $^{63}$O (*16) & critical slowing down at \\
	&  $T< 0.3 $K & (stretched exponential) &  &  7.5 K \\
	& (two components) (*1), & (*10) & Field-induced spin & (*23)\\
	& $^{13}$C $1/T_{1};\sim1/T^{1.5}$ at $T< 0.2$ K &  & freezing (*17) & \\
	& (stretched exponential) & & & \\
	& (*4) & & & \\ \hline
	\textbf{$\mu$SR} & No internal field at 0 T & & No internal field at 0 T  & Emergence of distributed \\
	& (*5,*6) & & (*18) & local fields below 6 K \\
	& & & & Quasi-static short-ranged \\
	& & & & spin freezing with slow dynamics \\
	& & & & (*24) \\ \hline
	\textbf{Neutron} & & & Powders &  \\
	& & &  $\sim$ gapless ($<$0.1 meV) (*11,*19) &  \\
	& & & Single crystal & \\
		& & & $\sim$ gapless ($<$0.25 meV) (*20) & \\
	& & & Continuum in dynamic structure factor (*11,*19,*20) & \\ \hline
	\textbf{References} & *1 Shimizu \textit{et al.}, 2003, & *7 Kato, 2014, & *11 Helton \textit{et al.}, 2007,& *21 Okamoto \textit{et al.}, 2007 \\
	& *2 Yamashita \textit{et al.}, 2008&  *8 Yamashita \textit{et al.}, 2011, & *12 Bert \textit{et al.}, 2007, & *22 Singh \textit{et al.}, 2013 \\
	& *3 Yamashita \textit{et al.}, 2009, & *9 Yamashita \textit{et al.}, 2010, & *13 de Vries \textit{et al.}, 2008, & *23 Shockley \textit{et al.}, 2015 \\
	& *4 Shimizu \textit{et al.}, 2006, &  *10 Itou \textit{et al.}, 2010 & *14 Olariu \textit{et al.}, 2008, & *24 Dally \textit{et al.}, 2014 \\
	& *5 Pratt \textit{et al.}, 2011, & & *15 Fu \textit{et al.}, 2015 & \\
	&  *6 Nakajima \textit{et al.}, 2012 & & *16 Imai \textit{et al.}, 2008 & \\
	& & & *17 Jeong \textit{et al.}, 2011, & \\
	& & & *18 Mendels \textit{et al.}, 2007 & \\
	& & & *19 de Vries \textit{et al.}, 2009 & \\
	& & & *20 Han \textit{et al.}, 2012 & \\ \hline
\end{tabular}
\label{Spin_liquid_summary}
\end{table}

\end{widetext}

 \subsection{Experimental summary}

   Due to intensive experimental studies, unconventional thermodynamic and magnetic properties that evoke spin liquids have been found in several materials with anisotropic triangular lattices, kagome lattices and hyperkagome lattices as seen above.
   These materials exhibit no indications of conventional magnetic ordering.
   Their magnetic and thermodynamic properties are summarized in Table \ref{Spin_liquid_summary}.
   It appears that the gapless nature is a property that a class of frustrated lattices constructed with triangles possesses,
   although the thermal conductivity of $\kappa$-(ET)$_{2}$Cu$_{2}$(CN)$_{3}$ suggested
   a tiny excitation gap three orders of magnitude smaller than $J$. A recent NMR work on herbertsmithite insists on gapped spin excitations, and anomalous quasi-static spin freezing has recently been revealed by $\mu$SR and NMR studies of the hyperkagome system.
   This feature and the successful observation of fractionalized excitations in a kagome lattice \cite{Han12} tempt ones to think about spinons as promising elementary excitations in spin liquids.
   How to detect the spinon Fermi surfaces, if they exist, is a focus---- smoking-gun experiments are awaited.

   As seen in Table  \ref{Spin_liquid_summary}, several experimental characteristics are seemingly inconsistent within given materials; understanding the apparently contradicting data in a consistent way requires clarification of the nature of the spin states.
   One of the key issues may be the randomness present in real materials. In particular, it has long been recognized that the effect of inevitable Zn/Cu admixtures in herbertsmithite has to be separated from the intrinsic magnetism. More recently, the issue of inhomogeneous quasi-static spin correlation with slow dynamics in the hyperkagome-lattice system has emerged as a consequence of disorder.
   Theoretically, it was proposed that as randomness is intensified, the 120-degree Neel order in the triangular-lattice Heisenberg model is changed to a sort of random singlets but not spin glass state. It is intriguing that randomness appears to enhance the quantum nature because the singlet is a purely quantum state \cite{Watanabe14,Shimokawa2015}.
   In the case of kagome lattices, it was theoretically suggested that disorder could lead to a valence-bond glass state\cite{Singh10} or a gapless spin liquid state \cite{Shimokawa2015,Kawamura2014}.
   Furthermore, a recent NMR experiment on an organic Mott insulator, i.e., $\kappa$-(ET)$_{2}$Cu[N(CN)$_{2}$]Cl, found that the antiferromagnetic ordering in the pristine crystal, when irradiated by X-rays, disappears.
   Spin freezing, spin gap and critical slowing down are not observed, but gapless spin excitations emerge, suggesting a novel role of disorder that brings forth a QSL from a classical ordered state \cite{Furukawa15b}.
   Whether the randomness is fatal or vital to the physics of a QSL is a non-trivial issue to be resolved.

  The development of new materials, although not addressed in this article, is under way. Among them is a new type of hydrogen-bonded $\kappa$-H$_{3}$(Cat-EDT-TTF)$_{2}$ with a triangular lattice of one-dimensional anisotropy \cite{Isono13} and  $\kappa$-(ET)$_{2}$Ag$_{2}$(CN)$_{3}$, an analogue of  $\kappa$-(ET)$_{2}$Cu$_{2}$(CN)$_{3}$ \cite{GSaito14}.
  Another compound with hyperkagome lattice structure, i.e., PbCuTe$_2$O$_6$, with Curie-Weiss temperature $\theta=-22$K is also proposed to be a spin liquid candidate \cite{Koteswararao14,Khuntia16}.
  The entanglement of additional degrees of freedom with quantum spins may be another direction for future studies; e.g., Ba$_{3}$CuSb$_{2}$O$_{9}$ is proposed to host a spin-orbital coupled liquid state \cite{HDZhou2011, Nakatsuji12}.

   It should be emphasized that the identification of QSL experimentally is a very important and challenging task. As a ``featureless" Mott insulator, there exists no simple magnetic order for identifying QSL states, and so far, there exists only indirect experimental evidence for mobile fermionic spinons in some candidate compounds as discussed above.

   To remedy this situation, theorists have proposed new experiments to identify QSLs through identifying nontrivial properties of spinons and gauge fields. For example, power law AC conductivity inside the Mott gap has been noted \cite{NgLee07}. A giant-magnetoresistance-like experiment was proposed to measure mobile spinons through oscillatory coupling between two ferromagnets via a QSL spacer \cite{Norman2009}.
   The thermal Hall effect in insulating quantum magnets was proposed as a probe for the thermal transport of spinons, where different responses were used to distinguish between magnon- and spinon- transports \cite{KNL2010}. Raman scattering was proposed as a signature to probe the $U(1)$ QSL state \cite{Ko10}.  It was also proposed that the spinon life time and mass as well as gauge fluctuations can be measured through a sound attenuation experiment \cite{YZhou2011}, and neutron scattering can be used to detect scalar spin chirality fluctuations in the kagome system \cite{Lee13}.
   Low energy electron spectral functions were evaluated for future ARPES experiments \cite{Tang13} and it was proposed that spin current flow through a metal-QSL-metal junction can be used to distinguish different QSLs \cite{Chen13}.
   More recently, it was suggested that there exists a long-life surface plasmon mode propagating along the interface between a linear medium and a QSL with spinon Fermi surface at frequencies above the charge gap,
   which can be detected by the widely used Kretschmann-Raether three-layer configuration \cite{Ma15}.

  However, there exists an important discrepancy between existing experiments and theories in some of the above experiments.

   1) Specific heat: Using the one-loop calculation supplemented by scaling analysis \cite{LeeNagaosa92,Polchinski94},
   it is found that the strong coupling between the $U(1)$ gauge field and spinon Fermi surface leads to $T^{2/3}$ correction
   to the temperature dependence of specific heat in $U(1)$ gauge theory. This predicted $T^{2/3}$ behavior has never been observed in experiments.
   Instead, linear, Fermi-liquid-like specific heat is found to exists in a wide range of temperatures in both organic materials ($\kappa$-ET and dmit).

   Some theories exsit that try to explain this missing singular $T^{2/3}$ specific heat. For instance, $Z_4$ and $Z_2$ spin liquid states with a spinon Fermi surface have been proposed \cite{Barkeshli13} as well as $Z_2$ spin liquid states with quadratic touched spinon bands \cite{Mishmash13}. However, all these proposals require fine-tuned parameters. A more natural way of explaining existing experiments is still missing.

   2) Thermal Hall effect: Katsura, Nagaosa and Lee \cite{KNL2010} have theoretically investigated the thermal Hall effect induced
   by the external magnetic field in a $U(1)$ spin liquid with a spinon Fermi surface and have predicted measurable electronic contributions.
   Their predicted sizable thermal Hall effect have never been observed in an experiment on dmit compounds \cite{MYamashita10}.
   This contradiction between experiment and theory remains unsolved, although an explanation that depends on fine-tuned parameters has been proposed \cite{Mishmash13}.

   3) Power law AC conductivity: A power law AC conductivity inside the Mott gap was proposed by Ng and Lee \cite{NgLee07}.
   Indeed, power law behavior $\sigma(\omega)\sim \omega^{\alpha}$ has been observed in both $\kappa$-ET \cite{Elsasser12} and Herbertsmithite \cite{Pilon13}.
   However, the power $\alpha$ observed in both compounds is smaller than predicted value, indicating that there exist more in-gap electronic excitations than those predicted in the $U(1)$ gauge theory.

   Thus, despite all the theoretical efforts, the understanding and finding of realistic ``smoking gun" evidence for QSLs remains the greatest challenge in the study of QSLs.

\section{Summary}

    In this article, we provide a pedagogical introduction to the subject of QSLs and review the current status of the field.  We first discuss the semi-classical approach to simple quantum antiferromagnets. We explain how it leads to the Haldane conjecture in one dimension and why it fails for frustrated spin models. We then focus on spin-$1/2$ systems with spin rotational symmetry and introduce the RVB concept and the slave-particle plus Gutzwiller-projected wavefunction approaches. We explain the technical difficulties associated with these approaches and why slave-particle approaches naturally lead to gauge theories for spin liquid states. The natures of $SU(2)$, $U(1)$ and $Z_2$ spin liquid states are explained, and the extensions of the approach to systems with spin-orbit coupling and $S>1/2$ systems are introduced. We explain that because of the intrinsic limitations of the analytical slave-particle approach, many alternative approaches to spin liquid states have been developed, both numerically and analytically. These approaches complement each other and often lead to exotic possibilities not covered by the simple fermionic slave-particle approach. The experimental side of the story is also introduced with a review of the properties of several candidate spin liquid materials, including anisotropic triangular lattice systems ($\kappa$-(ET)$_{2}$Cu$_{2}$(CN)$_{3}$ and EtMe$_{3}$Sb[(Pd(dmit)$_{2}$]$_{2}$), kagome lattice systems (ZnCu$_{3}$(OH)$_{6}$Cl$_{2}$) and hyperkagome lattice systems (Na$_{4}$Ir$_{3}$O$_{8}$). We note several outstanding difficulties with attempts to explain experimental results using existing theories. {\em These difficulties indicate that the field of QSLs is still wide open and immature and that important physics may still be missing in our present understanding of QSLs}.

   While keeping the article at an introductory level, we are not able to cover many important developments in the study of spin liquid states, and many technical details have been neglected, both theoretically and experimentally. For example, the important techniques of renormalization groups and conformal field theory are not addressed in this article. We also do not discuss in detail the many developments related to MPSs and/or PEPSs and the corresponding numerical DMRG technique, the understanding of spin systems with broken rotational symmetry following the discovery of the Kitaev state, and the spin liquid physics of $S>1/2$ systems.  The role of topology in spin liquid states is not touched upon except as it is relevant to examples of spin liquid states. These are rapidly evolving areas in which new discoveries are expected.

In the following section, we outline a few other topics that are neglected in this article but either have played important historical roles in the development of the field of QSLs or shed light on future research:

   {\em Quantum dimer models:} Quantum dimer models (QDMs) are a class of models defined in the Hilbert space of nearest neighbor valence bond (or dimer) coverings
   over a lattice instead of the spin Hilbert space \cite{Rokhsar88}. QDMs can be obtained in certain large-$N$ limits of $SU(N)$ or $Sp(N)$ antiferromagnets \cite{Read89b} and provide a simplified description of RVB states.
   This simplification allows researcher to proceed further in  analytical treatments because of the close relations that arise to classical dimer problems, Ising models and Z$_2$ gauge theory \cite{Kasteleyn61,Fisher61,Kasteleyn63,Moessner02,Misguich02a,Moessner03}.
   However, by construction, QMDs focus on the dynamics in the spin-singlet subspace and ignore spin-triplet excitations.
   Therefore, they are not directly relevant to spin systems in which the magnetic excitations are gapless.

   An advantage of QDMs is that some QDMs are exactly solvable \cite{Misguich02a,Yao12}. Thus, many issues related to QSLs that are difficult to address, such as spinon deconfinement, $Z_2$ vortices and topological order, can be addressed explicitly in QDMs.
   Interestingly, some spin-$1/2$ Hamiltonians give rise to sRVB ground states defined in the dimer Hilbert space when the relationship between the spin and dimer configurations is properly chosen \cite{Fujimoto05,Seidel09,Cano10}.
   Readers who are interested in further details on QDMs can refer to Chapter 5.5 in reference \cite{BookDiep} and Chapter 17 in reference \cite{BookLacroix}.

   {\em Chiral spin liquids:} QSL states that break the parity (P) and time-reversal (T) symmetries while conserving the spin rotational symmetry have been proposed by Kalmeyer and Laughlin \cite{Kalmeyer87,Kalmeyer89}. These states are called chiral spin liquids.

   Kalmeyer and Laughlin proposed that some frustrated Heisenberg antiferromagnets in 2D can be described by bosonic fractional quantum Hall wavefunctions. Soon afterward, Wen, Wilczek and Zee \cite{Wen89} introduced a generic method of describing chiral spin liquids.  They suggested that chiral spin states can be characterized in terms of the spin chirality $E_{123}=\vec{S}_1\cdot(\vec{S}_2\times\vec{S}_3)$,
  defined for three different spins, $\vec{S}_1$, $\vec{S}_2$ and $\vec{S}_3$. The expectation value of the spin chirality in fermionic RVB theory is given by
   $\langle E_{123}\rangle={1\over 2}\rm{Im} \langle\chi_{12}\chi_{23}\chi_{31}\rangle$,
   where the $\chi_{ij}$ are the short-range order parameters defined in Eq.~\eqref{MFparameter}.

  Exactly solvable Hamiltonians hosting both gapful chiral spin liquid states \cite{Laughlin89,Yao07,Schroeter07,Thomale09} and gapless chiral spin liquids \cite{Chua11} have been found.
  There is also numerical evidence for chiral spin liquids on some 2D frustrated lattices \cite{Sorella03,nielsen13,bauer13,He14,He14a,Gong14,Gong15,Zhu15}.
  It has been suggested that the statistics of spinons in these chiral spin liquid states can be non-Abelian; see, e.g., \cite{Yao07,Greiter09}.

   {\em Characterizing spin liquid states numerically:} Because of rapid advancements in the power of numerical approaches to spin models, the characterization of spin liquid states for specific spin models from numerical data has become a rapidly evolving field. In addition to the  MPS and/or PEPS approach and the corresponding numerical DMRG technique,  Tang and Sandvik developed a quantum Monte Carlo method of characterizing spinon size and confinement length in quantum spin systems, which allows the spinon confinement-deconfinement issue to be studied numerically \cite{TangSandvik13}. Another important achievement is the use of entanglement entropy to characterize  QSL states. Readers may consult reference \cite{Grover13} for a brief review.

   To conclude, the field of QSLs is still wide open, both theoretically and experimentally. The major difficulty in understanding QSLs is that they are intrinsically strongly correlated systems, for which no perturbative approach is available. Theorists have been using all of the available tools as well as inventing new theoretical tools to understand QSLs with the hope that novel emerging phenomena not covered by perturbative approaches can be uncovered. Thus far, there have been a few successes, and new experimental discoveries and theoretical ideas are rapidly emerging. However, a basic mathematical framework that can be used to understand QSLs systematically is still lacking. We expect that more new physics will be discovered in QSLs, posing a challenge to both theorists and experimentalists to construct a basic framework for the understanding of QSLs.

\begin{acknowledgements}
  Y.Z. and T.K.N. thank Patrick A. Lee, Zheng-Xin Liu, Naoto Nagaosa, Shaojin Qin, Zhaobin Su, Hong-Hao Tu, Tao Xiang, Xiao-Gang Wen, Zheng-Yu Weng, Guang-Ming Zhang, and Fu-Chun Zhang
  for their close collaboration on related issues over the years.
  K.K. is grateful to K. Miyagawa, Y. Shimizu, Y. Kurosaki, H. Hashiba, H. Kobashi, H. Kasahara, T. Furukawa, M. Maesato, G. Saito, F. Pratt, and M. Poirier for their collaboration on the topic of spin liquids.
  We benefited greatly from discussions with our colleagues Yan Chen, Yin-Chen He, Bruce Normand, Fa Wang, Cenke Xu, and Hong Yao.
  Y.Z. is supported by the National Key R\&D Program of China (No.2016YFA0300202), the National Basic Research Program of China under Grant No. 2014CB921201 and by the National Natural Science
  Foundation of China under Grant No. 11374256. He also wishes to acknowledge the hospitality of the Max Planck Institute for the Physics of Complex Systems in Dresden, where this review article was finalized.
  K.K. is partially supported by JSPS KAKENHI under Grant Nos. 20110002, 25220709, and 24654101, by the US National Science Foundation under Grant No. PHYS-1066293, and by the hospitality of the Aspen Center for Physics.
  T.K.N. acknowledges support from HKRGC through Grant No. 603913.

\end{acknowledgements}

\appendix
\section{Path integral for a single spin}
  We consider the path integral for a single spin $\mathbf{S}$ in a magnetic field $\mathbf{B}$ ($H=\mathbf{S}\cdot\mathbf{B}$) in the coherent state representation. Spin coherent states are defined as
  \[
     \mathbf{\hat{S}}|\mathbf{n}\rangle=S\mathbf{n}|\mathbf{n}\rangle,  \]
  where $\mathbf{\hat{S}}$ is the spin operator. The path integral can be derived by using the identity operator
  \begin{subequations}
  \label{app1}
  \begin{equation}
   \mathbf{I}=\left({2S+1\over4\pi}\right)\int d^3n\delta(\mathbf{n}^2-1)|\mathbf{n}\rangle\langle\mathbf{n}|=\int D{\mathbf{n}}|\mathbf{n}\rangle\langle\mathbf{n}|
   \end{equation}
   and the corresponding inner product
  \begin{equation}
   \langle\mathbf{n}_1|\mathbf{n}_2\rangle=e^{iS\Phi(\mathbf{n}_1,\mathbf{n}_2,\mathbf{n}_0)}\left({1+\mathbf{n}_1\cdot\mathbf{n}_2\over2}\right)^S,
   \end{equation}
   \end{subequations}
   where $\mathbf{n}_0$ is a fixed unit vector and is usually chosen to be $\mathbf{n}_0=\hat{z}$, $\Phi(\mathbf{n}_1,\mathbf{n}_2,\mathbf{n}_0)$ is the area of the spherical triangle with vertices $\mathbf{n}_1$, $\mathbf{n}_2$, and $\mathbf{n}_0$, and $S\Phi$ is the Berry's phase acquired by a particle traveling through a loop formed by the edges of the spherical triangle.

   The partition function $Z=e^{-\beta H}$ can be written as a path integral using the standard procedure:
   \begin{eqnarray}
   \label{app2}
   Z & = & \lim_{N_t\rightarrow\infty, \delta t\rightarrow 0}\left(e^{-\delta tH}\right)^{N_t}  \\ \nonumber
     & = & \lim_{N_t\rightarrow\infty, \delta t\rightarrow 0}\left(\Pi_{j=1}^{N_t}\int D\mathbf{n}_j\right)\left(\Pi_{j=1}^{N_t}
     \langle\mathbf{n}_j|e^{-i\delta tH}|\mathbf{n}_{j+1}\rangle\right),
     \end{eqnarray}
     with the periodic boundary condition $|\mathbf{n}(0)\rangle=|\mathbf{n}(\beta)\rangle$.

     In the limit $\delta t\rightarrow0$, we may approximate
     \begin{eqnarray}
   \label{app3}
   \langle\mathbf{n}_j|e^{-i\delta tH}|\mathbf{n}_{j+1}\rangle & \sim & \langle\mathbf{n}_j|\mathbf{n}_{j+1}\rangle-\delta t\langle\mathbf{n}_j|H|\mathbf{n}_{j+1}\rangle  \\ \nonumber
   & \sim & \langle\mathbf{n}_j|\mathbf{n}_{j+1}\rangle(1-\delta t {\langle\mathbf{n}_j|H|\mathbf{n}_{j+1}\rangle\over\langle\mathbf{n}_j|\mathbf{n}_{j+1}\rangle})  \\ \nonumber
   & \sim & e^{iS\Phi(\mathbf{n}_j,\mathbf{n}_{j+1},\mathbf{n}_0)}({1+\mathbf{n}_j\cdot\mathbf{n}_{j+1}\over2})^S  \\ \nonumber
   & & \times(1-\delta t S\mathbf{B}\cdot\mathbf{n}_t),
     \end{eqnarray}
  which is valid to the first order in $\delta t$. In deriving the last equality in Eq.~\eqref{app3}, we have made use of the result $\langle\mathbf{n}|\hat{\mathbf{S}}=\langle\mathbf{n}|\mathbf{n}$. Furthermore, we note that
  \begin{eqnarray}
  \label{app4}
  ({1+\mathbf{n}_j\cdot\mathbf{n}_{j+1}\over2})^S & \sim & e^{S\ln(1+{\delta t\over2}\mathbf{n}(t)\cdot\partial_t\mathbf{n}(t))_{t=t_j}}  \\ \nonumber
  & \sim & e^{S\delta t\partial_t[\mathbf{n}(t)]^2}=e^{(0)}
  \end{eqnarray}
  to leading order in $\delta t$. Therefore,
  \begin{eqnarray}
  \label{app5}
   \langle\mathbf{n}_j|e^{-i\delta tH}|\mathbf{n}_{j+1}\rangle & \sim & e^{iS\Phi(\mathbf{n}_j,\mathbf{n}_{j+1},\mathbf{n}_0)-\delta t S\mathbf{B}\cdot\mathbf{n}_t}
   \end{eqnarray}
  and
  \begin{eqnarray}
   \label{app6}
   Z & \sim & \int\mathbf{D}\mathbf{n}(t)e^{iS\Omega(\mathbf{n}(t))-S\int^{\beta}_0 dt\mathbf{B}\cdot\mathbf{n}(t)},
     \end{eqnarray}
   where $\int\mathbf{D}\mathbf{n}(t)=\lim_{N_t\rightarrow\infty, \delta t\rightarrow 0}\left(\Pi_{j=1}^{N_t}\int D\mathbf{n}_j\right)$ and
   \[ \Omega(\mathbf{n}(t))=\sum_j\Phi(\mathbf{n}_j,\mathbf{n}_{j+1},\mathbf{n}_0)\]
   is the total area on the surface of the unit sphere covered by the (closed) path swept out by the spin $\mathbf{n}(t)$ from $t=0$ to $t=\beta$.

     The classical action of the system in real time is given by
     \begin{subequations}
     \label{app7}
   \begin{equation}
   S_{cl}=S\Omega(\mathbf{n}(t))-S\int^{T}_0 dt\mathbf{B}\cdot\mathbf{n}(t),
   \end{equation}
   and the classical equation of motion ${\delta S_{cl}\over\delta'\mathbf{n}(t)}=0$ leads to the Euler equation of motion
   \begin{equation}
   \mathbf{n}\times\left((\mathbf{n}\times\partial_t\mathbf{n})-\mathbf{B}\right)=0,
   \end{equation}
   \end{subequations}
   where we have used the result that a small variation $\delta\mathbf{n}$ leads to a change in $\Omega(C[\mathbf{n}])$ that is given by
   \[ \delta\Omega[\mathbf{n(t)})=\int_0^{\beta}dt\delta\mathbf{n}(t)\cdot(\mathbf{n}(t)\times\partial_t\mathbf{n}(t)).  \]

\bibliographystyle{apsrmp4-1}
\bibliography{refall}

\end{document}